\documentclass{aa}
\usepackage[varg]{txfonts}
\usepackage{graphicx}
\usepackage{lscape}
\usepackage[breaklinks]{hyperref}
\usepackage{siunitx}
\usepackage{orcidlink}
\usepackage{lscape}
\usepackage{subcaption}
\usepackage{listings}

\renewcommand{\linenumbers}{}

\newcommand{\ur}[1]{\,\mathrm{#1}} 
\newcommand{\abs}[1]{\lvert #1\rvert }

\definecolor{cobalt}{rgb}{0.06, 0.2, 0.65}
\hypersetup{
  colorlinks,
  citecolor=cobalt,
  linkcolor=cobalt,
  urlcolor=cobalt,
}

\definecolor{dkgreen}{rgb}{0,0.6,0}
\definecolor{gray}{rgb}{0.5,0.5,0.5}
\definecolor{mauve}{rgb}{0.58,0,0.82}
\begin{document}

\title{Solar twins in Gaia DR3 GSP-Spec}
\subtitle{I. Building a large catalog of Solar twins with ages}

\author{
    Daisuke Taniguchi\inst{\ref{inst:TMU},\ref{inst:NAOJ}}\fnmsep\thanks{The Tokyo Center For Excellence Project, Tokyo Metropolitan University. }\orcidlink{0000-0002-2861-4069}
    \and 
    Patrick de Laverny \inst{\ref{inst:OCA}}\orcidlink{0000-0002-2817-4104}
    \and 
    Alejandra Recio-Blanco\inst{\ref{inst:OCA}}\orcidlink{0000-0002-6550-7377}
    \and 
    Takuji Tsujimoto\inst{\ref{inst:NAOJ}}\orcidlink{0000-0002-9397-3658}
    \and 
    Pedro A. Palicio\inst{\ref{inst:OCA}}\orcidlink{0000-0002-7432-8709}
}

\institute{
    Department of Physics, Tokyo Metropolitan University, 1-1 Minami-Osawa, Hachioji, Tokyo 192-0397, Japan \\
    \email{d.taniguchi.astro@gmail.com}\label{inst:TMU}
    \and 
    National Astronomical Observatory of Japan, 2-21-1 Osawa, Mitaka, Tokyo 181-8588, Japan\label{inst:NAOJ}
    \and 
    Universit\'e C\^ote d'Azur, Observatoire de la C\^ote d'Azur, CNRS, Laboratoire Lagrange, Bd de l'Observatoire, CS 34229, 06304 Nice cedex 4, France\label{inst:OCA}
}

\date{Received ?? ; accepted ??}

\abstract
{
Solar twins, stars whose stellar parameters (effective temperature $T_{\mathrm{eff}}$, surface gravity $\log g$, and metallicity [M/H]) are very close to the Solar ones, offer a unique opportunity to investigate Galactic archaeology with very high accuracy and precision. 
However, most previous catalogs of Solar twins contain only a small number of objects (typically a few tens), and their selection functions are poorly characterized. 
}
{
We aim at building a large catalog of Solar twins from \textit{Gaia} DR3 GSP-Spec, providing stellar parameters including ages determined using a model-driven, rather than data-driven, method, together with a well-characterized selection function. 
}
{
Using stellar parameters from the \textit{Gaia} DR3 GSP-Spec catalog, we selected Solar-twin candidates whose parameters lie within ${\pm }200\ur{K}$ in $T_{\mathrm{eff}}$, ${\pm }0.2$ in $\log g$, and ${\pm }0.1\ur{dex}$ in [M/H] of the Solar values. 
Candidates unlikely to be genuine Solar twins were removed using \textit{Gaia} flags and photometric constraints. 
We determined accurate ages for individual twins with a Bayesian isochrone-projection method, considering three combinations of parameters: $T_{\mathrm{eff}}$, [M/H], and either $\log g$, $M_{G}$, or $M_{K_{\mathrm{s}}}$. 
We also constructed a mock catalog to characterize the properties and selection function of our observed sample. 
}
{
Our final GSP-Spec Solar-twin catalog contains $6{,}594$ stars. 
The mock catalog consisting of $75{,}588$ artificial twins well reproduces the main characteristics of the observed catalog, especially for ages determined with $M_{G}$ or $M_{K_{\mathrm{s}}}$. 
To demonstrate the usefulness of our catalog, we compared chemical abundances [X/Fe] with age. 
We statistically confirmed the age--[X/Fe] relations for several species (e.g., Al, Si, Ca, and Y), demonstrating that trends previously identified in small but very high-precision samples persist in a much larger, independent sample. 
}
{
Our study bridges small high-precision Solar-twin samples and large data-driven ones by providing a large sample with model-driven stellar parameters and a quantified selection function, enabling demographic studies of Solar twins. 
}

\keywords{
    Stars: solar-type -- 
    Stars: fundamental parameters -- 
    Hertzsprung-Russell and C-M diagrams -- 
    Stars: abundances -- 
    Stars: statistics -- 
    Galaxy: evolution
}

\titlerunning{Solar twins in Gaia DR3 GSP-Spec. I}

\maketitle

\section{Introduction}\label{Sec:Intro}

Solar twins are defined as stars with stellar parameters (i.e., effective temperature $T_{\mathrm{eff}}$, surface gravity $\log g$, and metallicity [M/H]) very close to those of the Sun~\citep{Hardorp1978,CayreldeStrobel1981}\footnote{The level of similarity required, and whether other parameters (e.g., age, rotational velocity, or Li abundance) are included, depends on the specific purpose of a given Solar-twin study. }. 
By performing differential analyses between stars with very similar stellar parameters---Solar twins in our case---one can achieve extremely high accuracy in the measurements of stellar parameters and chemical abundances because systematic biases associated with $T_{\mathrm{eff}}$ and $\log g$ are reduced~\citep{Bedell2014,Nissen2018}. 
Such very precise and accurate stellar parameters, sometimes combined with precise parallaxes, can also yield accurate stellar ages through isochrone modeling~\citep[e.g.,][]{Nissen2015,Spina2016a}, even though Solar twins are main-sequence stars, for which isochrones of different ages are closely spaced in the Hertzsprung--Russell (HR) diagram. 

By exploiting these advantages, Solar twins have been used, for example, to determine the Solar properties such as its color~\citep{Holmberg2006,Casagrande2018}, to study planet--host interactions through photospheric chemical abundances~\citep{Israelian2004,Israelian2009,Melendez2009}, to construct chemical clocks~\citep{TucciMaia2016,Casali2020}, and to investigate the chemodynamical evolution of the Milky Way~\citep{Jofre2017,Bedell2018,Tsujimoto2021}. 
Among these applications, the ability to determine accurate stellar ages is particularly valuable for Galactic archaeology, where ages are crucial for tracing the history of the Milky Way but are often difficult to obtain~\citep[e.g.,][]{Ness2019,Magrini2023}. 

There have been great efforts to identify Solar twins. 
In early studies, Solar twins in a narrow sense, i.e., stars whose stellar parameters are identical to the Solar values within the errors, were explored. 
In that era, only a few very similar Solar twins were spectroscopically confirmed: 18~Sco~\citep{PortodeMello1997,Soubiran2004}, HD~143436~\citep{King2005}, HD~98618~\citep{Melendez2006}, HIP~100963~\citep{Takeda2007}, and HIP~56948~\citep{Melendez2007,Takeda2009}; see also the review by \citet{CayreldeStrobel1996}. 
Later, motivated by these identifications and also by the growing interest in understanding planet--host interactions, several dedicated surveys of Solar twins were conducted, yielding samples of several tens of Solar twins~\citep[e.g.,][]{Takeda2007,Melendez2009}. 
In particular, archival spectral data obtained with the HARPS spectrograph~\citep{Mayor2003}, originally collected for exoplanet detection via radial-velocity monitoring, have been extensively used to derive high-precision chemical abundances and ages of Solar twins through differential analysis of high signal-to-noise-ratio (S/N) stacked spectra~\citep[e.g.,][]{Nissen2015,Martos2025}. 
Furthermore, recent massive spectroscopic surveys are capable of identifying thousands of Solar twins. 
Nevertheless, only a few such large samples of Solar twins have so far been published: using GALAH DR3 by \citet{Walsen2024} and \citet{Lehmann2025} and \textit{Gaia} DR3 Radial Velocity Spectrometer (RVS) data by \citet{Rampalli2024}. 
In the present contribution, we aim at building a large catalog of Solar twins based on non-data-driven stellar parameters, with strict filtering and a well-characterized selection function. 

To obtain the largest possible catalog of high-quality Solar-twin candidates, the most effective approach is to mine the catalogs published by large spectroscopic surveys, at the expense of some loss in precision compared to dedicated surveys targeting individual objects. 
The largest such catalog in terms of sample size (by more than one order of magnitude compared to ground-based surveys) is the homogeneous General Stellar Parametrizer from Spectroscopy (GSP-Spec) catalog collected from space by the ESA/\textit{Gaia} mission and published as part of the third \textit{Gaia} data release \citep[DR3,][]{GaiaOverview,GaiaDR3,GaiaGSPSpecDR3}. 
GSP-Spec is based on the analysis of ${\sim }5.6$ million \textit{Gaia}/RVS stellar spectra for which the main atmospheric parameters are provided, together with the mean enrichment in $\alpha $-elements with respect to iron ([$\alpha $/Fe], in dex) and individual chemical abundances of up to $13$ elements (including two ions for iron). 
With the magnitude limit of $G{\lesssim }12$ for Solar-type stars with best GSP-Spec parameterization, GSP-Spec Solar twins extend out to distances of ${\sim }300\ur{pc}$. 
This reachable distance is about three times larger than that of high-precision dedicated spectroscopic surveys, which typically reach ${\sim }100\ur{pc}$. 
Hence, the volume covered, and therefore the expected number of objects, is increased by a factor of ${\sim }30$. 

This paper is organized as follows. 
We first select thousands of Solar twins found in the GSP-Spec catalog~(Sect.~\ref{sec:Data}). 
Then, by comparing their stellar parameters and \textit{Gaia} parallaxes with theoretical isochrones, we determine their ages and initial masses in a homogeneous and consistent way~(Sect.~\ref{sec:age}), we validate the resulting ages~(Sect.~\ref{sec:validation}), and we place our work in the context of previous Solar-twin studies~(Sect.~\ref{sec:agecomp}). 
Finally, we take a brief look at several applications of our catalog to assess its scientific potential~(Sect.~\ref{sec:chem}).

\section{Solar-twin selection}\label{sec:Data}

\subsection{GSP-Spec sample of Solar-twin candidates}\label{ssec:gaiadata}

To look for robustly parametrized Solar twins within the GSP-Spec catalog,
we first considered the ${\sim }2$ million stars whose first $13$ GSP-Spec quality flags (\texttt{flags\_gspspec}) are all equal to zero, refer to the quality of the $T_{\mathrm{eff}}$, $\log g$, [M/H]\footnote{Within GSP-Spec, [M/H] is a proxy for [Fe/H] abundance~\citep{GaiaGSPSpecDR3,RecioBlanco2024}. }, and [$\alpha $/Fe] derivations. 
They mostly depend on the quality of the input RVS spectra, and ``\texttt{0}'' means ``Best Quality'' \cite[see, Sect.~8 of][for more details]{GaiaGSPSpecDR3}. 
The seventh digit of the quality flag indicates the uncertainty of the parameterization; ``\texttt{0}'' means that the statistical uncertainties on $T_{\mathrm{eff}}$, $\log g$, [M/H], and [$\alpha $/Fe] are less than $100\ur{K}$, $0.2$, $0.1\ur{dex}$, and $0.05\ur{dex}$, respectively. 
We moreover selected stars with a very high-quality GSP-Spec goodness-of-fit (\texttt{logchisq\_gspspec}), $\log \chi ^{2}<-3.2$, ensuring the high quality of the parameterization. 

We also imposed the following criteria to analyze only stars with good astrometric solutions and to remove possible binary stars: renormalized unit weight error (RUWE) is smaller than $1.4$, \texttt{astrometric\_params\_solved} is \texttt{0}, \texttt{duplicated\_source} is \texttt{False}, and \texttt{non\_single\_star} is \texttt{0}. 
We confirmed that the \texttt{fidelity\_v2} index introduced by \citet{Rybizki2022} is larger than $0.5$ for all the selected Solar-twin candidates, indicating that the quality of the astrometric solution is good. 

We then calibrated $\log g$ and [M/H] for these stars, by following the relations as a function of $T_{\mathrm{eff}}$ provided in \citet{RecioBlanco2024}. 
The applied corrections imply changes of ${\sim }0.1\text{--}0.2$ for $\log g$ and ${<}0.01\ur{dex}$ for [M/H]. 
We further calibrated the stellar parameters so that they are well differential against the Sun (see Appendix~\ref{app:zeropoints})\footnote{We note that, throughout this paper, we shifted the zero-points in literature $T_{\mathrm{eff}}$ and $\log g$ values to our adopted scale, $T_{\mathrm{eff},\odot }=5777\ur{K}$ and $\log g_{\odot }=4.44$, if the reference Solar values are explicitly mentioned. }. 
In short, for $T_{\mathrm{eff}}$ and [M/H], we added offsets of $1\ur{K}$ and $0.062\ur{dex}$, respectively. 
For $\log g$, we applied Equation~\ref{Eq:logg_y}. 
Hereafter, we call the stellar parameters after the calibration described in Appendix~\ref{app:zeropoints} the ``calibrated'' parameters and use them unless otherwise specified, while we refer to the parameters before this second calibration as the ``original'' ones, though they have already been subjected to the first-stage calibration following \citet{RecioBlanco2024}. 

With the zero-point biases reduced, the $7{,}918$ Solar-twin candidates were defined as having their $T_{\mathrm{eff}}$, $\log g$, and [M/H] within ${\pm }200\ur{K}$, ${\pm }0.2$, and ${\pm }0.1\ur{dex}$ around the Solar values, respectively. 
In the following, we adopted the standard Solar values: $T_{\mathrm{eff},\odot }=5777\ur{K}$, $\log g_{\odot }=4.44$, and $\mathrm{[M/H]}_{\odot }=0.0\ur{dex}$. 
The typical parameter uncertainties are much smaller than the above-mentioned filtering: for these $7{,}918$ candidates, their median uncertainties are actually equal to $50\ur{K}$, $0.05$\footnote{During the calibration process described in Appendix~\ref{app:zeropoints}, the errors in $\log g$ shrank. The adopted broken-line relation implies that the largest shrinkage occurs in stars with higher $\log g$ (original $\log g\gtrsim 4.5$, or equivalently, calibrated $\log g\gtrsim 4.48$), where they were reduced to $a_{2}=0.133$ times their original error values. }, and $0.03\ur{dex}$ for $T_{\mathrm{eff}}$, $\log g$, and [M/H], respectively. 

\begin{figure*}
\centering 
\includegraphics[width=18cm]{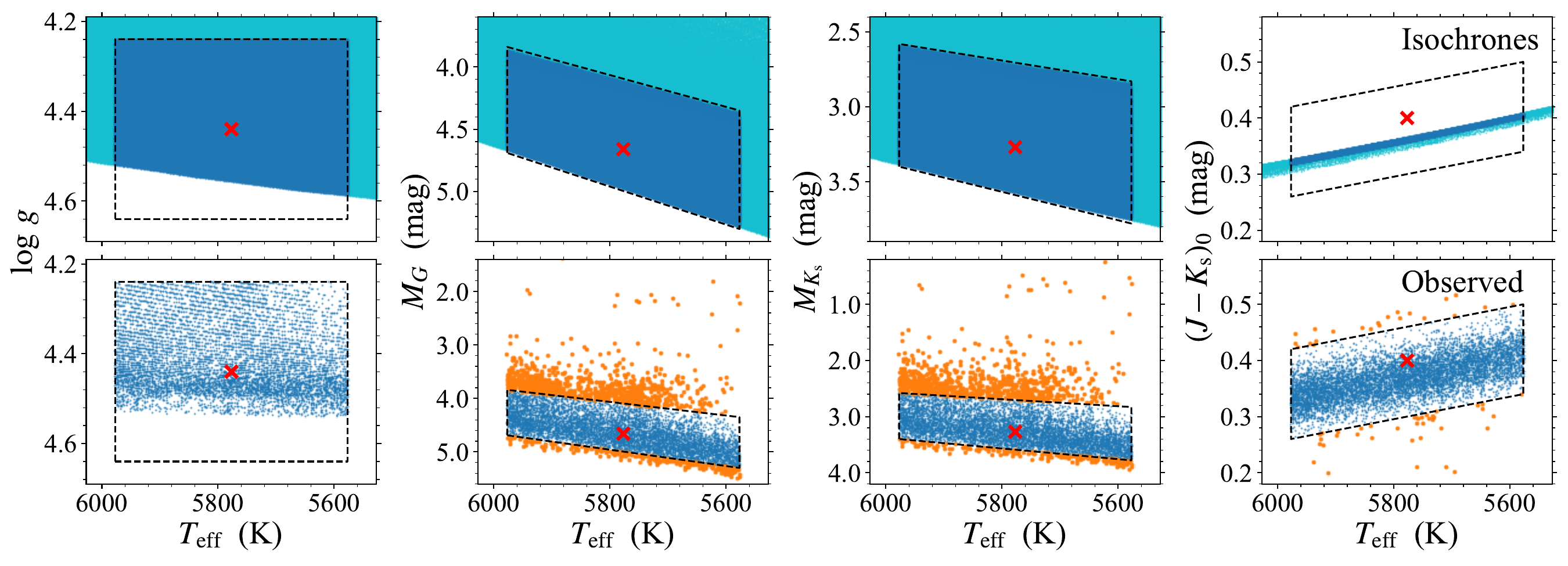}
\caption{Some parameters of theoretical isochrones (top) and observed Solar-twin candidates (bottom). 
The horizontal axes in all the panels represent $T_{\mathrm{eff}}$, while the vertical axes represent some other parameters ($\log g$, $M_{G}$, $M_{K_{\mathrm{s}}}$, and $(J-K_{\mathrm{s}})_{0}$). 
Cyan dots in the upper panels show the isochrone grid points satisfying $-0.1<\mathrm{[M/H]}_{\mathrm{curr}}<+0.1\ur{dex}$. 
Blue dots in the upper panels show a subset of them having $T_{\mathrm{eff}}$ and $\log g$ between $\pm 200\ur{K}$ and $\pm 0.2$ around the Solar values, which is the selection criterion of our Solar-twin candidates in Sect.~\ref{ssec:gaiadata} and is represented as a black dashed rectangle in the left panels. 
Black dashed trapezoids in the rest of the panels show the selection criteria for good Solar twins in Sect.~\ref{ssec:flag}. 
Dots in the lower panels show the parameters of the $7{,}775$ observed Solar-twin candidates with good distances and 2MASS data. 
Orange dots in the lower panels show a subset of them that do not satisfy the selection criteria (i.e., outside the black trapezoids). 
Red crosses show the reference Solar values provided in Sect.~\ref{ssec:SolarAge} and $(J-K_{\mathrm{s}})_{0}=0.40$~\citep{Willmer2018}. 
We note that several outliers of observed Solar-twin candidates fall outside the plotted range and are not shown in the figures. }
\label{fig:flag}
\end{figure*}

\subsection{Interstellar extinctions}\label{ssec:ISM}

Since the absolute magnitudes of all these Solar twins are necessary for their accurate age determination, we computed their interstellar extinctions in the \textit{Gaia} \textit{G}-band by comparing their \textit{Gaia} observed colors $(G_{\mathrm{BP}}-G_{\mathrm{RP}})$ to theoretical ones estimated from the \citet{Casagrande2021} relation between atmospheric parameters and color. 
For that relation, we adopted the ``original'' spectroscopic atmospheric parameters. 
These computations were performed for $1{,}000$ Monte-Carlo realizations, propagating the uncertainties on each atmospheric parameter and \textit{Gaia} magnitudes in the different bands. 
The adopted extinctions ($A_{G}$) are the medians of the Monte-Carlo distributions, and their associated uncertainties are half of the difference between the 84th and 16th percentiles, as it corresponds to a $1\sigma $ uncertainty for a normal distribution. 
The extinctions to our Solar twins are small, with the median and 95th percentile of $0.02$ and $0.15\ur{mag}$, respectively, in the \textit{G} band. 
We refer to de~Laverny et al. (in preparation) for more details on these extinction calculations and to \citet{Barbillon2025} for their validation in a Galactic and interstellar-medium context.

\subsection{Absolute magnitudes and de-reddened colors}\label{ssec:absmag}

We retrieved the \textit{Gaia} DR3 photometry (\texttt{phot\_g\_mean\_mag}, \texttt{phot\_bp\_mean\_mag}, and \texttt{phot\_rp\_mean\_mag}) of the Solar-twin candidates from the \textit{Gaia} Archive. 
Their 2MASS \textit{JHK}$_{\mathrm{s}}$ photometry~\citep{Cutri2003,Skrutskie2006} was extracted using the cross-matched result in the \texttt{gaiadr3.tmass\_psc\_xsc\_best\_neighbour} catalog. 
We then converted the \textit{G}-band extinction $A_{G}$ determined in Sect.~\ref{ssec:ISM} to the extinctions in the \textit{Gaia} EDR3 and 2MASS passbands using the \textsc{dustapprox} code~\citep{dustapprox}. 
Then, combining these extinctions with the distances from \citet{BailerJones2021}, we calculated the absolute magnitudes and de-reddened colors of the Solar-twin candidates.

\subsection{Flagging with absolute magnitudes}\label{ssec:flag}

In order to further clean our sample, we made use of the absolute magnitudes and de-reddened colors computed in Sect.~\ref{ssec:absmag} together with PARSEC isochrones retrieved in Sect.~\ref{ssec:PARSEC} and interpolated in Appendix~\ref{app:PARSECinterp}. 
From the $7{,}918$ Solar-twin candidates, we removed stars falling under any of the following conditions from the subsequent analysis: 
\begin{itemize}
\item Without a 2MASS entry in the \textit{JK}$_{\mathrm{s}}$ bands ($41$ stars)
\item Whose 2MASS counterpart is not uniquely matched, i.e., it also matched to another Solar-twin candidate ($2$ stars)
\item With the \texttt{Qflg} in the 2MASS \textit{JK}$_{\mathrm{s}}$ bands of \texttt{E}, \texttt{F}, \texttt{U}, \texttt{X}, or \texttt{Z} (i.e., poor photometry; $100$ stars). 
\end{itemize}
These conditions left $7{,}775$ stars with good distances and 2MASS data. 

We found that some of the Solar-twin candidates have absolute magnitudes and/or de-reddened colors that are far from those of the Sun. 
For example, some stars have absolute \textit{G}- and \textit{K}$_{\mathrm{s}}$-band magnitudes that are systematically higher (or lower) than the Solar values~(orange dots in lower middle panels of Fig.~\ref{fig:flag}). 
Such systematic deviations, regardless of wavelength, indicate either poor parallax, poor $\log g$ determination, and/or binarity. 
As another example, some other stars have their de-reddened $J-K_{\mathrm{s}}$ colors far from the Solar value~(orange dots in lower right panel of Fig.~\ref{fig:flag}). 
Regardless of the reason for such deviation, e.g., poorer image FWHM in 2MASS than in \textit{Gaia}, such stars should be treated with caution, given that we will use the 2MASS \textit{K}$_{\mathrm{s}}$ magnitude in one of our age-determination methods. 

To exclude stars having unexpected absolute magnitudes and/or de-reddened colors, we plotted $\log g$, $M_{G}$, $M_{K_{\mathrm{s}}}$, and $(J-K_{\mathrm{s}})_{0}$ against $T_{\mathrm{eff}}$ in the lower panels of Fig.~\ref{fig:flag}. 
In order to select Solar-twin candidates having good absolute magnitudes, we first gathered isochrone points having current metallicity $\mathrm{[M/H]}_{\mathrm{curr}}$, $T_{\mathrm{eff}}$, and $\log g$ within $\pm 0.1\ur{dex}$, $\pm 200\ur{K}$, and $\pm 0.2$ around the Solar values~(blue dots in the upper panels), which is equivalent to our selection criterion for Solar-twin candidates in Sect.~\ref{ssec:gaiadata}. 
Then, in the $T_{\mathrm{eff}}$ vs $M_{G}$ and $T_{\mathrm{eff}}$ vs $M_{K_{\mathrm{s}}}$ diagrams, we drew trapezoids that provide a near-minimal enclosure of the selected sample~(black-dashed trapezoids). 
Finally, the two trapezoids were used to remove $1{,}129$ suspicious candidates. 

In contrast to the absolute magnitudes of theoretical Solar twins, which show a spread of ${\sim }1\ur{mag}$, their $(J-K_{\mathrm{s}})_{0}$ colors follow a very tight $T_{\mathrm{eff}}$--color relation with a spread of ${\sim }0.01\ur{mag}$~(upper right panel of Fig.~\ref{fig:flag}). 
Hence, the observed color spread of Solar-twin candidates is dominated by the photometric error of ${\sim }0.03\ur{mag}$, which is larger than the intrinsic absolute-magnitude spreads. 
We also found that the Solar color of $0.40$~\citep{Willmer2018} deviates from the isochrones by $0.04\ur{mag}$. 
As such, it is practically impossible to select good Solar-twin candidates using the theoretical $(J-K_{\mathrm{s}})_{0}$ colors. 
Instead, we removed $52$ candidates by excluding stars that clearly lie outside the main locus in the diagram, using a rectangular selection box of $0.16\ur{mag}$ width chosen based on visual inspection. 

With all the conditions listed in this section considered, there remain $6{,}594$ good Solar twins suitable for the subsequent analysis.

\subsection{Orbital parameters}\label{ssec:kinematics}

Since orbital parameters of Solar twins are useful for discussing their origins, we computed their dynamical properties as in \citet{GaiaDR3Cartography} and \citet{Palicio2023}. 
In particular, we will consider the guiding radius, by numerically solving $\abs{L_{z}} = R_{\mathrm{g}} V_{\mathrm{circ}}(R_{\mathrm{g}})$, where $L_{z}$ is the angular momentum and $V_{\mathrm{circ}}(R)$ is the circular velocity at radii $R$. 
For these orbit determinations, we adopted the \textit{Gaia} coordinates, geometric distances from \citet{BailerJones2021}, \textit{Gaia} DR3 radial velocities \citep[$V_{\mathrm{rad}}$;][]{Katz2023}, and \textit{Gaia} EDR3 proper motions, together with a local standard of rest (LSR) velocity at the Sun's position equal to $V_{\mathrm{LSR}}^{}=238.5\ur{\si{km.s^{-1}}}$~\citep{Schonrich2010}, and a Galactic center distance of $R_{0}=8.249\ur{kpc}$~\citep{Gravity2020}. 
The determined orbital parameters are used in our companion papers~\citep[Papers~II and III;][]{Paper2,Paper3}.

\section{Age determination of the Solar twins}\label{sec:age}

We computed the ages of Solar twins having precise \textit{Gaia} stellar parameters
by adopting the so-called isochrone method, which has been widely used for dating stars including Solar twins~\citep[e.g.,][]{Jorgensen2005,Ramirez2014,Spina2018}. 
Basically, our methodology consists of an adaptation of those of \citet{Zwitter2010} and \citet{Kordopatis2023}, with optimization for Solar-twin ages. 
As we detail in this section, we determined the ages using three combinations of parameters: $T_{\mathrm{eff}}$, [M/H], and either $\log g$, \textit{G}-band absolute magnitude, or \textit{K}$_{\mathrm{s}}$-band absolute magnitude.

\subsection{PARSEC isochrone library}\label{ssec:PARSEC}

As the isochrone library used in the subsequent analysis and in Sect.~\ref{ssec:flag}, we retrieved PARSEC isochrones version 1.2S~\citep{Bressan2012,Chen2015} using the CMD~3.8 web interface\footnote{\url{https://stev.oapd.inaf.it/cgi-bin/cmd}}. 
We used the ``YBC+new Vega'' bolometric corrections~\citep{Chen2019,Bohlin2020} to obtain photometric magnitudes in the 2MASS~\citep{Cohen2003} and \textit{Gaia}~\citep{Riello2021} wavelength passbands. 

For the ages of the isochrones, we considered two grids of ages: a logarithmically evenly spaced age grid between $10^{8.0}\text{--}10^{8.95}\ur{yr}$ with $0.05\ur{dex}$ increments and a linearly evenly spaced age grid between $1\text{--}20\ur{Gyr}$ with $0.1\ur{Gyr}$ increments. 
The former was considered to densely populate the HR diagram with isochrone grids, given that isochrones of young massive stars move quickly on the HR diagram. 
The latter was considered to determine the ages and accompanying errors as precisely as ${\sim }0.1\ur{Gyr}$. 
The two grids are almost continuously connected at $1\ur{Gyr}$; in other words, the step size of the former grid at $1\ur{Gyr}$ of $0.05\ln 10\sim 0.1\ur{Gyr}$ is very similar to the interval in the latter grid. 

For the initial metallicity, we considered the range from $-0.5$ to $+0.5\ur{dex}$ with increments of $0.005\ur{dex}$\footnote{The zero-point in the initial-metallicity scale is assumed to be $(Z/X)_{\odot }=0.0207$ as suggested by \citet{Bressan2012}, who compiled literature Solar abundances from \citet{Grevesse1998}, \citet{Caffau2011}, and references therein. }. 
The interval of the initial-metallicity grid is equal to the minimum error in [M/H] of our GSP-Spec Solar-twin sample. 

In total, we retrieved $(20+191)\times 201=42411$ isochrone tracks. 
We only considered isochrone grid points with labels ``\texttt{0}'' (pre main sequence), ``\texttt{1}'' (main sequence), or ``\texttt{2}'' (subgiant branch) to reduce the data size of the isochrone library. 

In the subsequent analysis, we used the following quantities tabulated in the isochrones: age $\tau $, initial mass $M_{\mathrm{ini}}$, current surface metallicity $\mathrm{[M/H]}_{\mathrm{curr}}$\footnote{We used $Y_{\mathrm{S}}=0.24787$ and $Z_{\mathrm{S}}=0.01597$~\citep{Bressan2012}, and hence $(Z/X)_{\mathrm{S}}=0.02169$ to set the zero-point in the current metallicity $\mathrm{[M/H]}_{\mathrm{curr}}$, which is different from the zero-point in the initial metallicity $\mathrm{[M/H]}_{\mathrm{ini}}$, $(Z/X)_{\odot }=0.0207$. The use of $(Z/X)_{\mathrm{S}}$ as the zero-point is intended to ensure that the Solar $T_{\mathrm{eff}}$ and $\log g$ can be reproduced with the Solar-age, Solar-$\mathrm{[M/H]}_{\mathrm{curr}}$ model. }, $T_{\mathrm{eff}}$, $\log g$, and absolute magnitudes in the \textit{Gaia} EDR3 \textit{G}-band and 2MASS \textit{JHK}$_{\mathrm{s}}$-bands.

\subsection{Bayesian age determination}\label{ssec:ageprocedure}

As in several previous studies, we determined the age $\tau $, initial mass $M_{\mathrm{ini}}$, and initial metallicity $\mathrm{[M/H]}_{\mathrm{ini}}$ using a Bayesian isochrone projection method, weighting individual isochrone points. 
We used three sets of three observed quantities for the projection: $T_{\mathrm{eff}}$, $\mathrm{[M/H]}_{\mathrm{curr}}$\footnote{For the observed $\mathrm{[M/H]}_{\mathrm{curr}}$ and its error, we corrected for the effect of [$\alpha $/Fe] on isochrones using the equation in \citet{Salaris1993}. }, and one of the following three quantities: $\log g$, \textit{G}-band absolute magnitude, or \textit{K}$_{\mathrm{s}}$-band absolute magnitude. 
The use of $\mathrm{[M/H]}_{\mathrm{curr}}$, rather than $\mathrm{[M/H]}_{\mathrm{ini}}$, is necessary because atomic diffusion changes the surface metallicity of a star from its initial metallicity, and hence changes the determined ages with the amount depending on the evolutionary phase~\citep{ChristensenDalsgaard1996,Nissen2016,Dotter2017}. 
Atomic diffusion is indeed implemented in the PARSEC model used here, and the surface metallicity (i.e., common logarithm of the surface $Z/X$ plus an offset) decreased by ${\sim }0.1\ur{dex}$ almost steadily during the main-sequence lifetime of a $1M_{\odot }$ model. 

We calculated the weight $w_{i}$ of each isochrone grid point $i$ as 
\begin{equation}
w_{i}\equiv p_{i}^{\tau }p_{i}^{M_{\mathrm{ini}}}p_{i}^{\mathrm{[M/H]}_{\mathrm{ini}}}\exp \left(-\sum _{k}\frac{(\theta _{i,k}-\theta _{k}^{\mathrm{obs}})^{2}}{2{\sigma _{\theta _{k}^{\mathrm{obs}}}}^{2}}\right)\text{,}
\end{equation}
where $k$ labels the three observed quantities, $p_{i}^{\tau }$, $p_{i}^{M_{\mathrm{ini}}}$, and $p_{i}^{\mathrm{[M/H]}_{\mathrm{ini}}}$ represent factors corresponding to priors, $\theta _{i.k}$ represents the value for the quantity $k$ in the isochrone library, and $\theta _{k}^{\mathrm{obs}}$ and $\sigma _{\theta _{k}^{\mathrm{obs}}}$ represent the point estimate and accompanying error in the observed quantity $k$. 
Since Solar twins are located in a narrow region of stellar parameter space, we assumed flat priors in $\tau $, $M_{\mathrm{ini}}$, and $\mathrm{[M/H]}_{\mathrm{ini}}$. 
Thus, the factors $p_{i}^{\tau }$, $p_{i}^{M_{\mathrm{ini}}}$, and $p_{i}^{\mathrm{[M/H]}_{\mathrm{ini}}}$ were set to the increments of the corresponding values in the isochrones. 
For example, since we retrieved an isochrone library sampled on a linear grid in $\mathrm{[M/H]}_{\mathrm{ini}}$, the value of $p_{i}^{\mathrm{[M/H]}_{\mathrm{ini}}}$ is constant across all isochrone points. 
The calculated weights assigned to individual isochrone points represent the discrete posterior distribution of the output quantities ($\tau $, $M_{\mathrm{ini}}$, and $\mathrm{[M/H]}_{\mathrm{ini}}$). 

We note that in cases where the ``calibrated'' $\log g$ precision is smaller than $0.005$ ($46$ stars), we assigned an uncertainty of $0.005$ in calculating the weights, which corresponds to the minimum allowed error value in the ``original'' GSP-Spec catalog (i.e., those before the calibration in Appendix~\ref{app:zeropoints}), given that the \texttt{logg\_gspspec} values are rounded to two decimal places. 
The maximum $\log g$ step in our isochrone grid is ${\sim }0.003$~(Appendix~\ref{app:PARSECinterp}), which is smaller than the minimum assigned $\log g$ uncertainty of $0.005$, and is therefore sufficiently fine to accurately project $\log g$ onto $\tau $ and $M_{\mathrm{ini}}$. 
Similarly, we assigned an uncertainty of $0.005\ur{dex}$ to [M/H] values with reported uncertainties smaller than $0.005\ur{dex}$ ($4$ stars)\footnote{In our catalog of Solar twins, similar to the $\log g$ error, the errors in [M/H] and [$\alpha $/Fe] are quantized in $0.005$ increments. Thus, there are $4$ and $12$ stars with the [M/H] and [$\alpha $/Fe] errors of $0.000\ur{dex}$, respectively. Since we used the [M/H] value with the effect of [$\alpha $/Fe] on isochrones corrected, the error in the adopted (or corrected) ``[M/H]'' values are not in increments of $0.005$. }. 
Since the maximum step sizes of our isochrone grid are ${\sim }4\ur{K}$, $0.01\ur{mag}$, and $0.01\ur{mag}$ for $T_{\mathrm{eff}}$, $M_{G}$, and $M_{K_{\mathrm{s}}}$, respectively~(see, Appendix~\ref{app:PARSECinterp}), we adopted these values as uncertainties when the quoted observational errors were smaller ($0$, $2{,}745$, and $0$ stars, respectively). 

To obtain the point estimates of the output quantities, we generated $10^{6}$ Monte Carlo samples with replacement, where each sample $i$ was selected with a probability proportional to its weight $w_{i}$. 
Since the isochrone library is discrete, the Monte Carlo samples drawn from them are also discrete. 
To recover continuity, we added Gaussian noise to each sample with standard deviations of $0.1\ur{Gyr}$, $0.001\,M_{\odot }$, and $0.005\ur{dex}$ for $\tau $, $M_{\mathrm{ini}}$, and $\mathrm{[M/H]}_{\mathrm{ini}}$, respectively, reflecting typical isochrone increments. 
We then adopted the $50.0$, $15.8$, and $84.2$ percentiles as the median and $1\sigma $ uncertainties.

\section{Validation of the age determination}\label{sec:validation}

We publish the full catalog of Solar twins~(Table~\ref{table:catalog}) at CDS. 
Column descriptions are provided in Table~\ref{table:columndesc}. 
In this section, we validate the determined ages in two ways: by testing whether we recover the age of the Sun~(Sect.~\ref{ssec:SolarAge}) and by comparing our observed catalog to a mock Solar-twin sample~(Sect.~\ref{ssec:mock}). 
We also take a look at the determined age and $M_{\mathrm{ini}}$~(Sect.~\ref{ssec:plotparams}). 
We then discuss which of the three age estimates, i.e., those determined from $\log g$, $M_{G}$, and $M_{K_{\mathrm{s}}}$, should be preferred in practice~(Sect.~\ref{ssec:comp_threeage}).

\subsection{Age of the Sun}\label{ssec:SolarAge}

The age of the Sun has been measured precisely using both helioseismology~\citep{Dziembowski1999,Bonanno2002,Houdek2011} and radiometric dating of meteorites~\citep{Patterson1956,Amelin2002,Bouvier2010}. 
These studies concluded that the age of the Sun (or the Solar system) is ${\sim }4.5\text{--}4.6\ur{Gyr}$. 
Here, we apply our age-determination procedure to the Sun and check whether we can recover this age. 
We note that PARSEC isochrones version 1.2S were calibrated to reproduce Solar stellar parameters at an age of $4.593\ur{Gyr}$~\citep[Table~3 of][]{Bressan2012}, and thus our procedure should accurately recover the Solar age if our implementation is correct. 

As inputs for the Solar parameters, we adopted $M_{G,\odot }=4.66$~(see, \citealp{Creevey2023}, which adopt the Solar bolometric absolute magnitude $M_{\mathrm{bol},\odot }=4.74$ from \citealp{Prsa2016}) and $M_{K_{\mathrm{s}},\odot }=3.27$~\citep{Willmer2018}. 
We also used reference Solar values given in Sect.~\ref{ssec:gaiadata}, namely, $T_{\mathrm{eff},\odot }=5777\ur{K}$, $\log g_{\odot }=4.44$, $\mathrm{[M/H]}_{\mathrm{curr},\odot }=0.0\ur{dex}$, and $\mathrm{[\alpha /M]}_{\mathrm{curr},\odot }=0.0\ur{dex}$. 
The uncertainties in the input values were assumed to be $0.01\ur{mag}$ for $M_{G,\odot }$ and $M_{K_{\mathrm{s}},\odot }$, $4\ur{K}$ for $T_{\mathrm{eff},\odot }$, $0.005$ for $\log g_{\odot }$, $0.005\ur{dex}$ for $\mathrm{[M/H]}_{\mathrm{curr},\odot }$, as in Sect.~\ref{ssec:ageprocedure} for cases with very small errors. 

The recovered Solar age is summarized in Table~\ref{table:SunAge}. 
We found that the recovered Solar age is consistent with the literature value in all cases where $\log g$, $M_{G}$, or $M_{K_{\mathrm{s}}}$ is used as input. 
We also found that the recovered Solar $M_{\mathrm{ini}}$ is consistent with $1M_{\odot }$. 

As expected, the recovered Solar $\mathrm{[M/H]}_{\mathrm{ini}}$ is $0.08\ur{dex}$ (or $0.06\ur{dex}$ when adopting the zero-point used for $\mathrm{[M/H]}_{\mathrm{curr}}$), which is higher than the Solar surface value by ${\sim }10\sigma $. 
This offset reflects the decrease in surface metallicity with age in the $1M_{\odot }$ PARSEC model at a rate of ${\sim }0.015\ur{\si{dex.Gyr^{-1}}}$. 
This level of atomic diffusion has been supported by homogeneous spectroscopy of cluster stars; e.g., in the Solar-metallicity, Solar-age open cluster M67~\citep{BertelliMotta2018,Souto2019,Liu2019}. 

For comparison, we also determined the age of the Sun under the assumption that the observed [M/H] is equal to $\mathrm{[M/H]}_{\mathrm{ini}}$. 
In that case, the recovered age is $5.49^{+0.27}_{-0.27}$, $6.28^{+0.29}_{-0.30}$, or $6.09^{+0.28}_{-0.28}\ur{Gyr}$ when using $\log g$, $M_{G}$, or $M_{K_{\mathrm{s}}}$, respectively. 
As expected, these values deviate significantly from the true Solar age, demonstrating the importance of treating [M/H] and atomic diffusion consistently when determining ages. 

\begin{table}
\centering 
\caption{Recovered age, $M_{\mathrm{ini}}$, and $\text{[M/H]}_{\mathrm{ini}}$ of the Sun. }
\label{table:SunAge}
\scalebox{0.9}{
\begin{tabular}{c ccc}\hline \hline 
Input & Age & Initial mass & Initial metallicity \\
 & $\tau $ [Gyr] & $M_{\mathrm{ini}}$ [$M_{\odot }$] & $\mathrm{[M/H]}_{\mathrm{ini}}$ [dex]\tablefootmark{a} \\ \hline 
$(\mathrm{[M/H]}_{\mathrm{curr}},T_{\mathrm{eff}},\log g)$ & $4.53^{+0.23}_{-0.24}$ & $0.9984^{+0.0026}_{-0.0026}$ & $0.0828^{+0.0070}_{-0.0070}$ \\
$(\mathrm{[M/H]}_{\mathrm{curr}},T_{\mathrm{eff}},M_{G})$ & $4.47^{+0.25}_{-0.25}$ & $0.9985^{+0.0027}_{-0.0028}$ & $0.0821^{+0.0067}_{-0.0067}$ \\
$(\mathrm{[M/H]}_{\mathrm{curr}},T_{\mathrm{eff}},M_{K_{\mathrm{s}}})$ & $4.34^{+0.23}_{-0.23}$ & $0.9989^{+0.0027}_{-0.0027}$ & $0.0805^{+0.0067}_{-0.0067}$ \\
\hline 
\end{tabular}
}
\tablefoot{
\tablefoottext{a}{Zero-point of $\mathrm{[M/H]}_{\mathrm{ini}}$ is $(Z/X)_{\odot }=0.0207$, which is different from the zero-point of $\mathrm{[M/H]}_{\mathrm{curr}}$, $(Z/X)_{\mathrm{S}}=0.02169$. }
}
\end{table}

\begin{figure*}
\centering 
\includegraphics[width=18cm]{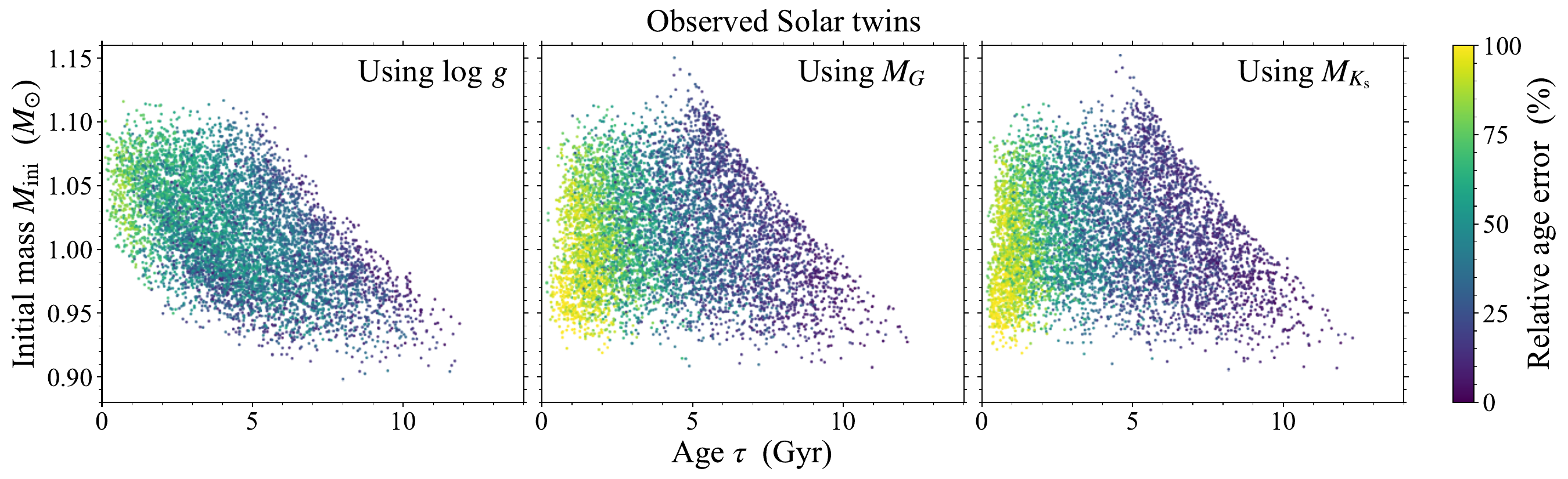}
\caption{Recovered parameters in the resulting observed Solar-twin catalog. Each panel shows a scatter plot between the determined age $\tau $ and initial mass $M_{\mathrm{ini}}$, color-coded by the relative age error. Left, middle, and right panels show the results obtained when using $\log g$, $M_{G}$, and $M_{K_{\mathrm{s}}}$, respectively. as the third input parameter in addition to $\mathrm{[M/H]}_{\mathrm{curr}}$ and $T_{\mathrm{eff}}$. }
\label{fig:AgeMini}
\end{figure*}

\begin{figure*}
\centering 
\includegraphics[width=18cm]{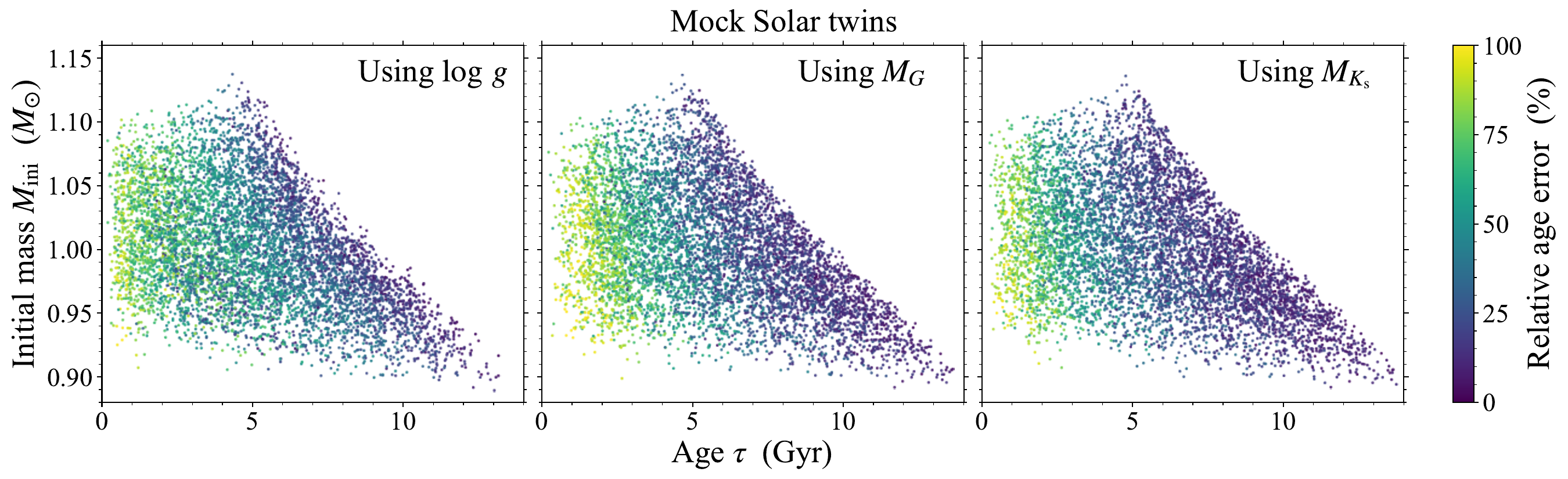}
\caption{Recovered parameters in the mock catalog of Solar twins. We show only random $6{,}594$ mock stars, i.e., the same number of stars as in the observed Solar-twin catalog. All the panels are defined in the same way as in Fig.~\ref{fig:AgeMini}. }
\label{fig:AgeMiniMock}
\end{figure*}

\subsection{Overview of the determined Solar-twin parameters}\label{ssec:plotparams}

Figure~\ref{fig:AgeMini} shows the scatter plots between the age and initial mass $M_{\mathrm{ini}}$ for our Solar-twin catalog. 
If the isochrone library retrieved from the CMD~3.8 web interface were used without further interpolation, stars would concentrate at discrete $M_{\mathrm{ini}}$ values, producing strip-like density patterns corresponding to the coarse initial-mass grid ($0.05M_{\odot }$ spacing). 
In contrast, thanks to the fine interpolation of the PARSEC isochrones described in Appendix~\ref{app:PARSECinterp}, no such artificial striping is visible in our results. 

Our Solar twins occupy a limited region in the age--$M_{\mathrm{ini}}$ plane in two respects. 
First, the $M_{\mathrm{ini}}$ values lie within ${\sim }0.9\text{--}1.1\,M_{\odot }$, reflecting the fact that we only analyzed Solar twins, i.e., main-sequence stars around the Solar $T_{\mathrm{eff}}$. 
Second, there are no stars in the upper-right region (older and more massive stars). 
Stars in that region would have already evolved off the main sequence; for example, the main-sequence lifetime of a $1M_{\odot }$ star is ${\sim }10\ur{Gyr}$. 

There is also a strong dependence of the relative age error on age and $M_{\mathrm{ini}}$. 
Older stars and stars with higher $M_{\mathrm{ini}}$ have smaller relative age errors. 
These stars lie more closely to the main-sequence turn-off, where isochrones move rapidly on the HR diagram, and hence ages can be more tightly constrained. 
Within this general trend, the decrease in relative age error toward older ages is clearly seen for the $M_{G}$-based and $M_{K_{\mathrm{s}}}$-based ages, but is less apparent for the $\log g$-based ages, especially at young ages, where the behavior appears more scattered. 
This difference may be related to the calibration of $\log g$ (Appendix~\ref{app:zeropoints}), in which the original--calibrated relation becomes nearly flat at high $\log g$ (i.e., for young stars). 
This flattening can reduce the propagated calibrated $\log g$ uncertainties for young stars and may affect the relative age errors derived from $\log g$.

\subsection{Comparison with mock Solar twins}\label{ssec:mock}

As we saw in Sect.~\ref{ssec:plotparams}, there are selection effects in our Solar-twin catalog. 
Any analysis based on the catalog needs to account for selection effects and the distribution of age errors. 
To facilitate this, we here constructed a mock Solar-twin catalog that mimics our observed selection function~(Sect.~\ref{sssec:createMock}), and then compared the observed and mock samples~(Sect.~\ref{sssec:compMock}).

\begin{figure*}
\centering 
\includegraphics[width=18cm]{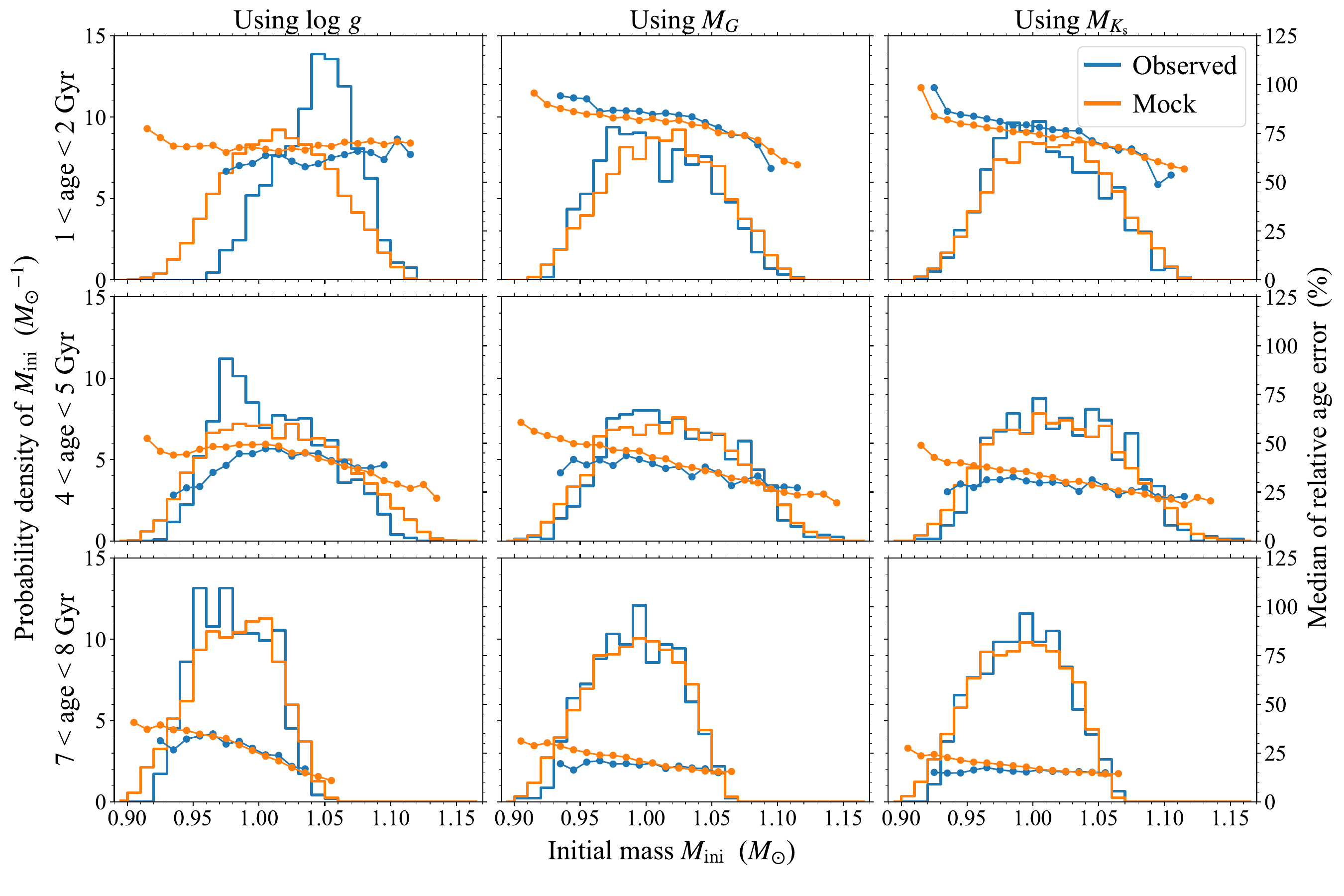}
\caption{Comparison between the observed (blue) and mock (orange) Solar-twin catalogs in different age bins. Top, middle, and bottom panels show results for twins with the determined ages between $1\text{--}2$, $4\text{--}5$, and $7\text{--}8\ur{Gyr}$, respectively. Left, middle, and right panels correspond to ages determined using $\log g$, $M_{G}$, and $M_{K_{\mathrm{s}}}$, respectively, as the third input parameter, together with $\mathrm{[M/H]}_{\mathrm{curr}}$ and $T_{\mathrm{eff}}$. In each panel, the histograms show the probability density of the initial mass $M_{\mathrm{ini}}$ (i.e., the normalized number of twins per $M_{\mathrm{ini}}$ bin), while line plots show the median relative age errors in each $M_{\mathrm{ini}}$ bins. }
\label{fig:Minihist}
\end{figure*}

\subsubsection{Creation of a mock Solar-twin sample}\label{sssec:createMock}

We constructed a mock sample of Solar twins as follows. 
First, we sampled stars across the age--$M_{\mathrm{ini}}$ grid points of the PARSEC isochrone library used in Sect.~\ref{sec:age} for age determination. 
At each grid point, stars were sampled from an uniform distribution in age and $\mathrm{[M/H]}_{\mathrm{ini}}$, with a density of $4\times 10^{8}\ur{\si{stars.Gyr^{-1}.dex^{-1}}}$. 
This corresponds to, for example, for ages ${>}1\ur{Gyr}$, where the grid spacing is $0.1\ur{Gyr}$ in age and $0.005\ur{dex}$ in $\mathrm{[M/H]}_{\mathrm{ini}}$~(Sect.~\ref{ssec:PARSEC}), $4\times 10^{5}$ stars per grid point. 

The initial mass $M_{\mathrm{ini}}$ of each star was sampled from the \citet{Kroupa2001} initial mass function (IMF). 
The distance $D$ to each star was sampled between $20\text{--}350\ur{pc}$ assuming an uniform underlying spatial density, corresponding to a probability distribution proportional to $D^{2}$. 
Other stellar parameters (i.e., $T_{\mathrm{eff}}$, $\log g$, $\mathrm{[M/H]}_{\mathrm{curr}}$, and absolute magnitudes in the \textit{G} and $K_{\mathrm{s}}$ bands) of each star were assigned by interpolating the PARSEC isochrones. 

The $\alpha $ abundance [$\alpha $/Fe] was assigned using the linear age--[Ca/Fe] relation from \citet{Bedell2018}, because GSP-Spec [$\alpha $/Fe] closely tracks [Ca/Fe]~\citep{GaiaGSPSpecDR3,RecioBlanco2024}. 
We then computed an ``observed'' [M/H] value, i.e., without correcting for the effect of [$\alpha $/Fe] on the isochrones using the equation from \citet{Salaris1993}. 

Sky coordinates were assigned assuming an uniform distribution over the celestial sphere. 
\textit{G}-band extinction $A_{G}$ was estimated using the 3D dust map of \citet{Leike2020} as implemented in the \texttt{dustmaps} code~\citep{Green2018,Green2019}, and the apparent \textit{G}-band magnitude was then computed. 
Because the 3D dust map of \citet{Leike2020} has a finite spatial coverage of $740\ur{pc}\times 740\ur{pc}\times 540\ur{pc}$, some mock stars between $270\text{--}350\ur{pc}$ fall outside the map and therefore have undefined $A_{G}$. 
For such stars, new sky coordinates were repeatedly assigned until $A_{G}$ could be estimated. 

Given the apparent \textit{G}-band magnitude, we then assigned S/N and observational uncertainties according to Equations~\ref{Eq:linearSNCoef} and \ref{Eq:LinearErrorCoef}, respectively, adding Gaussian scatter to reproduce the observed dispersion around these relations~(see, Appendix~\ref{app:mimickerror} for more details). 

Finally, we applied to the noise-added mock catalog the same selection criteria used for the real Solar twins. 
That is, we required $T_{\mathrm{eff}}$, $\log g$, and [M/H] to be close to the Solar values~(Sect.~\ref{ssec:gaiadata}), small errors in these parameters and in [$\alpha $/Fe]~(Sect.~\ref{ssec:gaiadata}), and consistency of $M_{G}$ and $M_{K_{\mathrm{s}}}$ as being Solar twins~(Sect.~\ref{ssec:flag}). 
In this way, selection effects such as the Malmquist bias are embedded in the mock sample. 

This procedure yielded a mock Solar-twin catalog consisting of $75{,}588$ artificial stars, more than ten times the size of the observed Solar-twin sample ($6{,}594$ stars). 
We note that the number of stars passing the selection depends on age. 
For example, the number of selected stars with ages between $10\text{--}11\ur{Gyr}$ is ${\sim }2.5$ times smaller than that with ages between $5\text{--}6\ur{Gyr}$, because the range of $M_{\mathrm{ini}}$values satisfying the Solar-twin selection criteria becomes narrower at older ages. 
We then determined the ages of these mock stars using exactly the same procedure as for the observed stars in Sect.~\ref{ssec:ageprocedure}. 

\begin{figure*}
\centering 
\includegraphics[width=18cm]{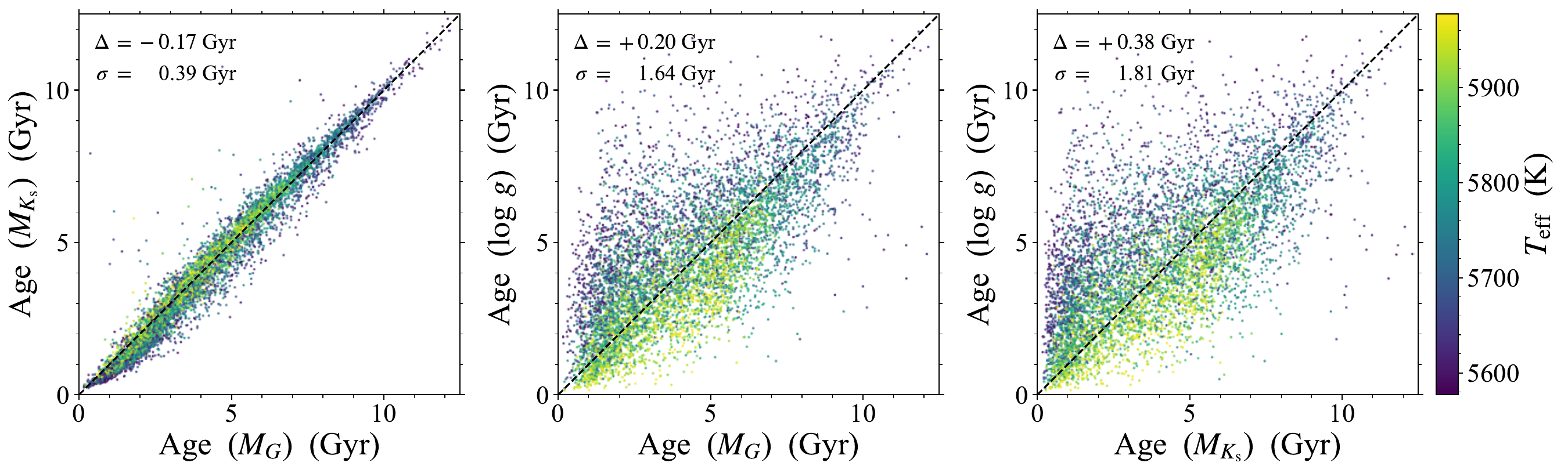}
\caption{Comparison between the Solar-twin ages determined with $\log g$, $M_{G}$, and $M_{K_{\mathrm{s}}}$, color-coded with $T_{\mathrm{eff}}$. }
\label{fig:Agecomp}
\end{figure*}

\subsubsection{Comparison of observed and mock Solar twins}\label{sssec:compMock}

Figure~\ref{fig:AgeMiniMock} shows the scatter plots between age and $M_{\mathrm{ini}}$ of for the mock Solar-twin catalog. 
In this figure, we display only $6{,}594$ stars randomly drawn from the full set of $75{,}588$ mock twins, so that the number of plotted stars matches that of the observed sample in Fig.~\ref{fig:AgeMini}. 
We found that the overall shape of the distribution and the dependence of relative age error on age and $M_{\mathrm{ini}}$ are very similar between the observed and mock samples, confirming the overall good accuracy in our Solar-twin catalog. 

The observed catalog contains fewer old stars (${\gtrsim }10\ur{Gyr}$) than the mock catalog. 
This difference likely reflects the underlying age distribution: we assumed a flat age distribution in the mock sample, whereas the observed sample may have a non-flat age distribution due to the star-formation history and radial migration. 
This topic is investigated in detail in our companion paper~\citep[Paper~II;][]{Paper2}. 

To assess our Solar-twin catalog more closely, we divided both the observed and mock samples into $1\ur{Gyr}$ age bins and examined the Solar-twin mass function, i.e., the number of Solar twins at each $M_{\mathrm{ini}}$ within a given age bin. 
Given the relatively narrow age bins, the mass function should be similar between the observed and mock samples as long as the assumed IMF and stellar-evolution model are accurate and the efficiency of radial migration does not depend strongly on stellar mass. 

Histograms in Fig.~\ref{fig:Minihist} show the Solar-twin mass function for the observed (blue) and mock (orange) samples. 
For the ages determined using $M_{G}$ or $M_{K_{\mathrm{s}}}$ (middle and right panels, respectively), the observed and mock histograms are nearly identical in shape. 
This agreement suggests that our determinations of age and $M_{\mathrm{ini}}$ based on $M_{G}$ or $M_{K_{\mathrm{s}}}$ are accurate. 
In contrast, the histograms for ages based on $\log g$ (left panels) show significant differences, especially at younger ages (top and middle panels). 
This discrepancy could be due to a residual systematic bias in $\log g$, which we will discuss in Sect.~\ref{ssec:comp_threeage}. 

We also examined the median relative age error in each $M_{\mathrm{ini}}$ bin for each age bin as a function of $M_{\mathrm{ini}}$ (line plots in Fig.~\ref{fig:Minihist}; again, blue for the observed sample and orange for the mock sample). 
At high mass ($M_{\mathrm{ini}}\gtrsim 0.95M_{\odot }$), the observed and mock age errors behave similarly, confirming that the error prescriptions in the mock sample well reproduce those in the observed sample. 
At lower masses, however, we found differences in the median relative age error of up to ${\sim }20\%$. 
Detailed comparisons between the observed and mock samples revealed that this difference arises from the dependence of metallicity on the distance present in the observed sample. 
In the observed catalog, there is a weak tendency for metal-rich stars (within our narrow metallicity range of ${\pm }0.1\ur{dex}$ around the Solar value) to lie at larger distances (and hence to have lower S/N and larger errors), regardless the direction, though the origin of this trend is unclear. 
No such trend appears in the mock sample. 
Because, for $M_{\mathrm{ini}}\lesssim 0.95M_{\odot }$, stars with higher metallicity do not satisfy the Solar-twin selection criteria due to the metallicity dependence of $T_{\mathrm{eff}}$, the lower-mass bins are dominated by lower-metallicity stars in the observed sample. 
These stars tend to lie at closer distances and therefore have smaller errors, resulting in smaller median relative age error for the observed sample than the mock sample at lower masses. 
This tendency should be kept in mind when analyzing our catalog.

\subsection{Comparison of the three age determinations}\label{ssec:comp_threeage}

Figure~\ref{fig:Agecomp} compares the ages of Solar twins determined using $\log g$, $M_{G}$, and $M_{K_{\mathrm{s}}}$. 
As shown in the left panel, the ages derived from $M_{G}$ and $M_{K_{\mathrm{s}}}$ agree very well. 
The mean bias and standard deviation are $-0.17$ and $0.39\ur{Gyr}$, respectively, both of which are much smaller than the typical age uncertainties of $2.8$ and $2.2\ur{Gyr}$ for the $M_{G}$- and $M_{K_{\mathrm{s}}}$-based ages. 
This is expected because both $M_{G}$ and $M_{K_{\mathrm{s}}}$ are tied to the bolometric luminosity $L_{\mathrm{bol}}$ through the bolometric corrections implemented in the isochrone library~\citep{Chen2019,Bohlin2020}, and hence both sets of ages are essentially determined by $T_{\mathrm{eff}}$, [M/H], and $L_{\mathrm{bol}}$. 
In other words, the close agreement between $M_{G}$- and $M_{K_{\mathrm{s}}}$-based ages supports the accuracy of the adopted bolometric corrections. 

A small number of outliers deviate from this one-to-one relation between $M_{G}$ and $M_{K_{\mathrm{s}}}$-based ages, most of them having $M_{K_{\mathrm{s}}}$-based ages larger than those from $M_{G}$. 
Though the exact cause is unclear, unresolved multiplicity in 2MASS likely plays a role. 
\textit{Gaia} photometry resolves these multiple systems that are unresolved in 2MASS due to its poorer spatial resolution. 
In such cases, the $K_{\mathrm{s}}$ magnitude could be biased by unresolved companions, making stars appearing brighter and, hence, older in the $M_{K_{\mathrm{s}}}$-based age estimates. 

Middle and right panels of Fig.~\ref{fig:Agecomp} show the ages based on $\log g$ against those based on $M_{G}$ and $M_{K_{\mathrm{s}}}$, respectively. 
In both panels, there is a large scatter around the one-to-one line ($1.64$ and $1.81\ur{Gyr}$, respectively), and the relation between the two age estimates depends on $T_{\mathrm{eff}}$. 
Since $\log g$ is related to $L_{\mathrm{bol}}$ via the Stefan-Boltzmann law, and we have already demonstrated the reliability of the bolometric corrections linking $L_{\mathrm{bol}}$ to $M_{G}$ and $M_{K_{\mathrm{s}}}$, any model-related uncertainties would affect all three age estimates in a similar way. 
Hence, the scatter could be due to a $T_{\mathrm{eff}}$-dependent systematic bias in $\log g$, which we could not fully correct in Sect.~\ref{ssec:gaiadata} and Appendix~\ref{app:zeropoints}. 

As discussed in this section and in Sect.~\ref{sssec:compMock}, a residual systematic bias in $\log g$ appears to remain despite our calibration efforts. 
This likely reflects the difficulty of determining spectroscopically $\log g$ accurately. 
Unlike absolute magnitudes, which can be determined accurately from photometry and astrometry when extinction is not severe, determinations of $\log g$ requires specific spectral features such as combinations of neutral and ionized lines from a specific element and/or the wings of very strong lines~\citep{Gray2005}. 
In our case, GSP-Spec $\log g$ is primarily constrained by the wings of the \ion{Ca}{ii} IR triplet, weak metallic lines, and molecular lines~\citep{RecioBlanco2016}, and the limited wavelength coverage associated to a medium spectral resolution of RVS makes the problem challenging. 
Indeed, we found a systematic trend between the original GSP-Spec $\log g$ estimates and literature $\log g$ values based on higher-resolution, wider-wavelength-coverage spectra~(Appendix~\ref{app:zeropoints}). 
If a sufficiently large calibration sample densely covering the $T_{\mathrm{eff}}$--$\log g$ plane would have been available, this bias could be removed more accurately, but with only a few dozen stars in common, the calibration remains not enough accurate when examining tiny effects. 

Given these difficulties in determining $\log g$, we do not recommend using our $\log g$-based ages for statistical analysis of the sample. 
In contrast, both the $M_{G}$- and $M_{K_{\mathrm{s}}}$-based ages appear sufficiently reliable for such applications. 
Either can be used, but $M_{G}$-based ages have the advantage that \textit{Gaia} offers higher spatial resolution, thereby reducing the risk of age overestimation due to unresolved companions. 
On the other hand, $M_{K_{\mathrm{s}}}$-based ages are less sensitive to extinction. 
Since in our companion paper, \citet[Paper~II]{Paper2} adopted the $M_{K_{\mathrm{s}}}$-based age for their analysis, for which selection effects are weaker, we will also use the $M_{K_{\mathrm{s}}}$-based ages in the subsequent discussion of this paper.

\section{Comparison with previous Solar-twin catalogs}\label{sec:agecomp}

In this section, we first compare our Solar-twin catalog with literature, in particular focusing on the number of twins in each sample~(Sect.~\ref{ssec:metrics}). 
Then, we compare different sets of stellar parameters, ages, and $M_{\mathrm{ini}}$, to further validate our catalog and to place this work in the context of previous Solar-twins studies~(Sect.~\ref{ssec:compliter}).

\subsection{Properties of the literature and this Solar-twin catalogs}\label{ssec:metrics}

\begin{table*}
\centering 
\caption{Solar twin catalogs adopted for comparison (incomplete manual compilation). }
\label{table:compsample}
\scalebox{0.80}{
\begin{tabular}{l c ccc cc cc}\hline \hline 
Paper & Spectrograph & \multicolumn{3}{c}{Typical error} & \multicolumn{2}{c}{\# samples} & \multicolumn{2}{c}{Available?} \\
 &  & $T_{\mathrm{eff}}$ [K] & $\log g$ & [M/H] & $N_{\mathrm{total}}$\tablefootmark{a} & $N_{\mathrm{twin}}$\tablefootmark{b} & Age & Abundances\tablefootmark{c} \\ \hline 
This work & \textit{Gaia} DR3 GSP-Spec & $51$ & $0.05$ & $0.03$ & $6594$ & $6594$ & Yes & Yes \\
\hline 
\citet{Lehmann2025}\tablefootmark{d} & GALAH DR3 & $38$ & $0.03$ & $0.02$ & $72288$ & $14571$ & Yes &  \\
\citet{Walsen2024}\tablefootmark{d} & GALAH DR3 & $4$ & $0.01$ & $0.00$ & $38320$ & $13132$ & \tablefootmark{e} & Yes \\
\citet{Rampalli2024}\tablefootmark{d} & \textit{Gaia} DR3 RVS & $61$ & $0.09$ & $0.04$ & $17412$ & $5347$ & Yes & Yes \\
\hline 
\citet{Sun2025a} & MIKE & $22$ & $0.05$ & $0.01$ & $25$ & $13$ & Yes & Yes \\
\citet{Carlos2025} & HARPS &  & $0.02$ & $0.01$ & $50$ & $10$ & Yes & Yes \\
\citet{Martos2025} & HARPS & $10$ & $0.02$ & $0.01$ & $99$ & $61$ &  & Yes \\
\citet{CarvalhoSilva2025} & HARPS & $7$ & $0.02$ & $0.01$ & $126$ & $7$ & Yes &  \\
\citet{Sun2025b} & MIKE & $20$ & $0.05$ & $0.01$ & $17$ & $7$ & Yes & Yes \\
\citet{Plotnikova2024} & HARPS & $9$ & $0.03$ & $0.01$ & $130$ & $48$ & Yes & Yes \\
\citet{Shejeelammal2024} & HARPS & $5$ & $0.02$ & $0.01$ & $233$ & $83$ & Yes & Yes \\
\citet{Rathsam2023} & HARPS & $6$ & $0.02$ & $0.01$ & $74$ & $17$ & Yes &  \\
\citet{Martos2023} & HARPS & $4$ & $0.01$ & $0.00$ & $118$ & $62$ & Yes &  \\
\citet{Lehmann2023}\tablefootmark{d} & HERMES & $77$ & $0.06$ & $0.04$ & $877$ & $99$ & Yes &  \\
\citet{Spina2021} & HARPS & $10$ & $0.03$ & $0.01$ & $60$ & $10$ & Yes & Yes \\
\citet{YanaGalarza2021} & Goodman/ARCES/2dcoude & $11$ & $0.03$ & $0.01$ & $129$ & $58$ & Yes &  \\
\citet{Nissen2020} & HARPS & $9$ & $0.02$ & $0.01$ & $72$ & $27$ & Yes & Yes \\
\citet{Liu2020} & HIRES & $14$ & $0.03$ & $0.01$ & $83$ & $7$ & Yes & Yes \\
\citet{Casali2020} & HARPS & $8$ & $0.02$ & $0.01$ & $560$ & $223$ & Yes & Yes \\
\citet{DelgadoMena2019} & HARPS &  &  &  & $1059$ & $93$ & Yes &  \\
\citet{LorenzoOliveira2018} & HARPS & $4$ & $0.01$ & $0.00$ & $82$ & $64$ & Yes &  \\
\citet{Adibekyan2018} & HARPS/UVES/FEROS & $26$ & $0.04$ & $0.02$ & $54$ & $17$ & Yes & Yes \\
\citet{Bedell2018} & HARPS &  &  &  & $79$ & $64$ & \tablefootmark{f} & Yes \\
\citet{Spina2018} & HARPS & $4$ & $0.01$ & $0.00$ & $79$ & $64$ & Yes & Yes \\
\citet{DelgadoMena2017} & HARPS & $28$ & $0.04$ & $0.02$ & $1059$ & $93$ &  & Yes \\
\citet{Reddy2017} & HARPS &  &  &  & $24$ & $22$ & Yes & Yes \\
\citet{Beck2017} & HERMES & $70$ & $0.15$ & $0.04$ & $18$ & $3$ & Yes &  \\
\citet{LopezValdivia2017} & CanHiS & $59$ & $0.27$ & $0.08$ & $38$ & $5$ &  & Yes \\
\citet{Spina2016a} & UVES & $3$ & $0.01$ & $0.00$ & $9$ & $9$ & Yes & Yes \\
\citet{Nissen2016} & HARPS & $6$ & $0.01$ & $0.01$ & $21$ & $18$ & Yes & Yes \\
\citet{Adibekyan2016} & HARPS/UVES & $22$ & $0.03$ & $0.02$ & $39$ & $9$ & Yes & Yes \\
\citet{TucciMaia2016} & MIKE & $6$ & $0.02$ & $0.01$ & $88$ & $75$ & Yes & Yes \\
\citet{Mahdi2016} & ELODIE & $10$ & $0.02$ & $0.01$ & $56$ & $40$ & Yes & Yes \\
\citet{Spina2016b} & HIRES & $10$ & $0.03$ & $0.01$ & $14$ & $14$ & Yes & Yes \\
\citet{Nissen2015} & HARPS & $6$ & $0.01$ & $0.01$ & $21$ & $18$ & Yes & Yes \\
\citet{Datson2015} & FEROS & $40$ & $0.07$ & $0.03$ & $148$ & $77$ &  &  \\
\citet{Gonzalez2015} & Sandiford & $50$ & $0.07$ & $0.04$ & $31$ & $6$ & Yes &  \\
\citet{Ramirez2014} & MIKE & $6$ & $0.02$ & $0.01$ & $88$ & $75$ & Yes &  \\
\citet{LopezValdivia2014} & B\&C & $50$ & $0.25$ & $0.06$ & $233$ & $34$ &  &  \\
\citet{Gonzalez2014} & Sandiford/2dcoude & $40$ & $0.06$ & $0.03$ & $37$ & $13$ & Yes & Yes \\
\citet{PortodeMello2014} & OPD/FEROS &  &  &  & $55$ & $29$ & Yes &  \\
\citet{Tsantaki2013} & HARPS & $23$ & $0.06$ & $0.02$ & $451$ & $38$ & Yes &  \\
\citet{Adibekyan2012} & HARPS &  &  &  & $1111$ & $93$ &  & Yes \\
\citet{daSilva2012} & CTIO &  &  &  & $25$ & $7$ & Yes & Yes \\
\citet{Sousa2011} & HARPS & $35$ & $0.05$ & $0.02$ & $582$ & $49$ &  &  \\
\citet{GonzalezHernandez2010} & HARPS/UVES/UES & $31$ & $0.04$ & $0.02$ & $11$ & $1$ &  & Yes \\
\citet{Ghezzi2010} & FEROS & $38$ & $0.13$ & $0.03$ & $262$ & $22$ & Yes &  \\
\citet{Baumann2010} & 2dcoude/HARPS & $39$ & $0.06$ & $0.03$ & $117$ & $65$ & Yes &  \\
\citet{Gonzalez2010} & 2dcoude & $33$ & $0.05$ & $0.02$ & $159$ & $18$ & Yes & Yes \\
\citet{Ramirez2009} & 2dcoude &  &  & $0.02$ & $64$ & $37$ &  & Yes \\
\citet{Neves2009} & HARPS &  &  &  & $451$ & $44$ &  & Yes \\
\citet{Pasquini2008} & FLAMES & $60$ &  &  & $59$ & $36$\tablefootmark{g} &  &  \\
\citet{Sousa2008} & HARPS & $24$ & $0.04$ & $0.02$ & $451$ & $44$ &  &  \\
\citet{Melendez2007} & 2dcoude & $36$ & $0.04$ & $0.02$ & $4$ & $4$ & Yes &  \\
\citet{Takeda2007} & HIDES &  &  &  & $118$ & $55$ & Yes &  \\
\citet{Melendez2006} & HIRES & $30$ & $0.03$ & $0.03$ & $2$ & $2$ & Yes & Yes \\
\citet{Laws2003} & 2dcoude/CTIO & $44$ & $0.07$ & $0.03$ & $31$ & $3$ & Yes &  \\
\citet{Gonzalez2001} & 2dcoude/CTIO & $39$ & $0.07$ & $0.03$ & $22$ & $1$ & Yes & Yes \\
\hline 
\end{tabular}
}
\tablefoot{
\tablefoottext{a}{Total number of targets analyzed. }
\tablefoottext{b}{Number of targets satisfying our Solar-twin criteria ($T_{\mathrm{eff}}$, $\log g$, and [M/H] within ${\pm }200\ur{K}$, ${\pm }0.2$, and ${\pm }0.1\ur{dex}$ around the Solar values, respectively). }
\tablefoottext{c}{We did not mark studies that provide abundances for only a very limited set of elements (e.g., Li only) as having abundances available. }
\tablefoottext{d}{Using a data-driven method for determining stellar parameters. }
\tablefoottext{e}{\citet{Walsen2024} adopted ages from GALAH DR3~\citep{Sharma2018,Buder2021}. }
\tablefoottext{f}{\citet{Bedell2018} adopted ages determined by \citet{Spina2018}. }
\tablefoottext{g}{We only considered $T_{\mathrm{eff}}$ in calculating $N_{\mathrm{twin}}$ for \citet{Pasquini2008} because they did not determine [Fe/H] and $\log g$ of their M67 stars. }
}
\end{table*}

Table~\ref{table:compsample} summarizes information on our catalog of Solar twins along with that of a (comprehensive but not complete) compilation of literature studies based on high-resolution spectroscopy. 
We mostly included studies that determined stellar parameters by themselves and that mainly focused on Solar twins (or at least Solar analogs). 

As shown in the table, most catalogs contain only a limited number of Solar twins (typically several tens), but in some of these catalogs the stellar parameters are measured with very high precision, if we consider the reported uncertainties. 
Many recent catalogs are based on archival spectra from the HARPS spectrograph~\citep{Mayor2003}, originally obtained for exoplanet searches. 
By stacking time-series spectra, HARPS data provide extremely high spectral resolution ($R=115{,}000$) and S/N of ${\sim}1{,}000$ over most of its wide $378\text{--}691\ur{nm}$ wavelength range, and hence extremely high precision in the derived stellar parameters and chemical abundances~\citep[e.g., $10\ur{K}$ in $T_{\mathrm{eff}}$, $0.02$ in $\log g$, and $0.01\ur{dex}$ in {[Fe/H]} and other abundances, in the work by][see Table~\ref{table:compsample} for precisions in other studies]{Martos2025}. 
Such small but highly precise and accurate samples are particularly valuable for identifying stars whose parameters must be determined with high accuracy relative to the Solar values. 
For example, ``true'' Solar twins, defined as stars whose parameters are not only close to those of the Sun, as in ordinary Solar twins, but effectively indistinguishable within very small uncertainties (e.g., ${\sim }10\ur{K}$ in $T_{\mathrm{eff}}$), have been sought for many years~\citep[e.g.,][]{PortodeMello1997,Melendez2007}. 
Another example is the search for Solar siblings, i.e., stars whose ages and chemical abundances are identical to those of the Sun, and which therefore could have formed in the same star cluster as the Sun~\citep[e.g.,][]{Adibekyan2018}. 
Since cluster-to-cluster dispersions (or azimuthal scatter) in chemical abundances are suggested to be of the order of a few hundredths of a dex~\citep[e.g.,][and references therein]{Freeman2002,Bellardini2021,Bhattarai2024}, searching for Solar siblings (or, in other words, carrying out strong chemical tagging) requires this level of precision and accuracy in abundance determinations. 
In general-purpose spectroscopic surveys with moderate S/N, in contrast to high-precision Solar-twin studies, the achievable precision and accuracy for individual stars are still insufficient for this purpose~\citep[e.g.,][]{Casamiquela2021,Ness2022}. 

In contrast, our work and a few recent studies have constructed large samples by homogeneously exploiting recent huge datasets from general-purpose spectroscopic surveys: \textit{Gaia} DR3 GSP-Spec in this study, \textit{Gaia} DR3 RVS in \citet{Rampalli2024}, and GALAH DR3 in \citet{Walsen2024} and \citet{Lehmann2025}. 
Stellar parameters and chemical abundances of ${\sim }1{,}700$ Solar twins from APOGEE DR16 were also used by \citet{Nibauer2021}. 
Such samples are generally limited in wavelength coverage, spectral resolution, and/or S/N, and thus the achievable precisions and accuracies in stellar parameters are lower. 
For example, in our case with \textit{Gaia} DR3 GSP-Spec, RVS spectra cover only $846\text{--}870\ur{nm}$ with $R=11{,}500$, and the median S/N per pixel (\texttt{rv\_expected\_sig\_to\_noise}) is $83$\footnote{Given the oversampling of \textit{Gaia} RVS, this translates into the median S/N per spectral resolution element of ${\sim }140$. }. 
As a result, the precision of the derived stellar parameters degrades to $51\ur{K}$ in $T_{\mathrm{eff}}$, $0.05$ in $\log g$, $0.03\ur{dex}$ in [M/H], and ${\sim }0.1\ur{dex}$ in other abundances. 

Nevertheless, the ${\sim }100$ times larger sample size ($6{,}594$ stars in our case), even with a lower precision, enables us to investigate the statistical properties of Solar twins, such as the age--abundance relation, with comparable significance, as demonstrated in Sect.~\ref{sec:chem}. 
Moreover, only such large samples allow us to study the density distributions of stellar parameters for Solar twins, including the probability density of $M_{\mathrm{ini}}$ at a given age (as shown in Fig.~\ref{fig:Minihist}) and the probability density of ages~\citep[presented in our companion paper, Paper~II;][]{Paper2}. 

To date, all previous large Solar-twin samples~\citep{Rampalli2024,Walsen2024,Lehmann2025} relied on data-driven approaches to determine stellar parameters and chemical abundances~(see, Appendix~\ref{app:review_literature}, for an in-depth review of their methods and results). 
As a result, they achieved internal precisions that are comparable to or higher than those in our GSP-Spec study, in which a traditional spectral-fitting technique was used. 
However, as a trade-off, some of their reported results may reflect correlations learned from the training data, since data-driven approaches infer, rather than directly measure, chemical abundances. 
Indeed, \citet{Rampalli2024} pointed out that their model inferred the abundances of C, O, Na, V, Al, and Y, even though no atomic lines of these elements with sufficient strengths are present in the \textit{Gaia} RVS spectra of Solar twins, and abundances of these elements are not included in the GSP-Spec catalog\footnote{Because CN molecular lines are present in the RVS wavelength range \citep{RecioBlanco2024}, and their strengths depend on the C, N, and O abundances through temperature-dependent molecular equilibrium \citep[see, e.g.,][] {Yong2015,Taniguchi2025}, the RVS spectra in principle contain information on the C and O abundances. Nevertheless, judging from Fig.~A1 in \citet{Rampalli2024}, their derived [C/H] and [O/H] abundances do not appear to rely strongly on CN lines. }. 
Further indications that data-driven inferences may be influenced by training-set correlations are discussed in Sects.~\ref{ssec:compliter} and \ref{ssec:ageabn_s}. 
Catalogs with model-driven abundances therefore offer a complementary way to validate data-driven results. 

In light of these previous large-sample studies of Solar twins, all of which rely on data-driven methods, our catalog based on \textit{Gaia} DR3 GSP-Spec aims to provide a large, homogeneous catalog in which all stellar parameters, age, and elemental abundances are determined from forward modeling based on stellar atmosphere and evolutionary models, rather than inferred solely through data-driven methods, thereby avoiding the propagation of trends learned from small training samples. 
In this way, we aim to provide a basis for confirming results previously obtained from small, high-precision samples (e.g., age--[X/Fe] relations) in a scaled, statistical manner, as demonstrated in Sect.~\ref{sec:chem}. 
Moreover, our characterization of the selection function (Sect.~\ref{ssec:mock}) facilitates analyses that are sensitive to selection effects (e.g., the interpretation of histograms). 
Taken together, our catalog offers model-driven parameters with a characterized selection function, serving as a complementary reference to both small high-precision samples and large data-driven samples. 

Beyond differences in the sample size and analysis methods, some of the previous studies adopted rather loose criteria on [M/H] when defining Solar twins. 
For example, \citet{Lehmann2025} allowed stars with any [Fe/H] to be included in their sample. 
Such a loose criterion on [M/H] helps to investigate trends with metallicity and to enlarge the sample size without introducing large systematic uncertainties in chemical abundances. 
However, our strict criterion on [M/H], namely ${\pm }0.1\ur{dex}$ around the Solar value, is essential for associating age with birth radius~\citep{Minchev2013,Plotnikova2024}. 
With such a criterion, the age--[X/Fe] relation for Solar twins can effectively provide information on the relation between birth radius and [X/Fe] at a given metallicity. 
Our moderately strict threshold on $\log g$ of $\pm 0.2\ur{dex}$ around the Solar value is also important for selecting only main-sequence stars, in contrast to some recent studies~\citep[e.g., $\pm 0.3\ur{dex}$ around the Solar value adopted by][]{Walsen2024,Rampalli2024}.

\subsection{Comparison of stellar parameters, ages, and initial mass}\label{ssec:compliter}

\begin{sidewaysfigure*}
\centering 
\includegraphics[width=24cm, page=26]{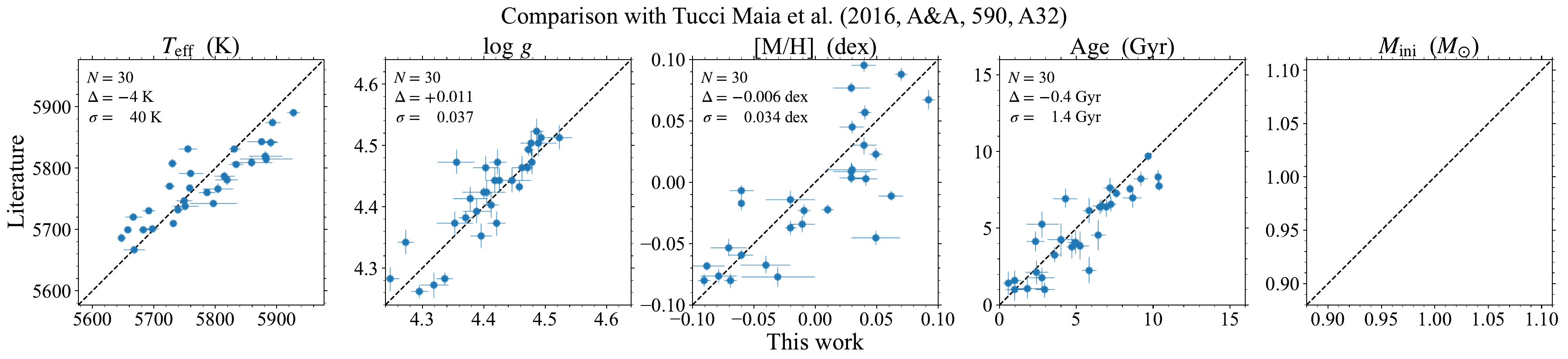}
\includegraphics[width=24cm, page=15]{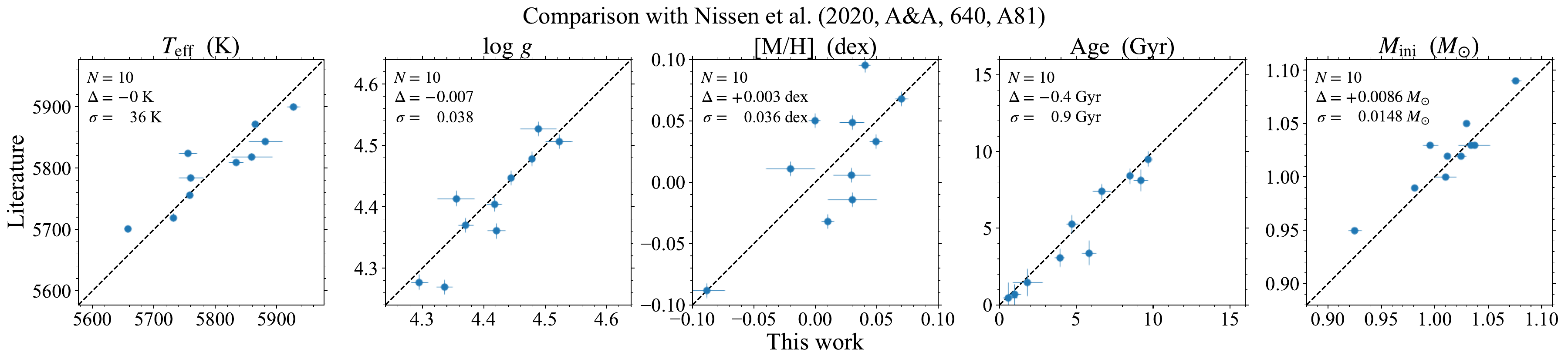}
\includegraphics[width=24cm, page=8]{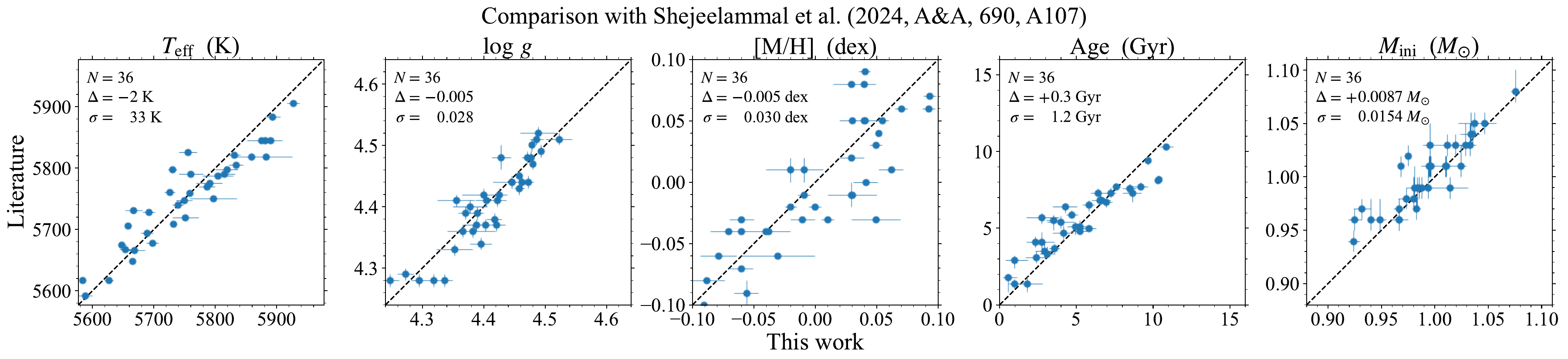}
\caption{Comparison of stellar parameters with literature~\citep{TucciMaia2016,Nissen2020,Shejeelammal2024}.  Each panel shows a comparison for $T_{\mathrm{eff}}$, $\log g$, [M/H] (or [Fe/H]), age, and $M_{\mathrm{ini}}$, from left to right. Three values are annotated: the number $N$ of stars in common between this work and literature, the mean difference $\Delta $ and the standard deviation $\sigma $ of the difference between this work and literature. Also available at Zenodo (\url{https://doi.org/10.5281/zenodo.XXXXXXXX}). }
\label{fig:compall1}
\end{sidewaysfigure*}

\begin{sidewaysfigure*}
\centering 
\includegraphics[width=24cm, page=10]{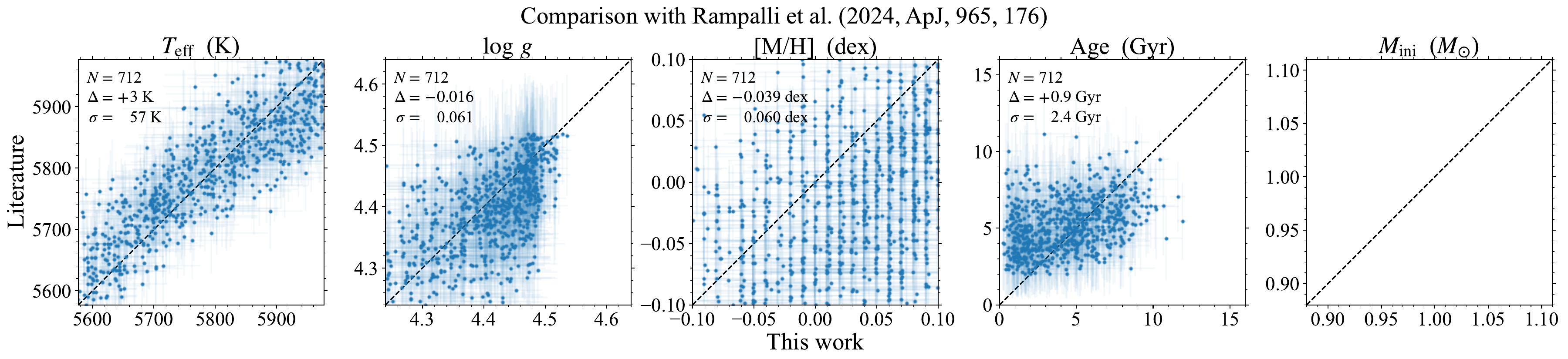}
\includegraphics[width=24cm, page=9]{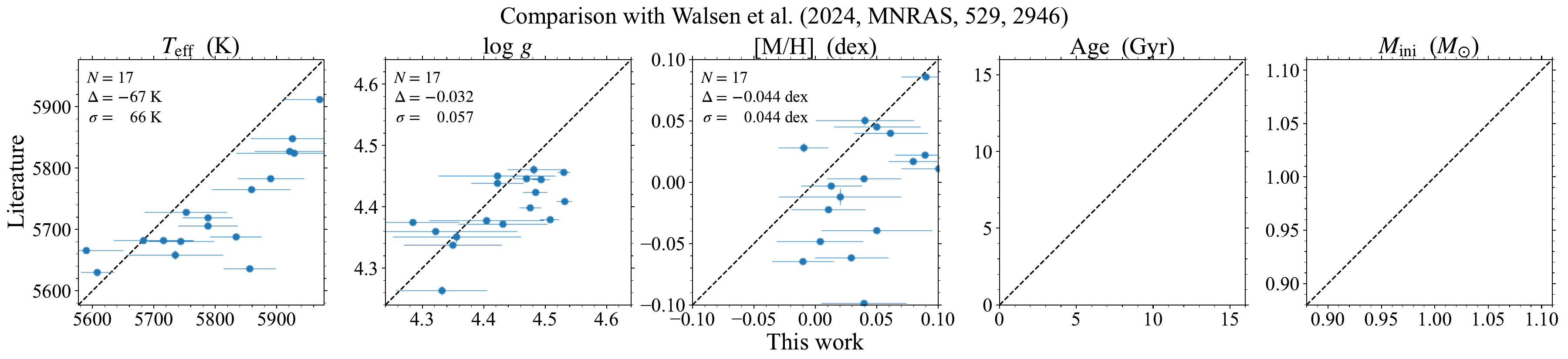}
\includegraphics[width=24cm, page=6]{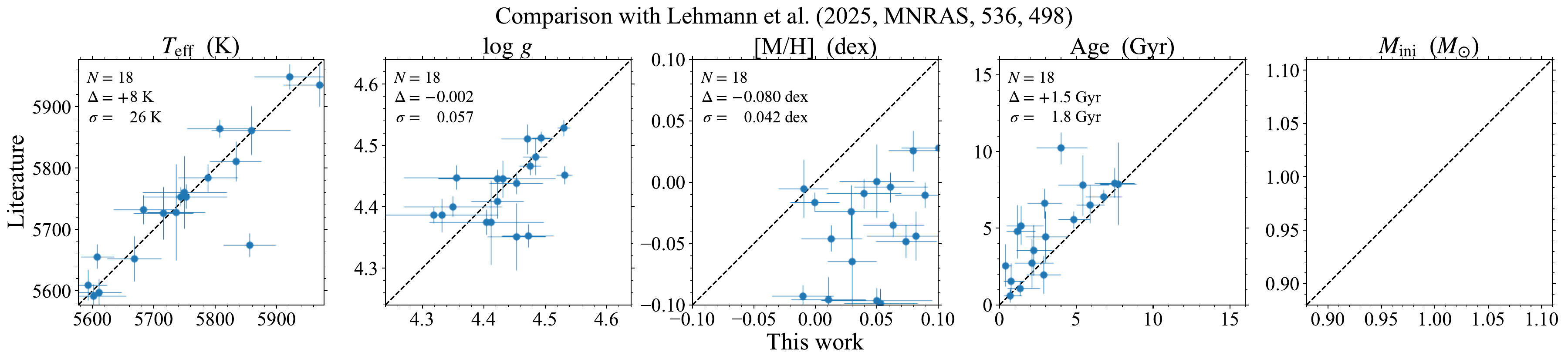}
\caption{Same as Fig.~\ref{fig:compall1} but for three other studies~\citep{Walsen2024,Lehmann2025,Rampalli2024}. }
\label{fig:compall2}
\end{sidewaysfigure*}

Figures~\ref{fig:compall1} and \ref{fig:compall2} show comparisons of stellar parameters ($T_{\mathrm{eff}}$, $\log g$, and [M/H] or [Fe/H]), age, and $M_{\mathrm{ini}}$ between this work and selected literature; Fig.~\ref{fig:compall1} for the studies used for the calibration of zero-points in Appendix~\ref{app:zeropoints}~\citep{TucciMaia2016,Nissen2020,Shejeelammal2024} and Fig.~\ref{fig:compall2} for the studies building large catalogs using spectra from general-purpose spectroscopic surveys~\citep{Rampalli2024,Walsen2024,Lehmann2025}\footnote{Comparisons to all the literature listed in Table~\ref{table:compsample} are available in Fig.~\ref{fig:compall} and at Zenodo (\url{https://doi.org/10.5281/zenodo.XXXXXXXX}). }. 
We plotted only stars whose three parameters from both this work and the literature satisfy the Solar-twin criteria, regardless of their errors. 
We note that we plot the calibrated stellar parameters in these figures, rather than the original parameters shown in Fig.~\ref{fig:literparams}. 

In Fig.~\ref{fig:compall1}, we found good agreement in all five parameters with those reported by \citet{TucciMaia2016}, \citet{Nissen2020}, and \citet{Shejeelammal2024}. 
The mean and standard deviation of the differences between our parameters and those from the literature are annotated in each panel. 
The agreement for the stellar parameters ($T_{\mathrm{eff}}$, $\log g$, and [M/H]) is expected, given that we calibrated our stellar parameters against these three studies. 
The agreement for age and $M_{\mathrm{ini}}$ supports the reliability of our isochrone projection. 
We found similar agreements with many other studies. 

In contrast, we found discrepancies with several other studies, most notably those based on large Solar-twin samples, namely \citet{Walsen2024}, \citet{Lehmann2025}, and \citet{Rampalli2024}, as shown in Fig.~\ref{fig:compall1}. 
In the studies by \citet{Walsen2024} and \citet{Lehmann2025}, [Fe/H] values are systematically $0.08$ and $0.04\ur{dex}$, respectively, lower than our [M/H], and hence the ages in \citet{Lehmann2025} are systematically $1.5\ur{Gyr}$ larger than ours. 
These systematic differences may be due to the fact that the two studies employed stellar parameters from GALAH DR3 and DR2, respectively, as training references for their data-driven analyses, which possibly have systematic zero-point biases, or at least zero points that differ from those in many high-precision Solar-twin studies\footnote{Ideally, the stellar parameters reported by \citet{Walsen2024} and \citet{Lehmann2025} would be directly compared with those from high-precision Solar-twin studies~\citep[e.g.,][]{TucciMaia2016,Nissen2020,Shejeelammal2024}. In practice, however, such a comparison is not feasible because the distance ranges of the Solar-twin samples differ substantially between these studies: the samples in \citet{Walsen2024} and \citet{Lehmann2025} are dominated by more distant stars, whereas high-precision Solar-twin studies focus on nearby targets to achieve very high S/N~(see, Fig.~XX). As a result, there are very few (if any) Solar twins in common between the two sets of studies, preventing a meaningful star-by-star comparison. }. 

In our comparison with \citet{Rampalli2024}, we found weak correlations between the two studies for $\log g$, [M/H], and age\footnote{Because the ages of \citet{Rampalli2024} were determined using $T_{\mathrm{eff}}$, [M/H], and $\log g$, they correlate more strongly with our $\log g$-based ages ($r=0.52$) than with our adopted $M_{K_{\mathrm{s}}}$-based ages ($r=0.36$), although the correlation remains modest in both cases. }, even though both studies are based on \textit{Gaia} DR3 RVS spectra. 
In particular, whereas our ages remain consistent with those from high-precision studies down to very young ages (${\lesssim }1\ur{Gyr}$; see Fig.~\ref{fig:compall1}), the ages reported by \citet{Rampalli2024} show an apparent lower limit of ${\sim }2\text{--}3\ur{Gyr}$. 
As a result, stars that we identify as very young Solar twins are generally assigned significantly older ages in their study, contributing to the weak overall correlation. 
Because most of the stars analyzed by \citet{Rampalli2024} do not overlap with high-precision Solar-twin samples, it remains unclear whether the observed scatter in stellar parameters and age between our results and those by \citet{Rampalli2024} can be fully explained by the combined statistical uncertainties, or whether additional systematic uncertainties are present.

\section{Age--chemical abundance trends}\label{sec:chem}

\begin{figure*}
\centering 
\includegraphics[width=18cm]{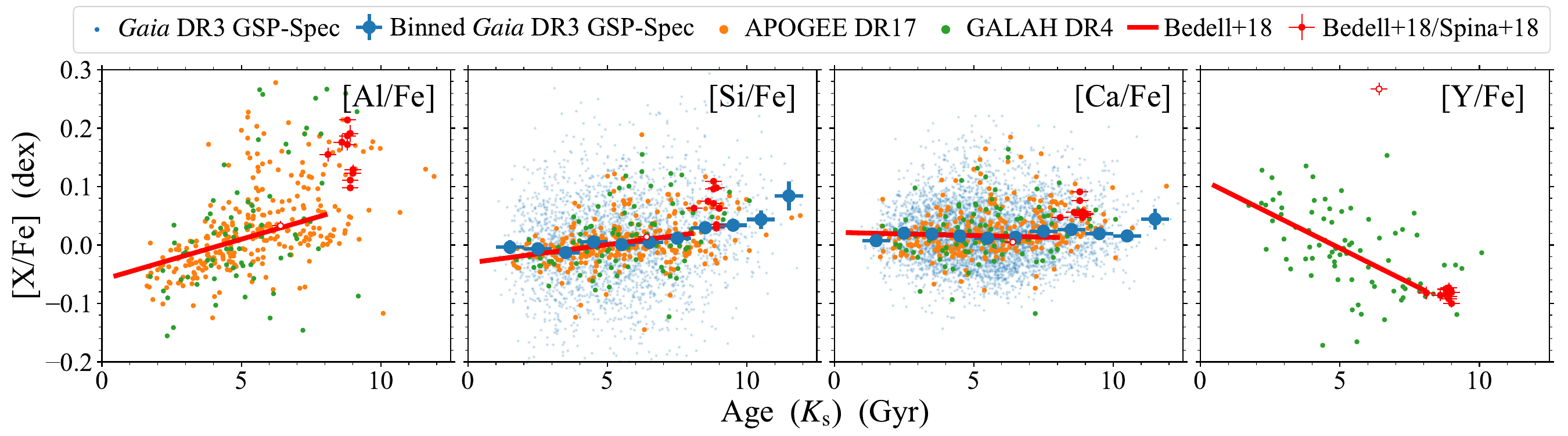}
\caption{Age--abundance relations for selected elements ([Al/Fe], [Si/Fe], [Ca/Fe], and [Y/Fe], from left to right). Blue dots show the GSP-Spec abundances of individual Solar twins, while blue filled circles with error bars indicate the medians of [X/Fe] within each $1\ur{Gyr}$ age bin. Orange and green dots represent abundances from APOGEE DR17~\citep{Majewski2017,Abdurrouf2022} and GALAH DR4~\citep{DeSilva2015,Buder2025}, respectively. Red thick line shows the linear age--abundance relation by \citet{Bedell2018} for comparison. Red filled circles and an open circle at $6.4\ur{Gyr}$ with error bars represent old $\alpha $-enhanced stars and an s-enhanced star, respectively, from \citet{Spina2018} and \citet{Bedell2018}. For each element from each survey, we applied a zero-point correction in [X/Fe] to make the abundance scale consistent with that of the age--[X/Fe] relation by \citet{Bedell2018}. The ages are determined using the $K_{\mathrm{s}}$-band magnitude, and only stars with relative age errors smaller than $50\%$ are included. For APOGEE DR17 and GALAH DR4, we include only stars with abundance errors below $0.05\ur{dex}$. We note that a small fraction of blue dots, as well as one [Al/Fe] measurement from APOGEE DR17 and one [Al/Fe] measurement from GALAH DR4, fall outside the plotted range and are therefore not shown. Plots for all the elements are presented in Fig.~\ref{fig:AgeAbnAll}. }
\label{fig:AgeAbn}
\end{figure*}

The precise and accurate chemical abundances of Solar twins have been used to address various astronomical questions, from planet--host interactions to the chemodynamical evolution of the Milky Way~(see, references in Sect.~\ref{Sec:Intro}). 
[X/Fe] abundances, rather than [X/H], are particularly useful for these purposes, because the surface [X/Fe] ratios for most elements change by only ${<}0.004\ur{dex}$ during main-sequence stellar evolution~\citep{Turcotte2002}, which is markedly smaller than the change in surface [X/H] of ${\sim }10\%$~\citep{ChristensenDalsgaard1996}. 
To briefly assess the usefulness of our catalog, we here show results obtained using our ages and chemical abundances. 

In our \textit{Gaia} DR3 GSP-Spec catalog, abundance measurements are available for only a fraction of Solar twins for most elements, with Ca being the exception, for which abundances are available for all twins. 
Restricting the analysis to stars with reliable abundance measurements, defined by the 14th--39th digits of the GSP-Spec Quality Flag corresponding to \texttt{XUpLim}${\leq }2$ and \texttt{XUncer}${\leq }1$, the resulting fractions are $0.70\%$ for [N/Fe], $3.4\%$ for [Mg/Fe], $49\%$ for [Si/Fe], $17\%$ for [S/Fe], $100\%$ for [Ca/Fe], $3.8\%$ for [Ti/Fe], $0.11\%$ for [Cr/Fe], $90\%$ for [Fe/M], and $1.5\%$ for [Ni/Fe], in order of atomic number. 
The GSP-Spec [X/Fe] abundances used throughout this work were calibrated following \citet{GaiaGSPSpecDR3,RecioBlanco2024}. 
To complement the chemical information of our catalog, we cross-matched the \textit{Gaia} DR3 \texttt{source\_id} of our Solar twins with those in APOGEE DR17~\citep{Majewski2017,Abdurrouf2022} and GALAH DR4~\citep{DeSilva2015,Buder2025} to obtain their abundance measurements, applying their recommended filtering. 
We also cross-matched these identifiers with the exoplanet data from NASA Exoplanet Archive~\citep{Christiansen2025}\footnote{\url{https://exoplanetarchive.ipac.caltech.edu/index.html}} retrieved on August 13, 2025~\citep[see, also][for a catalog of exoplanet host stars with \textit{Gaia} DR3 GSP-Spec]{deLaverny2025}. 
The retrieved APOGEE and GALAH abundances, as well as a flag indicating the presence or absence of confirmed exoplanets, are provided as part of our published Solar-twin catalog for reference~(Table~\ref{table:columndesc}). 

Figure~\ref{fig:AgeAbn} shows the relations between the derived stellar ages and elemental abundances\footnote{For all the \textit{Gaia} DR3 GSP-Spec, APOGEE DR17, and GALAH DR4 abundances, we added zero-point offsets in [X/Fe] so that, for elements analyzed by \citet{Bedell2018}, our Solar twins in the $0.5$--$8\ur{Gyr}$ range show no systematic offset from the linear age--[X/Fe] relations in that study, which were calibrated over a similar age interval. The added offsets are listed in Table~\ref{table:XFeoffsets}. }. 
Here we show results only for four elements (Al, Si, Ca, and Y). 
These four elements are selected on the grounds that (i) [Al/Fe] and [Y/Fe] are known to show tight and steep correlations with stellar age~\citep[e.g.,][]{Sharma2022}, and (ii) [Si/Fe] and [Ca/Fe] are measured for a relatively large number of stars with good precisions in our catalog. 
Results for the other elements available in our catalog are shown in Fig.~\ref{fig:AgeAbnAll} for reference. 
To ensure reliable age estimates, we include only stars with relative age uncertainties better than $50\%$. 
This cut, which naturally excludes most of the young Solar twins (${\lesssim }2\ur{Gyr}$) due to our selection effects (Sect.~\ref{ssec:mock}), is also beneficial because the stellar parameters of the youngest Solar twins could be less accurate owing to activity effects~\citep{YanaGalarza2019}. 

Age--abundance trends can serve as a diagnostic of the robustness and reliability of our age estimates, given that clear trends for Solar twins have already been reported by \citet[among others]{Bedell2018} using a small but high-precision sample. 
In addition, theoretical basis of these correlations can be understood within a framework that combines Galactic chemical evolution with radial migration of stars, in which the inner regions formed faster and became more metal-rich than the outer regions~\citep{Tsujimoto2021}. 
This view suggests that older Solar twins were born at smaller Galactocentric distances, i.e., closer to the Galactic center, because all Solar twins considered here have a common metallicity, nearly Solar. 
On the basis of this picture, we emphasize that age and birth radius are tightly coupled. 
Hence, our age--[X/Fe] relations are not relations defined at a fixed Galactocentric distance, nor, of course, at a fixed age. 

In the following, we discuss two topics regarding the age--[X/Fe] trends: slopes~(Sect.~\ref{ssec:ageabn_slope}) and s-enhanced Solar twins~(Sect.~\ref{ssec:ageabn_s}).

\subsection{Slopes of age--[X/Fe] trends}\label{ssec:ageabn_slope}

From Fig.~\ref{fig:AgeAbn}, we found that the age--[X/Fe] relations in our catalog, regardless of which abundance source is adopted (\textit{Gaia} DR3 GSP-Spec, APOGEE DR17, or GALAH DR4), show good agreement with the relations reported by \citet{Bedell2018} based on a small but very precise sample. 
In particular, for [Al/Fe] and [Y/Fe], whose age--abundance relations exhibit steep positive and negative slopes, respectively, our measurements reproduce the slopes obtained by \citet{Bedell2018} very well. 
These agreements support the reliability of our age determinations. 
Comparable agreement is also seen for most of the other elements shown in Fig.~\ref{fig:AgeAbnAll}. 
Although the uncertainties for individual stars in our catalog are relatively large, the consistency emerging from our much larger sample enables us to verify the age--[X/Fe] trends measured by \citet{Bedell2018} across a much larger Solar-twin sample. 

The exceptions to this behavior are the age--[X/Fe] relations for [S/Fe] and [Ni/Fe] from \textit{Gaia} DR3 GSP-Spec. 
Even though \citet{Bedell2018} presented positive age--[S/Fe] and [Ni/Fe] slopes, the age--[X/Fe] relations from GSP-Spec are rather flat for S and Ni. 
These flat trends from GSP-Spec are most likely due to selection effects (i.e., stars with small [X/Fe], and hence weak lines of element X, tend to lack [X/Fe] measurements), given that the lines from S and Ni are very weak in Solar-type stars in the RVS domain, although although the presence of unknown systematic errors in the GSP-Spec abundances cannot be entirely ruled out. 
Below, we examine this possibility for S and Ni in more detail. 

For [S/Fe], abundances from APOGEE DR17 indicate a positive relation with a slope of $+0.0082\pm 0.0023\ur{\si{dex.Gyr^{-1}}}$ (fitted using only Solar twins with ages younger than $8\ur{Gyr}$, to be consistent with the sample of \citet{Bedell2018}), which is in good agreement with the slope of $+0.0098\pm 0.0015\ur{\si{dex.Gyr^{-1}}}$ by \citet{Bedell2018}. 
GALAH DR4 does not provide [S/Fe]. 
Since S is an $\alpha $ element~\citep[e.g.,][and references therein]{Duffau2017,Perdigon2021}, the positive slope for [S/Fe] is expected, given that other $\alpha $ elements such as O and Mg also exhibit positive age--[X/Fe] trends. 
Nevertheless, because S sometimes exhibits behavior that differs from typical $\alpha $ elements~\citep[e.g., the negative radial {[S/Fe]} gradient seen in Cepheids;][]{Luck2018,daSilva2023}, caution is warranted when treating S as a standard $\alpha $ element. 
Taken together, we conclude that the age--[S/Fe] relation likely has positive relation, expected from its $\alpha $-element nature, and that the flat relation from \textit{Gaia} DR3 GSP-Spec is probably due to selection effects. 

For [Ni/Fe], when restricting ourselves to the stars plotted in Fig.~\ref{fig:AgeAbn} (i.e., stars with relative age errors smaller than $50\%$), APOGEE DR17 indicates a flat age--[X/Fe] relation within $1\sigma $, and GALAH DR4 indicates a possibly positive but nearly flat relation within $2\sigma $, seemingly contradicting the positive slope presented by \citet{Bedell2018}. 
However, when we remove the constraint on the relative age error, and hence include stars with very young ages (${\lesssim }2\ur{Gyr}$), we more or less recovered the age--[Ni/Fe] slope presented by \citet{Bedell2018} of $+0.0071\pm 0.0009\ur{\si{dex.Gyr^{-1}}}$, obtaining $+0.0043\pm 0.0006\ur{\si{dex.Gyr^{-1}}}$ from APOGEE DR17 and $+0.0061\pm 0.0009\ur{\si{dex.Gyr^{-1}}}$ from GALAH DR4 (again, including only stars with ages ${<}8\ur{Gyr}$). 
This is because, in APOGEE DR17 and GALAH DR4, stars with ages ${\lesssim }3\ur{Gyr}$ have systematically ${\sim }0.02\text{--}0.03\ur{dex}$ lower [Ni/Fe] than older stars. 
We also see a similar age--[Ni/Fe] trend in Fig.~3 of \citet{Bedell2018}, i.e., a nearly flat age--[Ni/Fe] relation for stars with ages ${\sim }2\text{--}8\ur{Gyr}$ and lower [Ni/Fe] for stars with ${\lesssim }2\ur{Gyr}$. 
In contrast, the \textit{Gaia} DR3 GSP-Spec age--[Ni/Fe] relation is continuously flat (or even possibly negative) over the entire $0\text{--}10\ur{Gyr}$ age range. 
Therefore, we conclude that the age--[Ni/Fe] relation likely has a flat trend, as in other Fe-peak elements, at ${\gtrsim }2\ur{Gyr}$, lower [Ni/Fe] at young ages, and that the flat \textit{Gaia} DR3 GSP-Spec relation is again probably due to selection effects. 

For several elements other than Ni (e.g., Mg, Co, and Sr), inspection of Fig.~3 of \citet{Bedell2018} suggests that the youngest Solar twins (${\lesssim }2\ur{Gyr}$) may exhibit slightly lower (or, for a few elements, slightly higher) [X/Fe] than would be expected from a linear extrapolation of the age--[X/Fe] relation of the older stars, as well as a larger dispersion than seen for the older stars, although this trend is not entirely clear. 
The apparent transition at an age of ${\sim }2\ur{Gyr}$, if real, coincides with the epoch of the recent star-formation peak~\citep{Mor2019,RuizLara2020,Paper2}. 
Such bursty star formation could have modified the relative contributions of type~Ia and core-collapse supernovae and other nucleosynthetic sources, thereby altering the [X/Fe] ratios of Solar twins younger than ${\sim }2\ur{Gyr}$~\citep[e.g.,][]{Tsujimoto1995,Johnson2021,Sun2025c}, since the X-to-Fe yield ratios are expected to differ among sources. 
Whether such bursts drive [X/Fe] upward or downward is rather complicated, it depends on several ingredients such as the differences in nucleosynthetic [X/Fe] yields among sources, the star-formation history, and the subsequent mixing of metals, and it is therefore not straightforward to predict the exact trend. 
The physical trigger of the recent star-formation peak also remains under debate. 
Nonetheless, it is intriguing that the epoch of ${\sim }2\ur{Gyr}$ ago coincides with the timing of possible external perturbations. 
For example, several studies have proposed that pericentric passages of the Sagittarius dwarf galaxy (Sgr) may have enhanced disk star formation~\citep[e.g.,][]{RuizLara2020,Annem2024}. 
As another example, a recent sudden drop in [Fe/H] has been observed and interpreted as evidence for a low-metallicity gas infall event~\citep{Spitoni2023,Palla2024}. 
Whatever the trigger of the bursty star formation, the resulting change in the relative contributions of nucleosynthetic sources might account for the departure of the young Solar twins from a simple linear age--[X/Fe] extrapolation. 
Aside from such physical scenarios for the departure, systematic [X/Fe] errors associated with stellar activity in the youngest stars~\citep{YanaGalarza2019} could provide another contribution. 

In light of the above discussion of [S/Fe] and [Ni/Fe], care should be taken when analyzing mean \textit{Gaia} DR3 GSP-Spec abundances at the very high precision level of a few-hundredths of a dex, even though with \textit{Gaia} DR3 GSP-Spec we succeeded in confirming the age--[X/Fe] trends by \citet{Bedell2018} for most elements. 
Larger, higher-precision samples are required to reach firm conclusions regarding the age-[X/Fe] trends. 
GSP-Spec abundances in \textit{Gaia} DR4 will be valuable in this context, given the expected increase in S/N thanks to an observing duration roughly twice as long as that of DR3. 
At the same time, characterizing activity-related systematics will also be important for interpreting abundances at young ages.

\subsection{s-process-enhanced stars}\label{ssec:ageabn_s}

Other key findings by \citet{Spina2018} and \citet{Bedell2018} from the age--[X/Fe] plots are (1)~higher [$\alpha $/Fe] ratios for most stars with ages ${\gtrsim }8\ur{Gyr}$ than expected from a linear trend defined by stars with ages ${\lesssim }8\ur{Gyr}$, and (2)~one star (HIP~64150) showing an anomalously high [s/Fe] ratio. 
Here we investigate the latter, s-enhanced stars in this section, while the former, $\alpha $-enhanced stars are investigated in our companion paper~\citep[Paper~III;][]{Paper3}. 

\citet{Spina2018} found that one star (HIP~64150) among their $79$ Solar twins shows an enhanced [s/Fe] ratio but not an enhanced [r/Fe]. 
As representative elements for the s- and r-processes, they measured $\text{[Y/Fe]}=+0.267$ and $\text{[Eu/Fe]}=+0.041\ur{dex}$, respectively. 
This [Y/Fe] value is $+0.31\ur{dex}$ higher than the linear age--[Y/Fe] relation by \citet{Bedell2018}, whereas [Eu/Fe] is only $0.01\ur{dex}$ smaller than the relation. 
Such enhancements in [s/Fe] with normal [r/Fe] have also been observed in two other Solar analogs, with $\text{[Y/Fe]}=+0.533\ur{dex}$ for HIP~10725\footnote{[Fe/H] of HIP~10725 is -0.173, which is much lower than our Solar-twin criterion (${\pm }0.1\ur{dex}$ around the Solar value). }~\citep{Schirbel2015} and $\text{[Y/Fe]}=+0.92\ur{dex}$ for HD~138004\footnote{[Fe/H] of HD~138004 is $-0.109\pm 0.011\ur{dex}$, slightly outside our Solar-twin criterion (${\pm }0.1\ur{dex}$ around the Solar value), but consistent with it within the uncertainty. Thus, we include this star in our discussion. }~\citep{Liu2020}. 
For HIP~64150 and HIP~10725, time variations in their radial velocities, most likely due to unseen white-dwarf companions, have been found~\citep[respectively]{Schirbel2015,dosSantos2017}, and pollution from a former AGB companion was suggested. 
For the last star, HD~138004, a radial-velocity variation of ${\sim }10\ur{\si{km.s^{-1}}}$ has been reported~\citep{Butler2017,TalOr2019}. 
Combined with the astrometric acceleration measurements by \citet{Brandt2018,Brandt2021}, this variation implies a companion with a mass of $0.658M_{\odot }$~\citep{An2025}. 
The same technique was also applied to HIP~64150, yielding a companion mass of $0.603M_{\odot }$~\citep{An2025}. 
We searched for other s-enhanced stars in the literature listed in Table~\ref{table:compsample}, but no clear additional cases were found. 
If we relax the conditions on the stellar parameters for being Solar twins, several probable s-enhanced Solar analogs ($2$, $4$, and $1$ stars) appear in the samples from \citet{daSilva2012}, \citet{Liu2020}, and \citet{Shejeelammal2024}, respectively, though these objects have large ages (${\gtrsim }10\ur{Gyr}$). 
Interestingly, \citet{Rampalli2024} published the [Y/Fe] abundances for $17,412$ Solar twins (in their definition), but none of these stars has [Y/Fe] larger than $3\sigma $ (${\simeq }0.2\ur{dex}$) from their age--[Y/Fe] relation. 
This lack of s-enhanced stars might be related to the data-driven nature of their analysis and the lack of detectable s-process lines in \textit{Gaia} RVS spectra of Solar twins with low or moderate s-process enhancements (see Fig.~1 of \citealp{Contursi2023} and Fig.~2 of \citealp{Contursi2024} for Ce and Nd, respectively). 

Using our catalog of Solar twins, we searched for s-enhanced twins using GALAH DR4 [Y/Fe] abundances. 
We found that none of the $80$ Solar twins plotted in Fig.~\ref{fig:AgeAbnAll} has a zero-point corrected [Y/Fe] value that is $0.2\ur{dex}$ or more above the age--[Y/Fe] relation by \citet{Bedell2018}. 
There is one star (Gaia DR3~5365738724921515776) with a zero-point-corrected [Y/Fe] of $0.153\pm 0.022\ur{dex}$, which is $0.199\ur{dex}$ higher than the relation by \citet{Bedell2018}, but even in this case the enhancement in [Y/Fe] is modest. 
The absence (or presence of only one) s-enhanced star among our $80$ stars could simply be due to small-number statistics, given that there is only one such star among the $79$ Solar twins of \citet{Bedell2018}. 
However, it could also reflect our removal of possible binary stars as much as possible using \textit{Gaia} flags and photometric constraints. 
To assess this possibility, we examined the absolute [Y/Fe] values for all Solar-twin candidates with GALAH~DR4 [Y/Fe] abundances, without applying the astrometric and photometric criteria used to remove binaries in our final sample. 
We found that $3$ out of $286$ such stars (${\sim }1\%$) have [Y/Fe] values exceeding $0.2\ur{dex}$, a fraction comparable to that found by \citet{Bedell2018}, although this comparison is based on absolute [Y/Fe] values rather than offsets from the age--[Y/Fe] relation. 
This interpretation is also partly supported by the fact that the s-enhanced star HD~138004 is included in our parent \textit{Gaia} sample but removed by its RUWE value and \texttt{non\_single\_star} flag, whereas another s-enhanced star, HIP~64150, passes our selection.

\section{Summary}\label{sec:summary}

This study constructed a large catalog of Solar twins with isochrone ages using model-driven, rather than data-driven, \textit{Gaia} DR3 GSP-Spec stellar parameters ($T_{\mathrm{eff}}$, $\log g$, [M/H], and [$\alpha $/Fe]). 

We selected candidate Solar twins with high-quality GSP-Spec parameterization~(Sect.~\ref{ssec:gaiadata}). 
We required two quality criteria for GSP-Spec: all the $13$ first GSP-Spec Quality Flags equal to zero, and the goodness-of-fit satisfies $\log \chi ^{2}<-3.2$. 
With these constraints, the uncertainties in the stellar parameters are smaller than $100\ur{K}$ in $T_{\mathrm{eff}}$, $0.2$ in $\log g$, $0.1\ur{dex}$ in [M/H], and $0.05\ur{dex}$ in [$\alpha $/Fe]. 
We calibrated the stellar parameters in two steps (first using the relations provided by \citet{RecioBlanco2024}, and then applying the corrections described in Appendix~\ref{app:zeropoints}) to place them on a scale consistent with high-resolution, high-precision spectroscopic results~\citep{TucciMaia2016,Nissen2020,Shejeelammal2024}. 
We then selected $7{,}918$ Solar-twin candidates whose $T_{\mathrm{eff}}$, $\log g$, and [M/H] fall within ${\pm }200\ur{K}$, ${\pm }0.2$, and ${\pm }0.1\ur{dex}$ of the Solar values, respectively. 
We filtered out stars with inaccurate parameterization and possible non-single stars as much as possible using RUWE, a few \textit{Gaia} flags (\texttt{astrometric\_params\_solved}, \texttt{duplicated\_source}, and \texttt{non\_single\_star}) and photometric constraints (Sects.~\ref{ssec:gaiadata} and \ref{ssec:flag} and Fig.~\ref{fig:flag}). 
Applying these constraints yielded our final Solar-twin sample consisting of $6{,}594$ stars. 

We determined ages, initial masses $M_{\mathrm{ini}}$, and initial metallicities $\text{[M/H]}_{\mathrm{ini}}$ for individual Solar twins with a Bayesian isochrone-projection~(Sects.~\ref{ssec:PARSEC} and \ref{ssec:ageprocedure}). 
In this step, we employed the PARSEC isochrones version 1.2S~\citep{Bressan2012,Chen2015} obtained with the CMD~3.8 web interface. 
Because the native isochrone grid is too sparse to determine ages accurately for Solar twins, we interpolated it to construct a sufficiently dense grid~(Appendix~\ref{app:PARSECinterp}). 
Assuming flat priors on age, $M_{\mathrm{ini}}$, and $\text{[M/H]}_{\mathrm{ini}}$, we derived posterior estimates using three observable triplets: $(T_{\mathrm{eff}}, \log g, \mathrm{[M/H]}_{\mathrm{curr}})$, $(T_{\mathrm{eff}}, M_{G}, \mathrm{[M/H]}_{\mathrm{curr}})$, and $(T_{\mathrm{eff}}, M_{K_{\mathrm{s}}}, \mathrm{[M/H]}_{\mathrm{curr}})$. 

We validated the resulting ages in two ways~(Sect.~\ref{sec:validation}). 
First, we applied our age-determination procedure to the Sun~(Sect.~\ref{ssec:SolarAge} and Table~\ref{table:SunAge}). 
We confirmed that the recovered age and $M_{\mathrm{ini}}$ agree with literature values of $4.5\text{--}4.6\ur{Gyr}$ and $1M_{\odot }$ for all three age-determination scheme. 
Second, we constructed a mock Solar-twin sample consisting of $75{,}588$ artificial stars~(Sect.~\ref{sssec:createMock} and Appendix~\ref{app:mimickerror}) and compared its statistical properties with those of the observed sample~(Sect.~\ref{sssec:compMock}). 
Comparisons of the age--$M_{\mathrm{ini}}$ distributions~(Figs.~\ref{fig:AgeMini} and \ref{fig:AgeMiniMock}) and the distribution of $M_{\mathrm{ini}}$ at a fixed age~(Fig.~\ref{fig:Minihist}) show good consistency between the observed and mock catalogs for ages based on $M_{G}$ or $M_{K_{\mathrm{s}}}$. 
This fact supports the reliability of the GSP-Spec parameterization, our age determination, and the mock construction. 
In contrast, we found inconsistencies for ages derived from $\log g$, likely reflecting the difficulty of determining enough accurate $\log g$ from the limited wavelength range of the \textit{Gaia} RVS. 
Taken together with direct comparisons among the three age estimates~(Fig.~\ref{fig:Agecomp}) we concluded that both $M_{G}$- and $M_{K_{\mathrm{s}}}$-based ages are reliable and hence suitable for statistical analyses. 
We adopted the $M_{K_{\mathrm{s}}}$-based ages in the subsequent analysis~(Sect.~\ref{ssec:comp_threeage}). 

To place our catalog in the context of a decade of Solar-twin studies, we compared our Solar-twin catalog with earlier catalogs~(Sect.~\ref{sec:agecomp}). 
A comparison of basic metrics~(Sect.~\ref{ssec:metrics} and Table~\ref{table:compsample}) shows that some previous high-resolution Solar-twin samples contain only several tens of stars but achieve extremely high precision in $T_{\mathrm{eff}}$, $\log g$, [Fe/H], and abundances. 
In contrast, our catalog and a few other recent works based on large spectroscopic surveys provide samples that are about two orders of magnitude larger but have worse precision in individual stellar parameters. 
While previous large Solar-twin catalogs ~\citep{Rampalli2024,Walsen2024,Lehmann2025} rely on data-driven methods, our catalog is built from model-driven stellar parameters, ages, $M_{\mathrm{ini}}$, and chemical abundances. 
The reasoning behind this choice is to avoid propagating trends learned from training samples (i.e., previous small-scale datasets) and to verify the results obtained from past high-precision studies using our lower-precision but substantially larger sample. 
We adopted a strict criterion of ${\pm }0.1\ur{dex}$ around Solar [M/H] to ensure that our Solar twins are near-Solar-metallicity main-sequence stars and that their ages can be tightly linked to their birth radii. 
Moreover, our construction of the mock sample is valuable for characterizing the selection function of our sample and, consequently, for investigating the probability distributions of stellar parameters~\citep[e.g., for investigating the underlying age distribution in our companion paper, Paper~II;][]{Paper2}. 
By comparing our stellar parameters, ages, and $M_{\mathrm{ini}}$ with values from high-precision literature samples~(Sect.~\ref{ssec:compliter}), we found good agreement with many previous differential Solar-twin studies~(Fig.~\ref{fig:compall1}), but discrepancies relative to some large, data-driven catalogs~(Fig.~\ref{fig:compall2}). 

To showcase the scientific potential of our catalog, we investigated age--[X/Fe] relations~(Sect.~\ref{sec:chem}). 
For this purpose, we combined GSP-Spec, APOGEE DR17, and GALAH DR4 abundances to compare the resulting age--[X/Fe] trends with those from high-precision studies by \citet{Bedell2018} and \citet{Spina2018}~(Figs.~\ref{fig:AgeAbn} and \ref{fig:AgeAbnAll}). 
We discussed the age--[X/Fe] relations in two regards: slopes and s-enhanced stars. 
Focusing first on the slopes of the age--[X/Fe] relations~(Sect.~\ref{ssec:ageabn_slope}), we recovered the classical positive age--[Al/Fe] and negative age--[Y/Fe] trends, supporting the reliability of our age determinations. 
For most elements, the age--[X/Fe] slopes agree well with those derived from a small high-precision sample by \citet{Bedell2018}, supporting these relations in a more statistical regime. 
Nevertheless, we found discrepancies for S and Ni measured by \textit{Gaia} DR3 GSP-Spec, for which selection effects or systematic errors may have altered the GSP-Spec abundances. 
We also pointed out that age--[X/Fe] relations are not necessarily linear and that the age distribution of the Solar-twin sample can modify the measured slopes, potentially even changing their signs. 
In addition, we identified a possible sudden drop in [Ni/Fe] (and in some other elements) around $2\ur{Gyr}$ ago. 
Second, we searched for s-process-enhanced Solar twins using GALAH DR4 [Y/Fe] measurements~(Sect.~\ref{ssec:ageabn_s}). 
We found no clear examples (at most one candidate) among the $80$ Solar twins with available [Y/Fe]. 
We will also investigate the $\alpha $-enhanced Solar twins in our companion paper~\citep[Paper~III;][]{Paper3}. 

Beyond the analyses presented in this study, our large, homogeneous, and well-characterized catalog provides a complementary resource to both high-precision small samples and large data-driven catalogs. 
It enables further investigations that benefit from model-driven stellar parameters with a quantified selection function. 
In our companion paper~\citep[Paper~II;][]{Paper2}, we will use our catalog to examine the age distribution of Solar twins, and the implied efficiency of radial migration of the Sun. 
Our final catalog of $6{,}594$ Solar twins is published online at CDS (see Table~\ref{table:columndesc} for a description of the catalog columns).

\begin{acknowledgements}
This work has been supported by the Tokyo Center For Excellence Project, Tokyo Metropolitan University. 
DT acknowledges financial support from JSPS Research Fellowship for Young Scientists and accompanying Grants-in-Aid for JSPS Fellows (23KJ2149). 
PdL and ARB acknowledge partial funding from the European Union's Horizon 2020 research and innovation program under SPACE-H2020 grant agreement number 101004214 (EXPLORE project). 
TT acknowledges the support by JSPS KAKENHI Grant No. 23H00132. 
\\
This work presents results from the European Space Agency~(ESA) space mission \textit{Gaia}. \textit{Gaia} data are being processed by the \textit{Gaia} Data Processing and Analysis Consortium~(DPAC). Funding for the DPAC is provided by national institutions, in particular the institutions participating in the Gaia MultiLateral Agreement (MLA). The Gaia mission website is \url{https://www.cosmos.esa.int/gaia}. The Gaia archive website is \url{https://archives.esac.esa.int/gaia}. 
This publication makes use of data products from the Two Micron All Sky Survey, which is a joint project of the University of Massachusetts and the Infrared Processing and Analysis Center/California Institute of Technology, funded by the National Aeronautics and Space Administration and the National Science Foundation. 
\end{acknowledgements}

\bibliographystyle{aa} 
\bibliography{ref}

\begin{appendix}

\section{Zero-point corrections of GSP-Spec stellar parameters}\label{app:zeropoints}

\begin{figure*}
\centering 
\includegraphics[width=18cm]{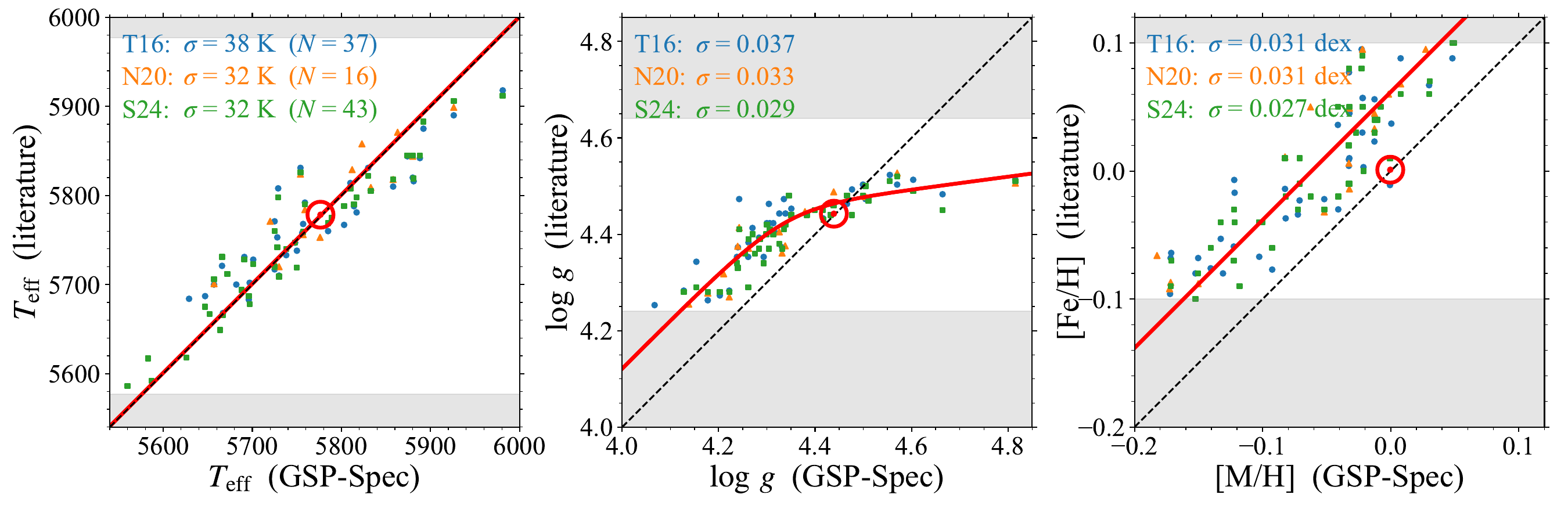}
\caption{Our \textit{Gaia} DR3 GSP-Spec and literature stellar parameters of Solar twins. GSP-Spec parameters were already calibrated following \citet{RecioBlanco2024}. Blue circles, orange triangles, and green squares represent comparisons with \citet[T16; $37$ stars]{TucciMaia2016}, \citet[N20; $16$ stars]{Nissen2020}, and \citet[S24; $43$ stars]{Shejeelammal2024}, respectively. Typical quoted errors are ${\sim }5\text{--}10\ur{K}$ in $T_{\mathrm{eff}}$, $0.02$ in $\log g$, and $0.005\text{--}0.01\ur{dex}$ in [M/H], for both GSP-Spec and literature parameters. Only stars with literature parameters satisfying our Solar-twin criteria are shown (i.e., stars in gray-shaded areas are excluded). The red open marker indicates the parameters of the Sun. Red thick lines show our adopted relations to convert GSP-Spec parameters to the literature scales. The scatters around the adopted relations are annotated in each panel. }
\label{fig:literparams}
\end{figure*}

The accuracy of the GSP-Spec stellar parameters is generally sufficient for many types of analyses. 
Even though, for Solar-twin studies, which require very high zero-point accuracy relative to the Solar values, the GSP-Spec zero-point accuracy is insufficient. 
In fact, comparisons with differentially derived Solar-twin stellar parameters have revealed systematic zero-point biases~\citep{Fouesneau2023,Gaia2023golden}. 
To quantify such biases in our dataset, we compared our GSP-Spec stellar parameters with those determined by \citet{TucciMaia2016}, \citet{Nissen2020}, and \citet{Shejeelammal2024}. 
These three studies are extensions and reanalyses of former studies by \citet{Ramirez2014}, \citet{Nissen2015,Nissen2016}, and \citet{Spina2018,Martos2023,Rathsam2023}, respectively. 
We arbitrarily selected these particular studies, among several that also meet the criteria below, for three reasons. 
First, they employed line-by-line differential analyses relative to the Solar spectrum, ensuring reliable zero points. 
Second, they derived atmospheric parameters with high precision using high-quality spectra, i.e., S/N of ${\gtrsim }500$ using the MIKE~\citep{Bernstein2003} or HARPS~\citep{Mayor2003} high-resolution spectrographs. 
Third, their samples contain a sufficiently large overlap of several tens of stars with ours. 

Figure~\ref{fig:literparams} compares our GSP-Spec stellar parameters with those from these three studies. 
Only stars meeting all of the following conditions are included: 
\begin{itemize}
\item their literature parameters (not GSP-Spec values) satisfy our Solar-twin criteria, i.e., $T_{\mathrm{eff}}$, $\log g$, and [M/H] within ${\pm }200\ur{K}$, ${\pm }0.2$, and ${\pm }0.1\ur{dex}$ of the Solar values; 
\item they satisfy the quality assessments described in Sect.~\ref{ssec:gaiadata}; 
\item they have S/N (\texttt{rv\_expected\_sig\_to\_noise}) ${>}80$ and reduced $\chi ^{2}$ (\texttt{logchisq\_gspspec}) ${<}-3.7$, ensuring best precision in GSP-Spec. 
\end{itemize}
We note that, the literature targets lie near the bright end of our GSP-Spec sample, and thus the third criterion on the precision of the GSP-Spec parameters excludes only one star. 

The three comparisons show excellent agreement in $T_{\mathrm{eff}}$, given the scatter. 
Similar levels of agreement are found in comparisons with other studies not shown in the figure. 
In contrast, clear residual biases remain in GSP-Spec $\log g$ and [M/H], despite our prior calibration of these parameters using the relations presented by \citet{RecioBlanco2024}, which were designed to match the $\log g$ and [M/H] scales to APOGEE DR17~\citep{Abdurrouf2022} and GALAH DR3~\citep{Buder2021} catalogs. 

In order to ensure that the GSP-Spec stellar parameters are strictly differential with respect to the Solar values, we further calibrated GSP-Spec stellar parameters. 
For $T_{\mathrm{eff}}$ and [M/H], we calculated the mean of the difference between GSP-Spec and literature parameters, and added these offsets, $1\ur{K}$ in $T_{\mathrm{eff}}$ and $0.062\ur{dex}$ in [M/H], to the GSP-Spec parameters. 

For $\log g$, we found that GSP-Spec $\log g$ values for stars with literature $\log g \sim 4.5$ exhibit a broad distribution extending up to ${\sim }4.8$. 
The cause of this discrepancy is unclear, but it is possibly related to model-grid issues in GSP-Spec and/or in the literature. 
Given that the three literature sources employed high-resolution and high-S/N spectra and that the trends between GSP-Spec and literature $\log g$ are similar among the three studies, we decided to fit the GSP-Spec and literature $\log g$ data with a smoothly connected broken line. 
The choice of relying on the literature values is supported by the fact that there is a good linear relation between $\log g$ in the three literature sources and bolometric luminosity (and hence evolutionary $\log g$). 
The obtained relation is 
\begin{align}
&y_{1} = (x - 4.44) + b_{1} \label{Eq:logg_y1} \\
&y_{2} = a_{2}(x - 4.44) + b_{2} \label{Eq:logg_y2} \\
&y - 4.44 = -\frac{1}{\alpha }\ln (e^{-\alpha y_{1}} + e^{-\alpha y_{2}})\text{,} \label{Eq:logg_y}
\end{align}
where $x$ and $y$ represent the GSP-Spec parameters with the calibration described by \citet{RecioBlanco2024} and the literature parameters, respectively, Equations~\ref{Eq:logg_y1} and \ref{Eq:logg_y2} correspond to the lines for smaller and larger $\log g$, respectively, and Equation~\ref{Eq:logg_y} represents a LogSumExp-type smooth minimum function with the parameter $\alpha $ arbitrary chosen as $20$. 
Fitting to the data points shown in Fig.~\ref{fig:literparams} yields $a_{2}=0.133$, $b_{1}=0.121$, and $b_{2}=0.032$. 
The fitted line is shown in the figure as the red thick line. 
Using Equation~\ref{Eq:logg_y}, we calibrated our GSP-Spec $\log g$ values and their errors.

Another possible approach, which we have not adopted here, would be to rely on the relative $\log g$ values of GSP-Spec, and simply add a $0.104$ offset to the GSP-Spec $\log g$ values, ignoring the broken relation seen in Fig.~\ref{fig:literparams}. 
This alternative might be supported by the seemingly linear relations with unit slope between GSP-Spec $\log g$ and those determined by \citet{Gonzalez2010}, \citet{Gonzalez2014}, \citet{PortodeMello2014}, and \citet{Spina2021} up to $\log g{\sim }4.5\text{--}4.6$.

\section{Further interpolation of the isochrone library}\label{app:PARSECinterp}

To accurately recover the age (and $M_{\mathrm{ini}}$) of a Solar twin, the HR diagram and the Kiel diagram need to be densely covered with isochrone grid points, with a spacing comparable to or smaller than the observational uncertainties. 
The uncertainties of [M/H], $T_{\mathrm{eff}}$, $\log g$, $M_{G}$, and $M_{K_{\mathrm{s}}}$ in our final Solar-twin catalog are $0.01\text{--}0.06\ur{dex}$, ${\sim }20\text{--}100\ur{K}$, $0.01\text{--}0.1$, $0.003\text{--}0.05$, and $0.02\text{--}0.03$, respectively, for most stars ($5\%$ and $95\%$ percentiles). 
The steps in the isochrone library should ideally be comparable to, or finer than, the minimum values of these precision ranges (e.g., finer than $20\ur{K}$). 
The condition for [M/H] is satisfied, given that we retrieved the isochrone library with $0.005\ur{dex}$ increments in $\mathrm{[M/H]}_{\mathrm{ini}}$. 
Hence, we here check the increments of the other four parameters. 

Our selection of Solar twins covers both main-sequence stars and old main-sequence turn-off stars. 
For turn-off stars, our grid with $0.1\ur{Gyr}$ increments retrieved from the CMD~3.8 web interface provides sufficiently high density on the $T_{\mathrm{eff}}$ vs $\log g$ Kiel diagram. 
Indeed, for the $10\ur{Gyr}$ isochrone, the age difference of $0.1\ur{Gyr}$ at the turn-off corresponds to a $T_{\mathrm{eff}}$ difference of $5\ur{K}$, which is a few times finer than the best $T_{\mathrm{eff}}$ precision. 
Also, the $\log g$ spacing around the turn-off is small, ${\lesssim }0.05$, owing to the fine $M_{\mathrm{ini}}$ increment of ${\lesssim }0.01M_{\odot }$ around there, which is comparable to the upper $5\%$ percentile of the $\log g$ precision of $0.045$. 
Still, the $M_{G}$ and $M_{K_{\mathrm{s}}}$ spacing around the turn-off of ${\sim }0.1\ur{mag}$ is an order worse than the typical precision of our photometric data. 

In contrast, the $M_{\mathrm{ini}}$ increments in the retrieved grid for main-sequence stars are not fine enough for $T_{\mathrm{eff}}$, $M_{G}$, and $M_{K_{\mathrm{s}}}$. 
The $M_{\mathrm{ini}}$ increments for main-sequence stars range from ${<}0.01$ to $0.05M_{\odot }$. 
The resulting $\log g$ step at $1M_{\odot }$ is up to ${\sim }0.03$, which is sufficient for recovering ages. 
However, given $\abs{\partial T_{\mathrm{eff}}/\partial M_{\mathrm{ini}}}\sim 3500\ur{K/M_{\odot }}$ along main-sequence isochrones, the $T_{\mathrm{eff}}$ step becomes ${>}150\ur{K}$ in the worst cases, which is several times coarser than the typical observational $T_{\mathrm{eff}}$ precision. 
Similarly, gradients of $M_{G}$ and $M_{K_{\mathrm{s}}}$ of $5\text{--}8$ and $3\text{--}7\ur{mag/M_{\odot }}$, respectively, correspond to steps of ${\sim }0.4\ur{mag}$ in the worst cases. 
These coarse $T_{\mathrm{eff}}$, $M_{G}$ and $M_{K_{\mathrm{s}}}$ steps are particularly evident in young isochrones with ages ${\lesssim }1.5\ur{Gyr}$, where $M_{\mathrm{ini}}$ increments are typically $0.05M_{\odot }$. 
Thus, if we were to determine ages and $M_{\mathrm{ini}}$ using the retrieved isochrone grid as it is, the likelihood would tend to place most of the weight on the nearest $M_{\mathrm{ini}}$ grid points, even when the true value lies between them, thereby pulling the inferred $M_{\mathrm{ini}}$ toward those grid points and introducing systematic biases. 
Once $M_{\mathrm{ini}}$ is biased in this way, the simultaneously inferred age is biased as well, despite the fine age grid. 

In order to overcome the issue of the coarse grid, we added isochrone points with $0.001M_{\odot }$ increments, by linearly interpolating existing isochrone grid points. 
The resulting maximum steps of $T_{\mathrm{eff}}$, $\log g$, $M_{G}$, and $M_{K_{\mathrm{s}}}$ along the isochrones are ${\sim }4\ur{K}$, $0.003$, $0.01\ur{mag}$, and $0.01\ur{mag}$, respectively.

\section{Mimicking the errors in stellar parameters}\label{app:mimickerror}

\begin{figure}
\centering 
\includegraphics[width=9cm]{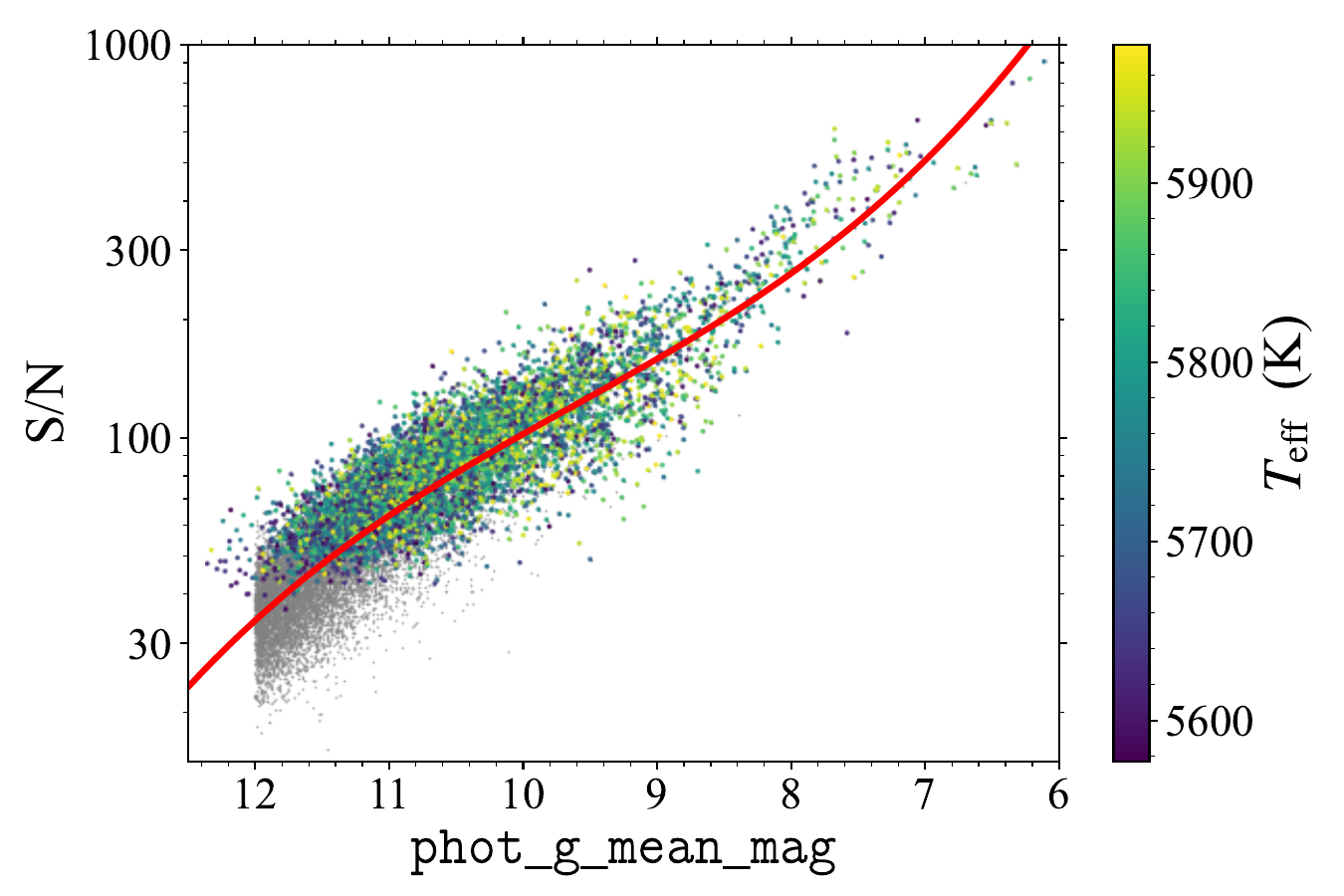}
\caption{Dependence of S/N on the \textit{G}-band magnitude. Dots color-coded by $T_{\mathrm{eff}}$ show our final $6{,}594$ Solar twins, while gray dots represent $27{,}782$ stars without imposing the condition on the seventh digit of the GSP-Spec Quality Flags on parameterization uncertainties. Red thick curve shows the cubic polynomial relation between the \textit{G}-band magnitude and the common logarithm of S/N, obtained by fitting the gray data points. }
\label{fig:SN_cTeff}
\end{figure}

\begin{figure*}
\centering 
\includegraphics[width=18cm]{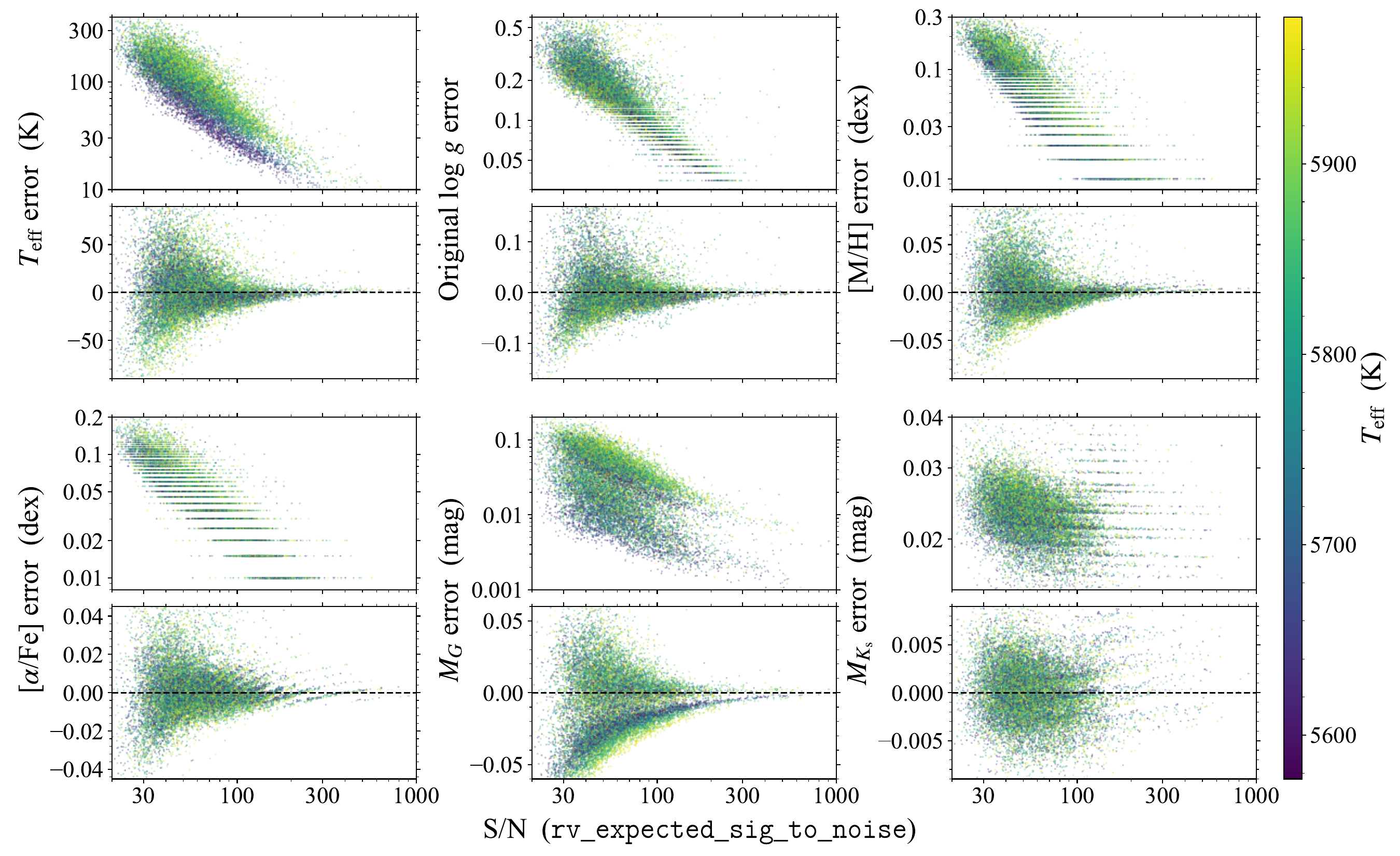}
\caption{Dependence of the errors in stellar parameters on S/N and $T_{\mathrm{eff}}$. }
\label{fig:literparams_cTeff}
\end{figure*}

\begin{figure*}
\centering 
\includegraphics[width=18cm]{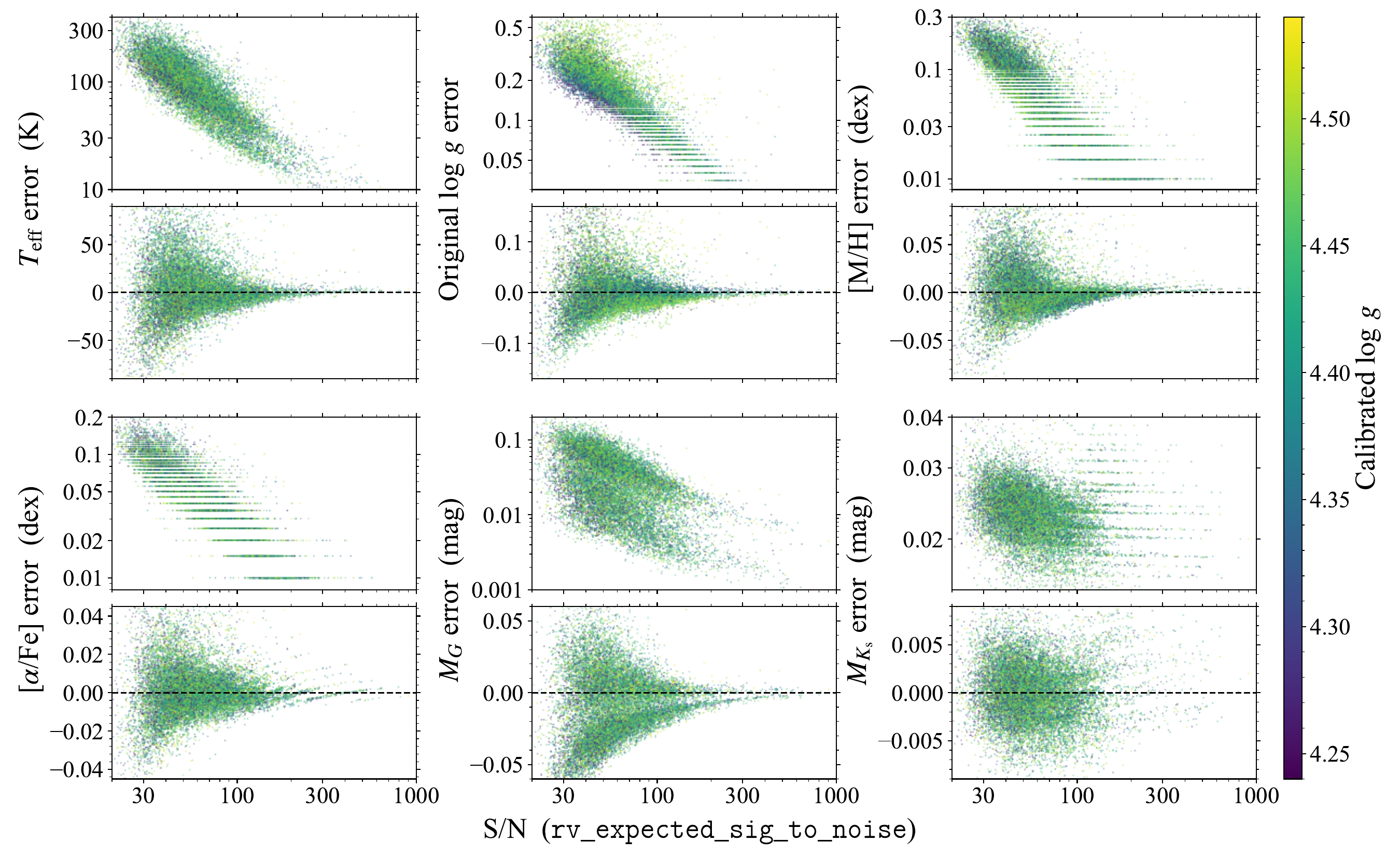}
\caption{Same as Fig.~\ref{fig:literparams_cTeff}, but color-coded by calibrated $\log g$. }
\label{fig:literparams_clogg}
\end{figure*}

\begin{figure*}
\centering 
\includegraphics[width=18cm]{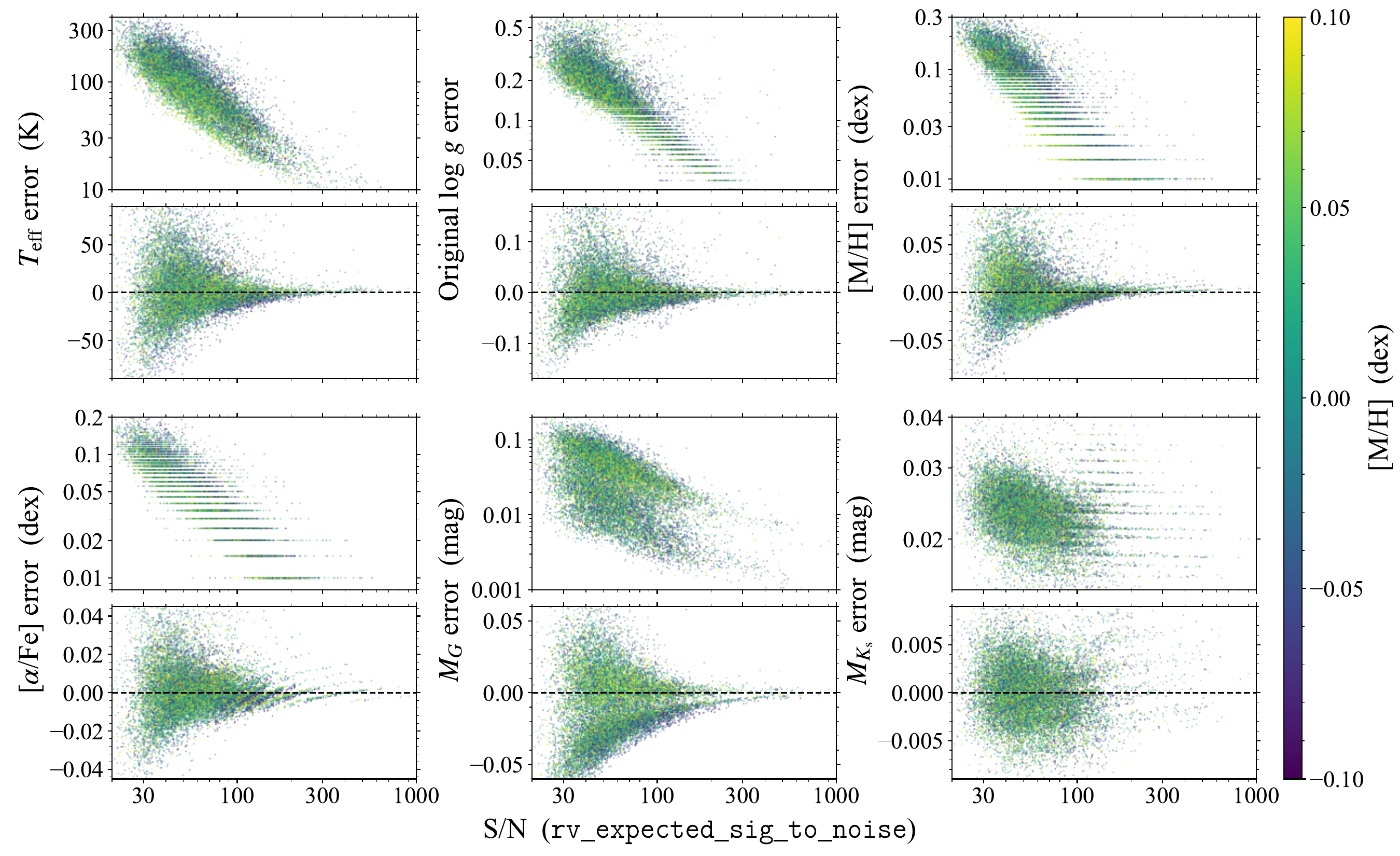}
\caption{Same as Fig.~\ref{fig:literparams_cTeff}, but color-coded by [M/H]. }
\label{fig:literparams_cMeta}
\end{figure*}

\begin{table*}
\centering 
\caption{Linear coefficients in Equation~\ref{Eq:LinearErrorCoef} to estimate the errors in Solar-twin parameters as a function of S/N and some stellar parameters. }
\label{table:CoefsError}
\begin{tabular}{l ccccccc}\hline \hline
Parameter & $a_{0}$ & $a_{\mathrm{S/N}}^{}$ & $a_{T_{\mathrm{eff}}}^{}$ & $a_{{T_{\mathrm{eff}}}^{2}}^{}$ & $a_{\log g}^{}$ & $a_{\mathrm{[M/H]}}^{}$ & RMS \\ \hline
$T_{\mathrm{eff}}$ (K) & $+4.0288$ & $-1.1752$ & $+0.0747$ & $-0.0194$ & $+0.0116$ & $-0.0753$ & $0.1142$ \\
Original $\log g$ & $+1.0109$ & $-1.0101$ & $+0.0113$ & $-0.0109$ & $+0.0547$ & $-0.0378$ & $0.1108$ \\
{[M/H]} (dex) & $+1.3692$ & $-1.4653$ & $+0.0379$ &  & $-0.0142$ & $-0.0928$ & $0.1477$ \\
{[$\alpha $/Fe]} (dex) & $+0.7201$ & $-1.1574$ & $+0.0176$ &  & $-0.0102$ & $-0.0449$ & $0.1109$ \\
$M_{G}$ (mag) & $+0.4061$ & $-1.0211$ & $+0.0623$ &  & $+0.0103$ & $-0.0350$ & $0.1722$ \\
$M_{K_{\mathrm{s}}}$ (mag) & $-1.4829$ & $-0.0870$ & $+0.0074$ &  &  &  & $0.0660$ \\
\hline
\end{tabular}
\end{table*}

In Sect.~\ref{sssec:createMock}, we constructed a mock catalog of Solar twins, by mimicking the observed stellar parameters, for which errors had to be assigned. 
To guide this assignment, we first investigated how S/N depends on magnitude and stellar parameters. 
We then examined how the errors in six parameters ($T_{\mathrm{eff}}$, $\log g$, [M/H], [$\alpha $/Fe], $M_{G}$ and $M_{K_{\mathrm{s}}}$) in our Solar-twin catalog depend on the stellar parameters ($T_{\mathrm{eff}}$, $\log g$, and [M/H]) and on S/N. 
We note that here we use the ``original'' values (i.e., those before the calibration in Appendix~\ref{app:zeropoints}) for the $\log g$ error, while we use the ``calibrated'' values for the other parameters including $\log g$ itself. 
We also note that for [M/H] and its error, we use the values before correcting for the effect of [$\alpha $/Fe]. 

For the sample in this section, we considered $27{,}782$ \textit{Gaia} DR3 stars that (1)~have the first $13$ GSP-Spec Quality Flags, except for the seventh one, equal to zero, (2)~have parameters that pass our Solar-twin selections in Sects.~\ref{ssec:gaiadata} and \ref{ssec:flag}, and (3)~satisfy $G<12$. 
These conditions are intended to avoid imposing any threshold on parameterization uncertainties, thereby allowing stars with large uncertainties to be included. 

Figure~\ref{fig:SN_cTeff} shows the relation between the \textit{G}-band magnitude and the common logarithm of S/N. 
We see no dependence of this relation on $T_{\mathrm{eff}}$, as shown by the colored dots (i.e., our final $6{,}594$ Solar twins) in the figure. 
We also see no dependence on $\log g$ or [M/H], though these are not shown in the figure. 
In contrast, we see a clear lower boundary of S/N around $50$ in the final sample, which is the result of the seventh digit of GSP-Spec Quality Flag (i.e., upper thresholds on the parameterization uncertainties of $T_{\mathrm{eff}}$, $\log g$, [M/H], and [$\alpha $/Fe]). 
To properly account for this selection criterion, we fitted the gray dots in Fig.~\ref{fig:SN_cTeff} (i.e., the sample without imposing the thresholds on parameterization uncertainties, namely the sample defined in this section) and obtained the relation, 
\begin{equation}
\log _{10}(\mathrm{S/N}) = - 0.0071388 G^{3} + 0.205971 G^{2} - 2.17006 G + 10.2518\text{,}\label{Eq:linearSNCoef}
\end{equation}
for which the RMS scatter around the relation is $0.0960$. 

With the $G$--S/N relation obtained, to mimic the relations between S/N and parameterization uncertainties, we show vertically arranged pairs of panels plotting the errors in the six parameters for our sample in Fig.~\ref{fig:literparams_cTeff}. 
The upper panel of each pair shows the relation between S/N (\texttt{rv\_expected\_sig\_to\_noise} from the \textit{Gaia} DR3 catalog) and the error. 
As can be seen, there are good linear relations between the logarithms of the horizontal and vertical axes for most panels, indicating power-law relations between the S/N and error values. 
There are two exceptions: the errors in $M_{G}$ and $M_{K_{\mathrm{s}}}$. 
For $M_{G}$, the error is in many cases dominated by the error in the extinction $A_{G}$, which is determined using \textit{Gaia} photometry and GSP-Spec stellar parameters. 
Hence, the error in $M_{G}$ mainly depends on $A_{G}$ and the errors in GSP-Spec stellar parameters. 
The latter correlates well with S/N, as illustrated in the other panels, but the former depends only weakly on S/N through the distance. 
As a result, the trend between S/N and the $M_{G}$ error has a large scatter. 
For $M_{K_{\mathrm{s}}}$, the error is mostly dominated by the error in the 2MASS photometry, and hence depends only weakly on S/N. 

The dependence of the errors on other parameters ($T_{\mathrm{eff}}$, $\log g$, and [M/H]) is illustrated by the colors of the dots in Figs.~\ref{fig:literparams_cTeff}, \ref{fig:literparams_clogg}, and \ref{fig:literparams_cMeta}, respectively. 
There is clear dependence of the errors of most parameters on $T_{\mathrm{eff}}$, and some of them also (in most cases weakly) depend on $\log g$ and [M/H] on top of the dependence on $T_{\mathrm{eff}}$. 
Again, the error of $M_{K_{\mathrm{s}}}$ is an exception, which depends only weakly on $T_{\mathrm{eff}}$ but does not on $\log g$ nor [M/H]. 

To model the dependence of the errors on S/N and stellar parameters, we fitted the common logarithms of the errors as linear functions of the form, 
\begin{align}
&\log _{10}(\mathrm{Error}) = a_{0} + a_{\mathrm{S/N}}^{}\log _{10}(\mathrm{S/N}) + a_{T_{\mathrm{eff}}}^{}\frac{T_{\mathrm{eff}} - T_{\mathrm{eff},\odot}}{100\ur{K}} \notag \\
&\quad + a_{{T_{\mathrm{eff}}}^{2}}^{}\left(\frac{T_{\mathrm{eff}} - T_{\mathrm{eff},\odot}}{100\ur{K}}\right)^{2} + a_{\log g}^{}\frac{\log g-\log g_{\odot }}{0.1} + a_{\mathrm{[M/H]}}^{}\frac{\mathrm{[M/H]}}{0.1\ur{dex}}\text{,} \label{Eq:LinearErrorCoef}
\end{align}
using least-squares minimization. 
We fixed some coefficients to zero when visual inspection showed no clear dependence of an error on a parameter, and when allowing the coefficients to vary produced negligible improvements in the root-mean-squares (RMS) of the fitting residuals. 
The resulting coefficients and RMS values of the fits are summarized in Table~\ref{table:CoefsError}. 
We used Equations~\ref{Eq:LinearErrorCoef} and \ref{Eq:linearSNCoef}, together with their scatter, to assign errors to mock Solar twins.

\section{Review of previous large Solar-twin catalogs}\label{app:review_literature}

In recent years, a few large catalogs of Solar twins have been published using spectra from large surveys. 
\citet{Rampalli2024} analyzed \textit{Gaia} RVS spectra of Solar twins published in DR3~\citep{GaiaRVSDR2,GaiaDR3} with a data-driven approach using \texttt{The Cannon} algorithm~\citep{Ness2015} to infer the stellar parameters and chemical abundances of $15$ elements for $5{,}347$ Solar twins (in our definition). 
They used high-resolution stellar parameters and chemical abundances of $34$ Solar twins from \citet{Hinkel2014}, \citet{Brewer2016}, \citet{Brewer2018}, and \citet{Bedell2018} as the reference during the training, among which most of the training stars are from \citet{Brewer2016} and \citet{Brewer2018}. 
Among the four studies, only \citet{Bedell2018} focused exclusively on Solar twins, but given that \citet{Brewer2016} and \citet{Brewer2018} analyzed their high-resolution spectra with a differential analysis method, their parameters might be as reliable as those in studies focusing specifically on Solar twins. 
They achieved high internal precisions of $0.03\text{--}0.07\ur{dex}$ for many elements, supporting the effectiveness of the data-driven \texttt{The Cannon} method. 
Then, \citet{Rampalli2024} estimated isochrone ages using the \texttt{isoclassify} code~\citep{Huber2017,Berger2020}, with $T_{\mathrm{eff}}$, $\log g$, and [Fe/H] as input parameters, though they mentioned that they do not report their ages ``as robust results for individual stars''. 
By plotting condensation temperature $T_{\mathrm{c}}$--abundance trends, they independently confirmed that the Sun is depleted in refractory elements compared to their sample of Solar twins, regardless of whether those stars host terrestrial, close-in giant, or giant planets, or no planets at all. 

There are also two large catalogs of Solar twins based on GALAH DR3~\citep{Buder2021} data. 
As one of these two studies, \citet{Walsen2024} applied the data-driven \texttt{The Cannon} algorithm~\citep{Ness2015} to GALAH DR3 spectra to derive high-precision stellar parameters and $14$ elemental abundances for a sample of $13{,}132$ Solar twins (again, in our definition). 
They used stellar parameters and abundances from GALAH DR3 of $150$ high-S/N stars, for which a sky-flat Solar spectrum was used to determine the zero-point in the chemical abundances, as the reference, though it is not clearly stated which parameters have been corrected for these zero-points. 
They also used GALAH DR3 abundances together with their derived \texttt{The Cannon} abundances, to see whether there are any differences between the results obtained with the two abundance catalogs. 
They adopted ages tabulated in GALAH DR3~\citep{Sharma2018,Buder2021}, though \citet{Buder2021} and \citet{Walsen2024} noted a few limitations in the GALAH DR3 ages, for example, the overestimation of the Solar age by $1.26\ur{Gyr}$. 
With these datasets, they compared the age--[X/Fe] relations for their sample to those from \citet{Bedell2018}. 
They claimed that some elements (Mg, Al, Si, Ca, Sc, Ti, and Zn) show consistent trends between the two samples, whereas some others (Na, Mn, Ni, Cu, Y, and Ba) show inconsistent trends, and argued that the discrepancies could be due, for example, to differences in precision and accuracy, distance range, or selection effects. 
At the same time, we suggest that the discrepancies that they found could be at least partly attributed to differences in the age range (${\sim }0\text{--}8\ur{Gyr}$ in \citet{Bedell2018} vs ${\sim }3\text{--}12\ur{Gyr}$ in \citet{Walsen2024}) and in the metallicity range (${\pm }0.1\ur{dex}$ around the Solar value by \citet{Bedell2018} vs ${\pm }0.3\ur{dex}$ by \citet{Walsen2024}). 
They then constructed phylogenetic trees from [X/Fe] ratios to investigate the evolution of Galactic-disk stars. 
From the phylogenetic tree constructed from the combination of low- and high-eccentricity Solar twins, they found that there are two distinct clans that differ in eccentricities, metallicities, and chemical-clock relations (i.e., the age--[Y/Mg] relation). 

In the other study using GALAH DR3, \citet{Lehmann2025} applied the equivalent-width-based data-driven \texttt{EPIC} algorithm~\citep{Lehmann2022} to derive precise $T_{\mathrm{eff}}$, $\log g$, and [Fe/H] for a large sample of Solar twins ($14,571$ stars in our definition). 
They employed the stacked GALAH DR2 spectra~\citep{Zwitter2018}, whose stellar parameters had been determined using \texttt{The Cannon} by \citet{Buder2018}\footnote{We note that, unlike GALAH DR2, where \texttt{The Cannon} was used, \citet{Buder2021} chose not to use \texttt{The Cannon} (or any other data-driven approaches) for GALAH DR3 because such methods can suffer from signal aliasing and learn unphysical correlations between input data and output labels. }, as the reference training sample. 
They achieved a factor of $2\text{--}4$ higher precision (i.e., smaller errors) in the parameterization compared to the original GALAH pipeline. 
They then estimated isochrone ages with the \texttt{SAMD} code~\citep{Sahlholdt2020} and used these ages together with orbital parameters to investigate the chemodynamic evolution of nearby moving groups defined as Solar twins having a common angular momentum $L_{z}$. 
In particular, they examined the age--metallicity relation for each moving group, without imposing the usual restriction of nearly Solar metallicity, to study radial migration within a moving group.

\section{Additional tables and figures}\label{app:figures}

\begin{table}
\centering 
\caption{Final solar-twin catalog (available at CDS). }
\label{table:catalog}
\end{table}

\begin{figure*}
\centering 
\includegraphics[width=18cm]{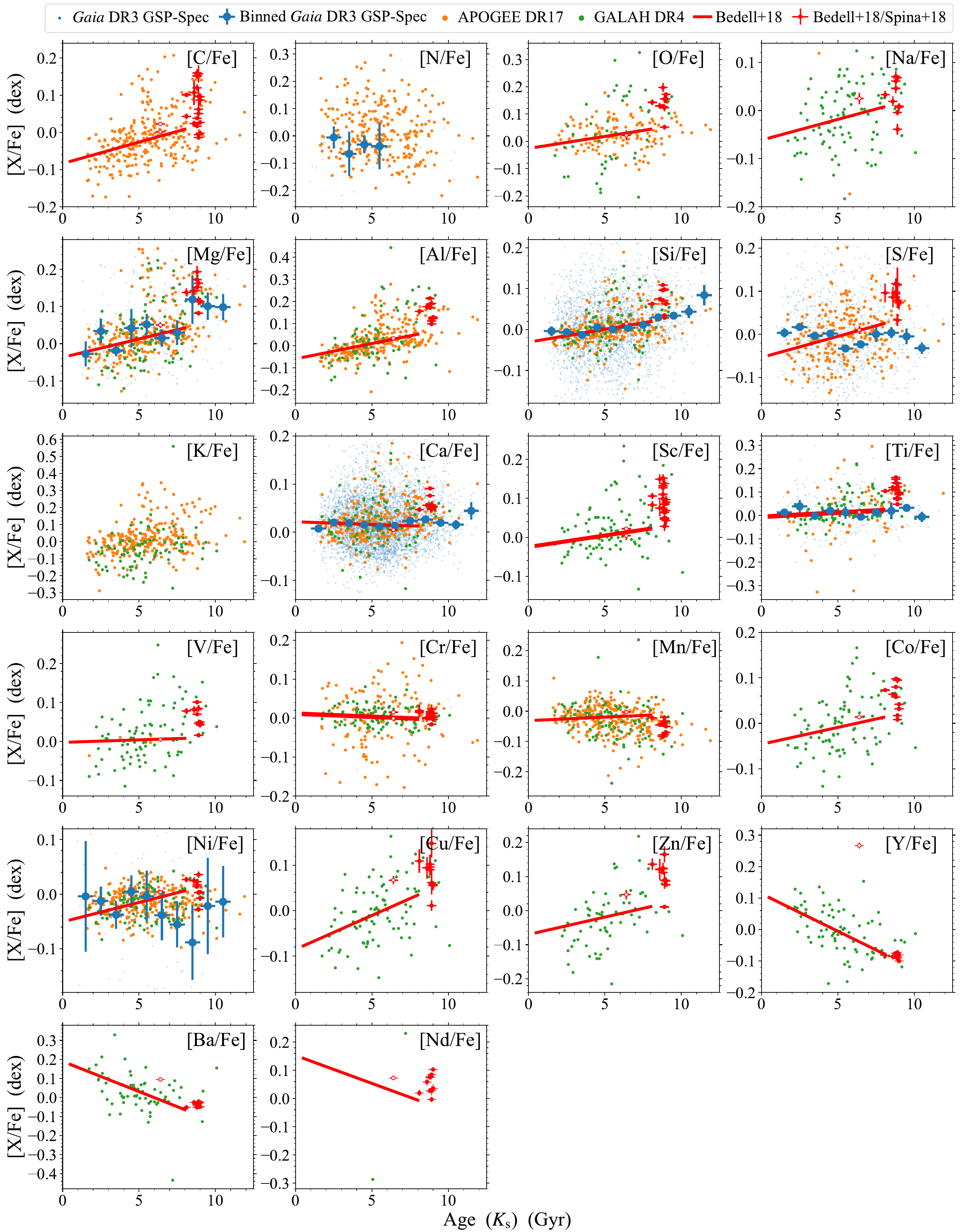}
\caption{Age--abundance relations for all the elements. Symbols are the same as in Fig.~\ref{fig:AgeAbn}. }
\label{fig:AgeAbnAll}
\end{figure*}

\begin{table*}
\centering 
\caption{Column description of the published Solar-twin catalog. }
\label{table:columndesc}
\begin{tabular}{lcl}\hline \hline 
Column & Unit & Description \\ \hline 
\texttt{GDR3\_source\_id} & --- & \textit{Gaia} DR3 \texttt{source\_id} \\
\texttt{2MASS} & --- & 2MASS designation \\
\texttt{RAdeg} & deg & Right ascension in \textit{Gaia} DR3 \\
\texttt{DEdeg} & deg & Declination in \textit{Gaia} DR3 \\
\texttt{Teff} & K & Calibrated GSP-Spec effective temperature, $T_{\mathrm{eff}}$ \\
\texttt{logg} & (cgs) & Calibrated GSP-Spec surface gravity, $\log g$ \\
\texttt{[M/H]} & dex & Calibrated GSP-Spec current surface metallicity, $\mathrm{[M/H]}_{\mathrm{curr}}$ \\
\texttt{[alpha/M]} & dex & GSP-Spec $\alpha $ abundance, [$\alpha $/Fe] \\
\texttt{A0} & mag & Monochromatic extinction, $A_{0}$ \\
\texttt{BJ21\_rgeo} & pc & Geometric distance from \citet{BailerJones2021}\tablefootmark{a} \\
\texttt{GMag} & mag & Absolute magnitude in the \textit{G} band of \textit{Gaia} DR3 \\
\texttt{BPMag} & mag & Absolute magnitude in the \textit{BP} band of \textit{Gaia} DR3 \\
\texttt{RPMag} & mag & Absolute magnitude in the \textit{RP} band of \textit{Gaia} DR3 \\
\texttt{JMag} & mag & Absolute magnitude in the \textit{J} band of 2MASS \\
\texttt{HMag} & mag & Absolute magnitude in the \textit{H} band of 2MASS \\
\texttt{KsMag} & mag & Absolute magnitude in the \textit{Ks} band of 2MASS \\
\texttt{Age(logg)} & Gyr & Age determined with \texttt{logg} \\
\texttt{Mini(logg)} & $M_{\odot }$ & Initial mass $M_{\mathrm{ini}}$ determined with \texttt{logg} \\
\texttt{[M/H]ini(logg)} & dex & Initial metallicity $\text{[M/H]}_{\mathrm{ini}}$ determined with \texttt{logg} \\
\texttt{Age(GMag)} & Gyr & Age determined with \texttt{GMag} \\
\texttt{Mini(GMag)} & $M_{\odot }$ & Initial mass $M_{\mathrm{ini}}$ determined with \texttt{GMag} \\
\texttt{[M/H]ini(GMag)} & dex & Initial metallicity $\text{[M/H]}_{\mathrm{ini}}$ determined with \texttt{GMag} \\
\texttt{Age(KsMag)} & Gyr & Age determined with \texttt{KsMag} \\
\texttt{Mini(KsMag)} & $M_{\odot }$ & Initial mass $M_{\mathrm{ini}}$ determined with \texttt{KsMag} \\
\texttt{[M/H]ini(KsMag)} & dex & Initial metallicity $\text{[M/H]}_{\mathrm{ini}}$ determined with \texttt{KsMag} \\
\texttt{ecc} &  & Eccentricity \\
\texttt{Lz} & Sun & Vertical angular momentum \\
\texttt{Rg} & kpc & Guiding radius \\
\texttt{Rapo} & kpc & Apocenter distance \\
\texttt{Rperi} & kpc & Pericenter distance \\
\texttt{Zmax} & kpc & Maximum vertical distance \\
\texttt{[X/Fe]} & dex & Calibrated X abundance in \textit{Gaia} DR3 GSP-Spec\tablefootmark{a} \\
\texttt{APOGEEDR17\_FE\_H} & dex & Iron abundance in APOGEE DR17~\citep{Majewski2017,Abdurrouf2022} \\
\texttt{APOGEEDR17\_X\_FE} & dex & X abundance in APOGEE DR17\tablefootmark{a} \\
\texttt{GALAHDR4\_fe\_h} & dex & Iron abundance in GALAH DR4~\citep{DeSilva2015,Buder2025}\tablefootmark{a} \\
\texttt{GALAHDR4\_x\_fe} & dex & X abundance in GALAH DR4\tablefootmark{a} \\
\texttt{exoplanet} & --- & Known exoplanet in NASA Exoplanet Archive~\citep{Christiansen2025}?\tablefootmark{a} \\
\hline 
\end{tabular}
\tablefoot{
We recommend using age, $M_{\mathrm{ini}}$, and $\text{[M/H]}_{\mathrm{ini}}$ determined from $M_{K_{\mathrm{s}}}$. 
\tablefoottext{a}{These literature values are provided for quick analyses using this data set. Interested users must check the up-to-date data and cite original papers. }
}
\end{table*}

\clearpage 

\begin{table}
\centering 
\caption{Added offsets to [X/Fe] abundances. }
\label{table:XFeoffsets}
\begin{tabular}{c ccc}\hline \hline
Element & GSP-Spec & APOGEE & GALAH \\ \hline
{[C/Fe]} &  & $+0.011$ &  \\
{[N/Fe]} & ---\tablefootmark{a} & ---\tablefootmark{a} &  \\
{[O/Fe]} &  & $-0.059$ & $-0.120$ \\
{[Na/Fe]} &  & $+0.103$ & $-0.018$ \\
{[Mg/Fe]} & $+0.031$ & $-0.015$ & $-0.044$ \\
{[Al/Fe]} &  & $-0.125$ & $-0.027$ \\
{[Si/Fe]} & $-0.023$ & $-0.075$ & $-0.030$ \\
{[S/Fe]} & $+0.025$ & $+0.026$ &  \\
{[K/Fe]} &  & ---\tablefootmark{a} & ---\tablefootmark{a} \\
{[Ca/Fe]} & $+0.012$ & $+0.034$ & $+0.025$ \\
{[Sc/Fe]} &  &  & $-0.027$ \\
{[Ti/Fe]} & $+0.026$ & $+0.143$ & $+0.013$ \\
{[V/Fe]} &  & $+0.014$ & $-0.017$ \\
{[Cr/Fe]} & $+0.006$ & $+0.145$ & $+0.004$ \\
{[Mn/Fe]} &  & $-0.036$ & $+0.006$ \\
{[Co/Fe]} &  &  & $-0.035$ \\
{[Ni/Fe]} & $+0.028$ & $-0.041$ & $-0.038$ \\
{[Cu/Fe]} &  &  & $-0.021$ \\
{[Zn/Fe]} &  &  & $-0.027$ \\
{[Y/Fe]} &  &  & $+0.065$ \\
{[Ba/Fe]} &  &  & $+0.082$ \\
{[Nd/Fe]} &  &  & $-0.035$ \\
\hline
\end{tabular}
\tablefoot{
\tablefoottext{a}{The age--[X/Fe] relation is not available for these elements in \citet{Bedell2018}. }
}
\end{table}


\clearpage 

\begin{figure*}
\centering 
\includegraphics[width=18cm, page=1]{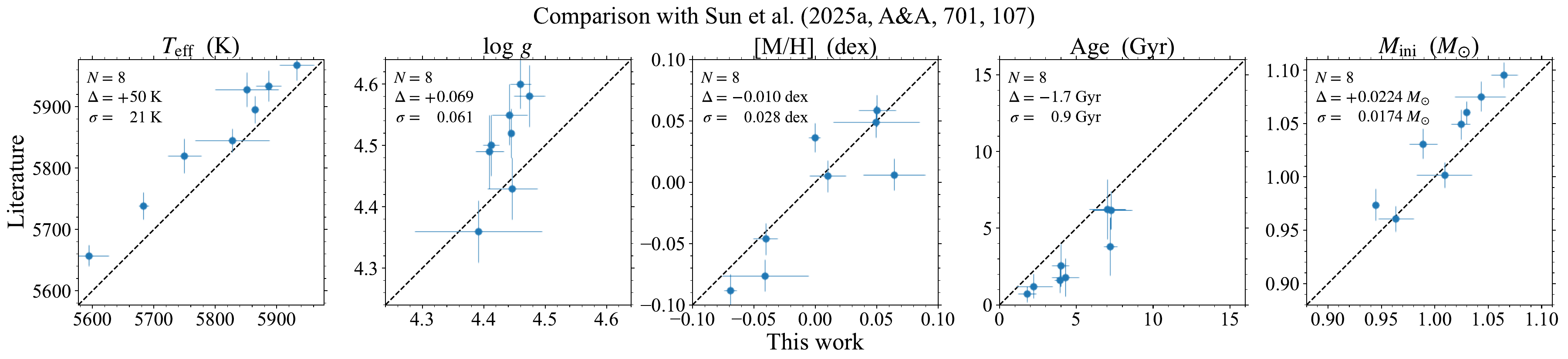}
\includegraphics[width=18cm, page=2]{literSolarTwins_comp.pdf}
\includegraphics[width=18cm, page=3]{literSolarTwins_comp.pdf}
\includegraphics[width=18cm, page=4]{literSolarTwins_comp.pdf}
\includegraphics[width=18cm, page=5]{literSolarTwins_comp.pdf}
\caption{Comparison of stellar parameters with literature. Also available at Zenodo (\url{https://doi.org/10.5281/zenodo.XXXXXXXX}). }
\label{fig:compall}
\end{figure*}

\begin{figure*}
\ContinuedFloat 
\centering 
\includegraphics[width=18cm, page=6]{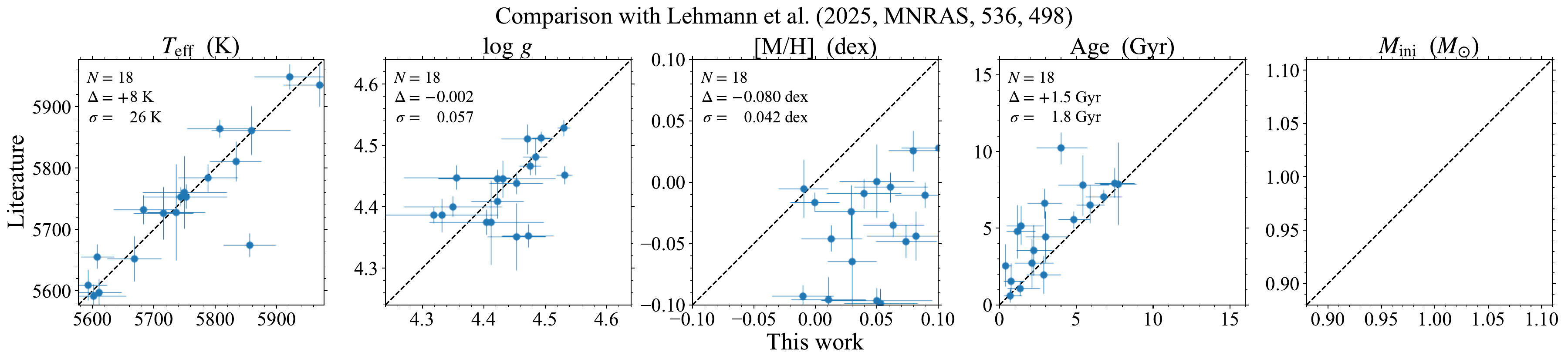}
\includegraphics[width=18cm, page=7]{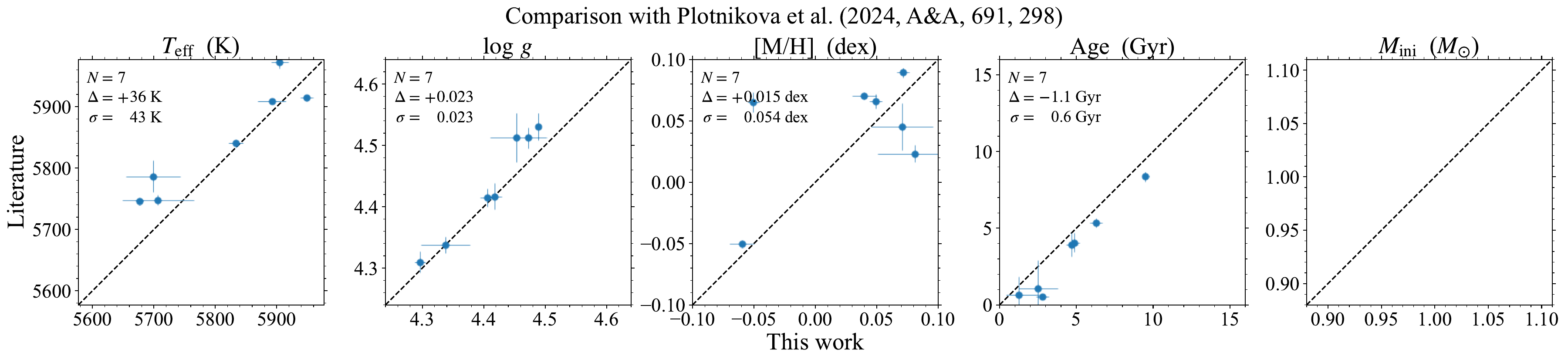}
\includegraphics[width=18cm, page=8]{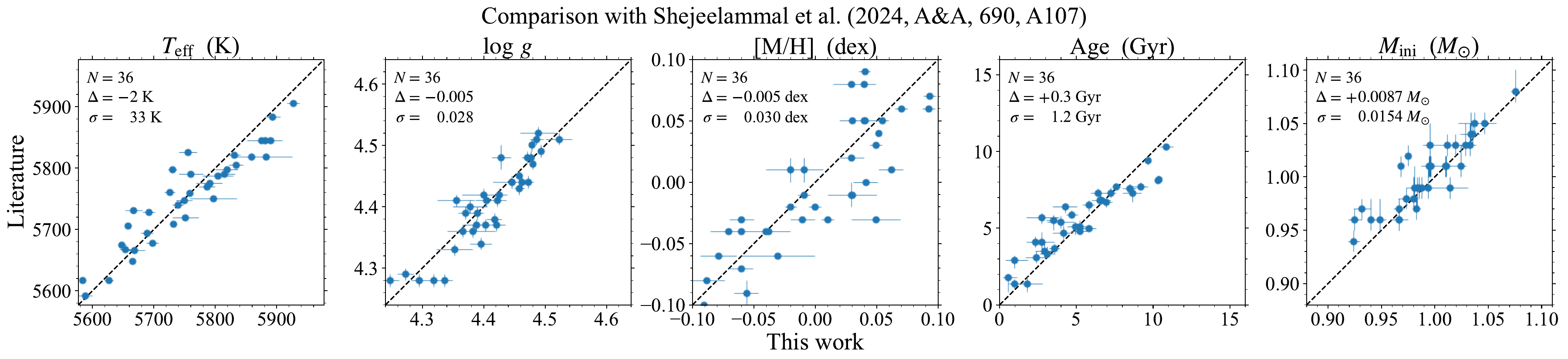}
\includegraphics[width=18cm, page=9]{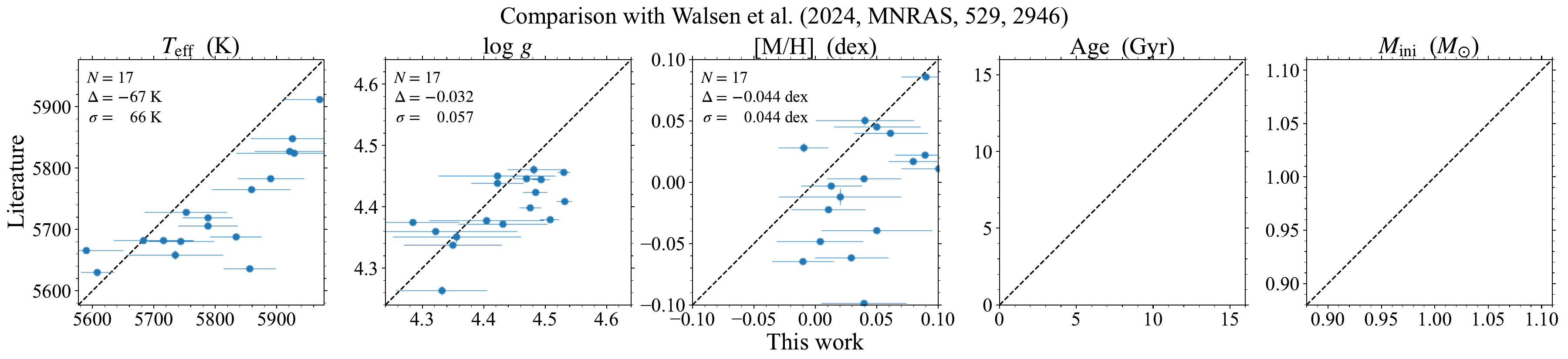}
\includegraphics[width=18cm, page=10]{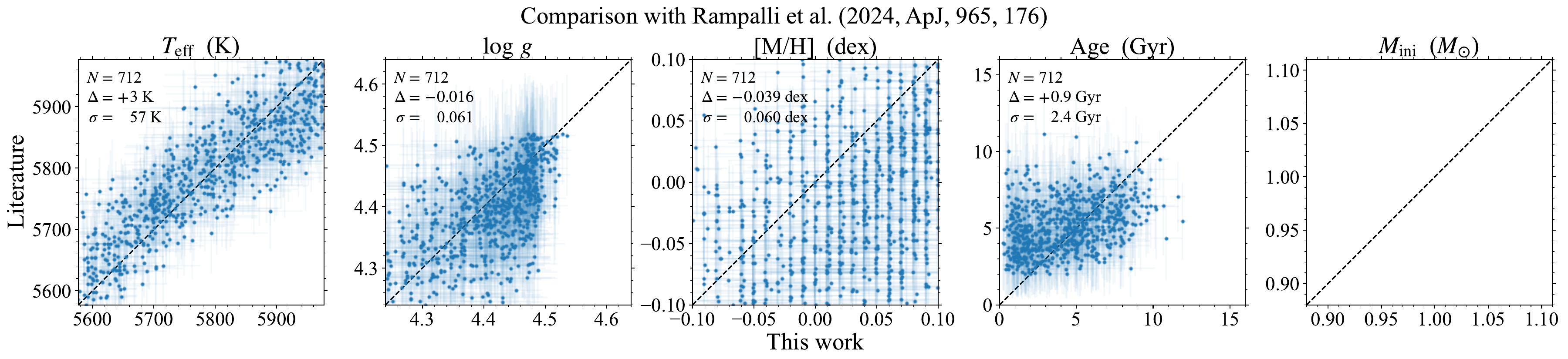}
\caption{Continued. }
\end{figure*}

\begin{figure*}
\ContinuedFloat 
\centering 
\includegraphics[width=18cm, page=11]{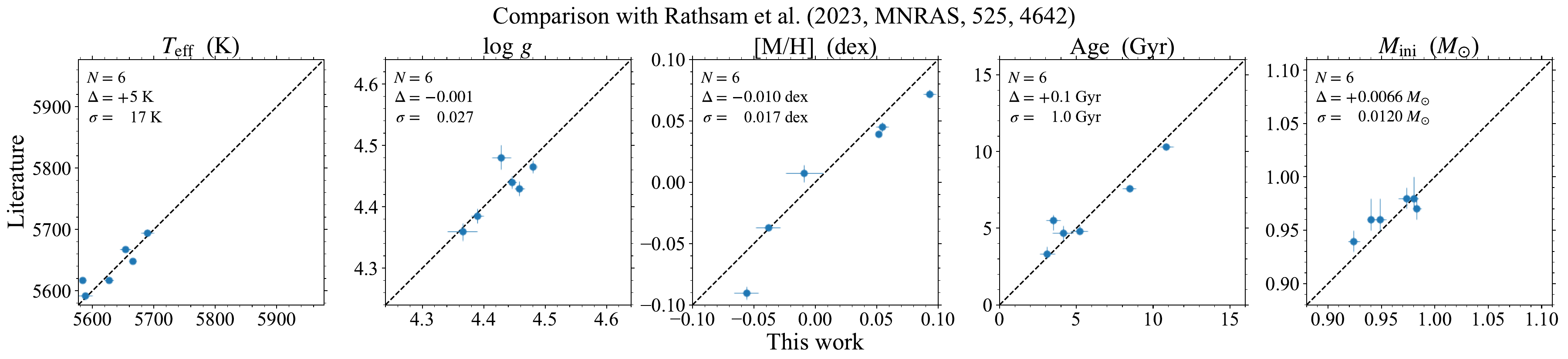}
\includegraphics[width=18cm, page=12]{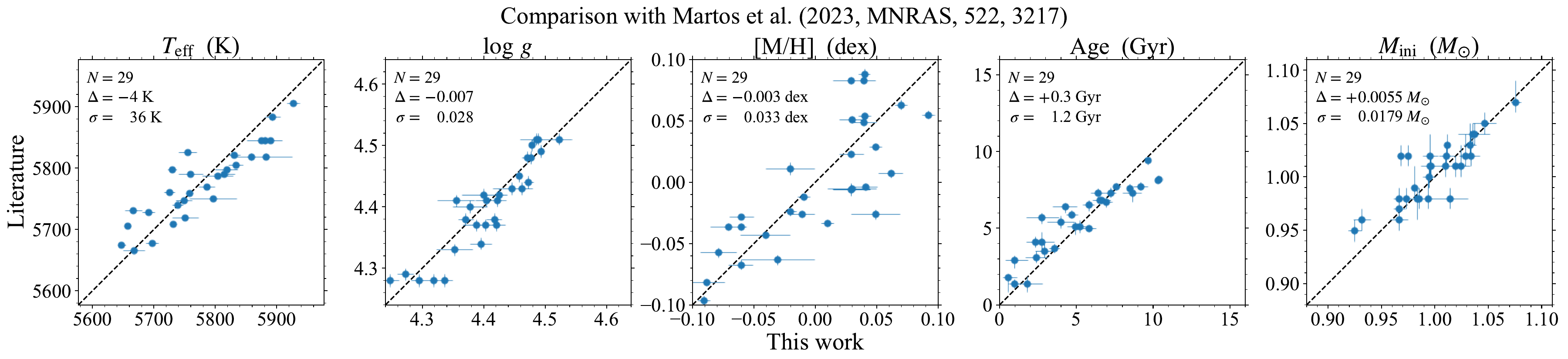}
\includegraphics[width=18cm, page=13]{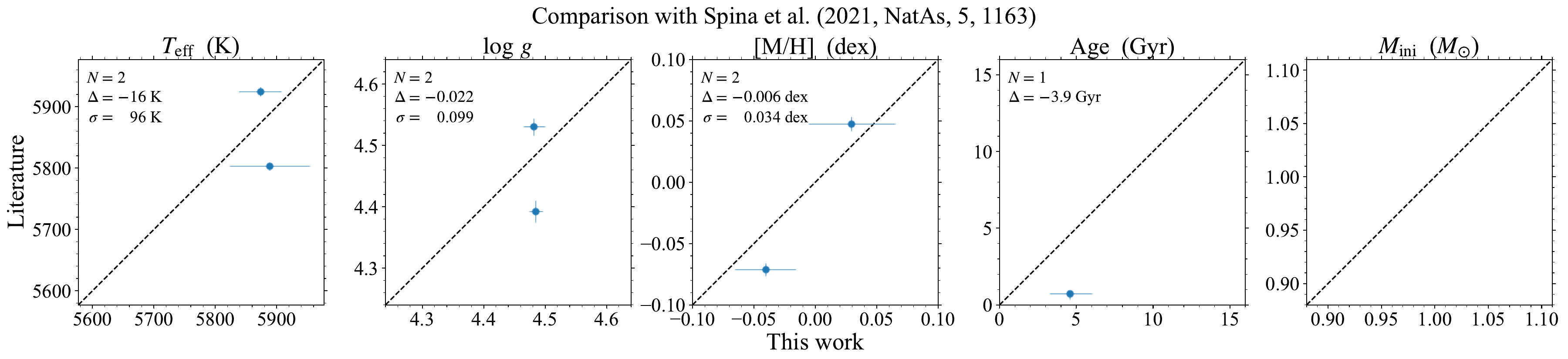}
\includegraphics[width=18cm, page=14]{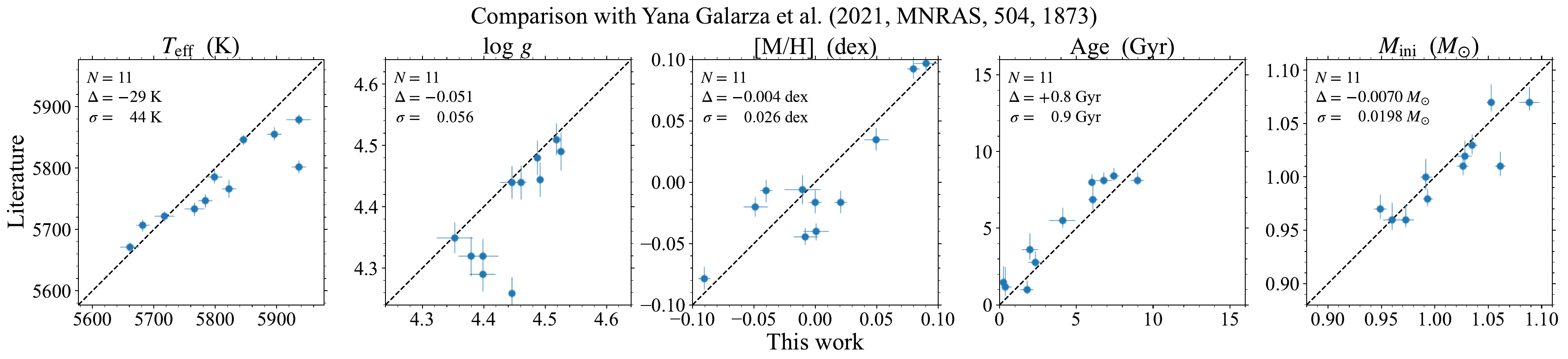}
\includegraphics[width=18cm, page=15]{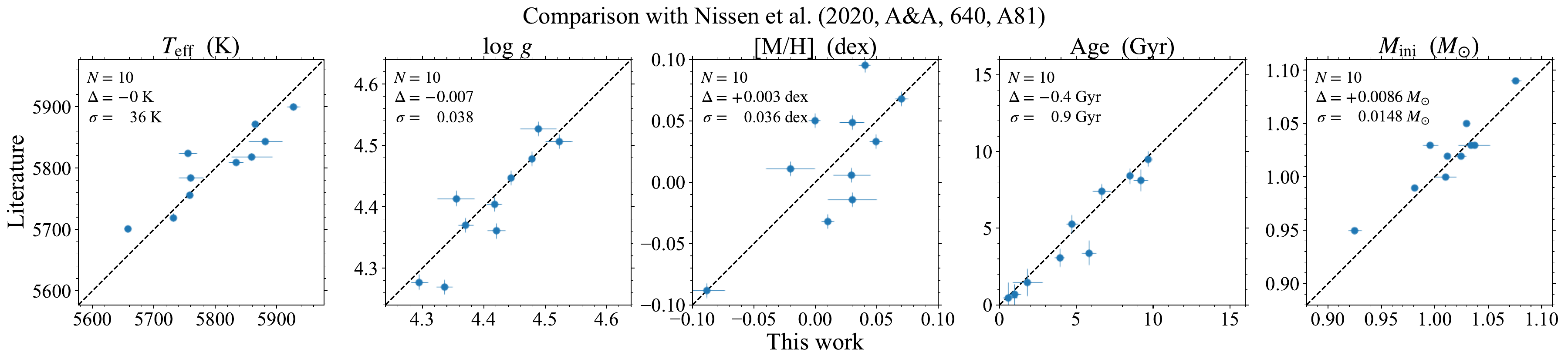}
\caption{Continued. }
\end{figure*}

\begin{figure*}
\ContinuedFloat 
\centering 
\includegraphics[width=18cm, page=16]{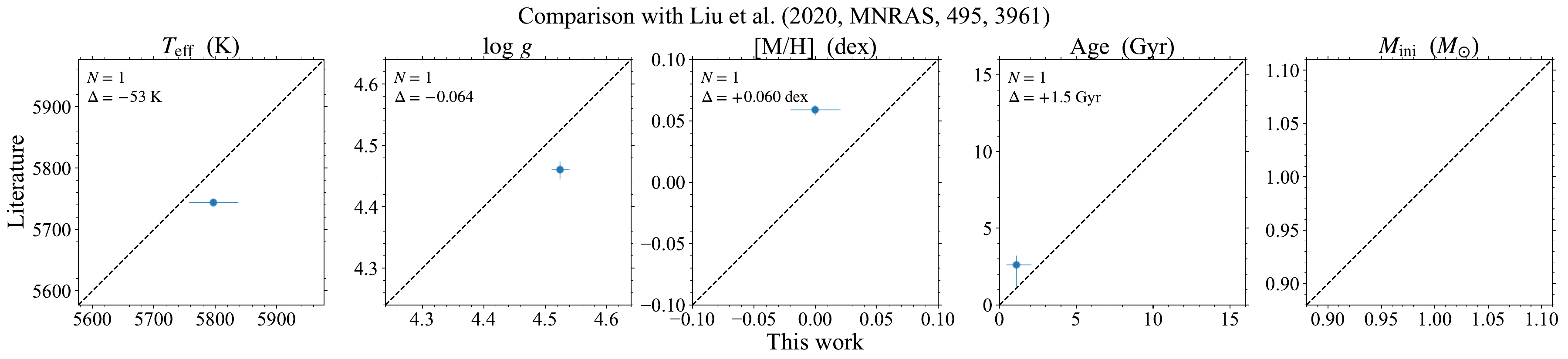}
\includegraphics[width=18cm, page=17]{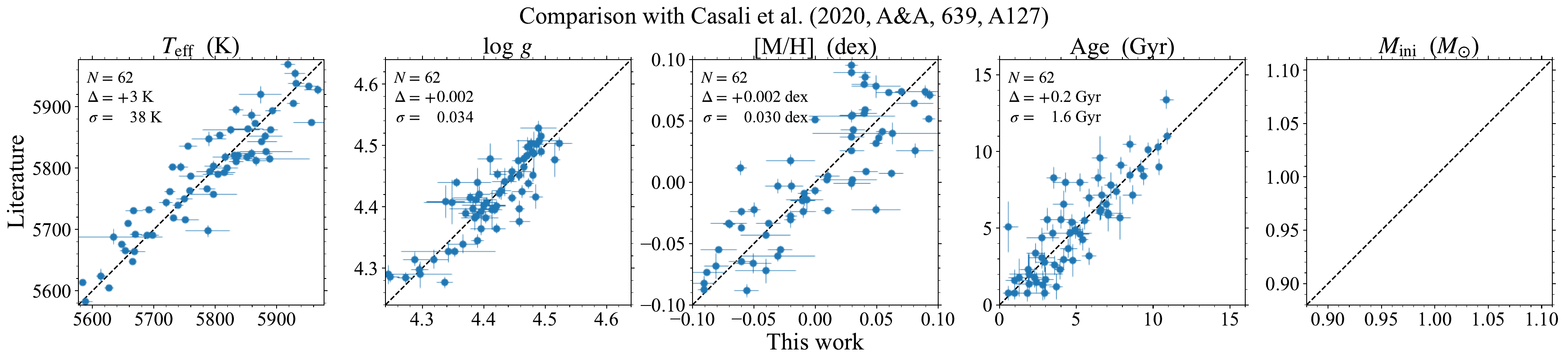}
\includegraphics[width=18cm, page=18]{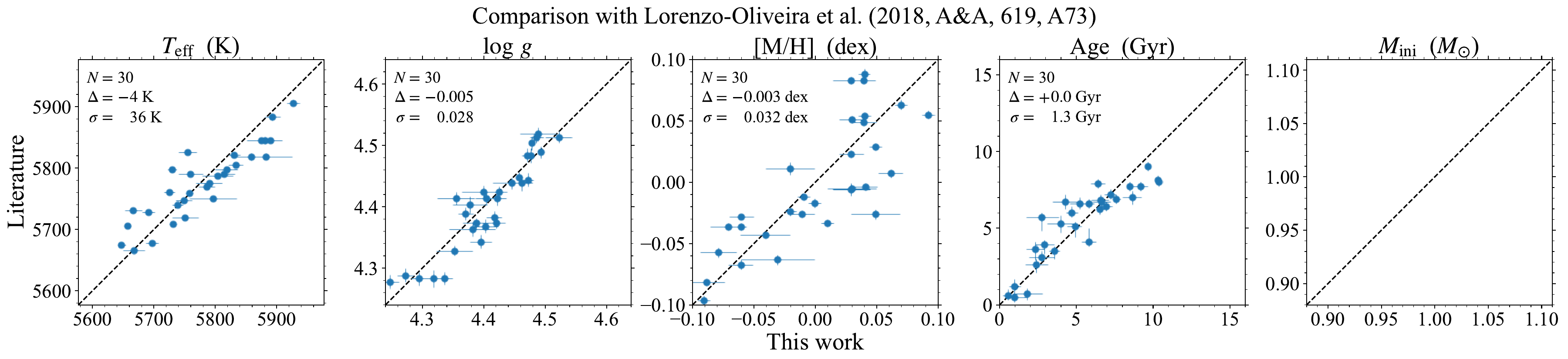}
\includegraphics[width=18cm, page=19]{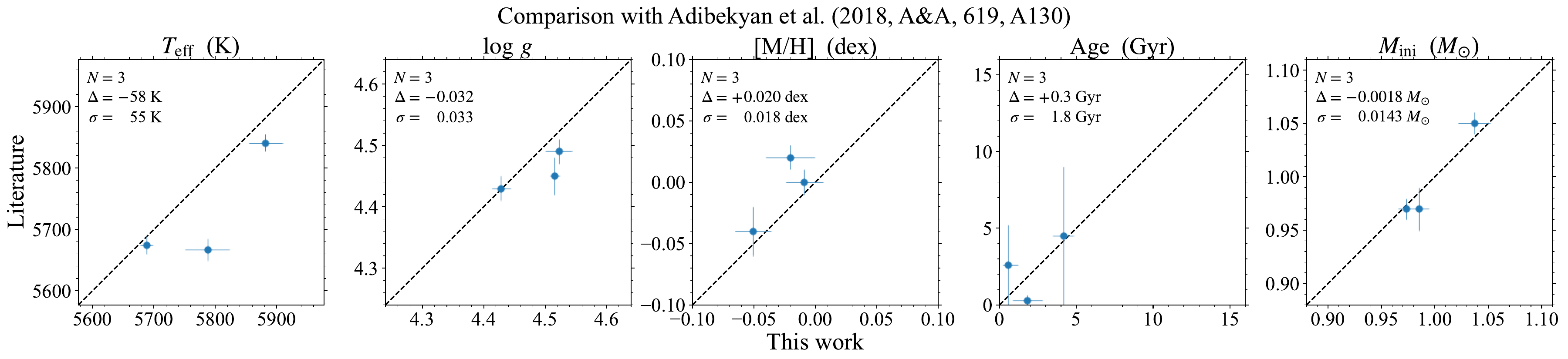}
\includegraphics[width=18cm, page=20]{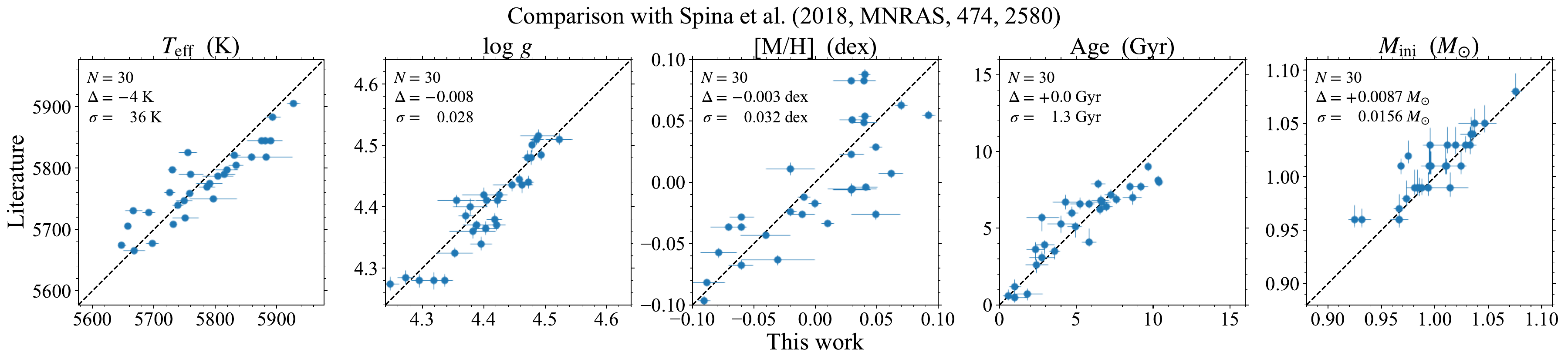}
\caption{Continued. }
\end{figure*}

\begin{figure*}
\ContinuedFloat 
\centering 
\includegraphics[width=18cm, page=21]{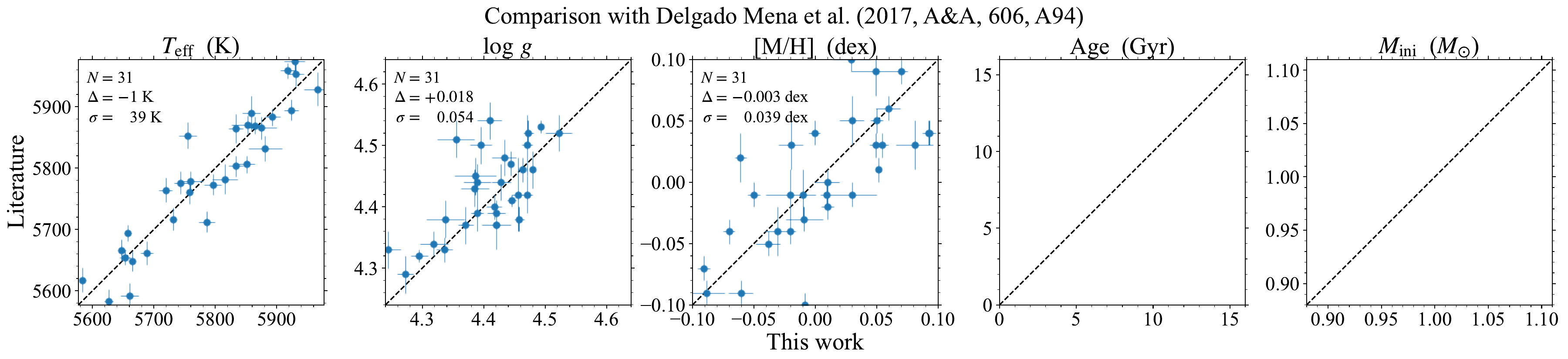}
\includegraphics[width=18cm, page=22]{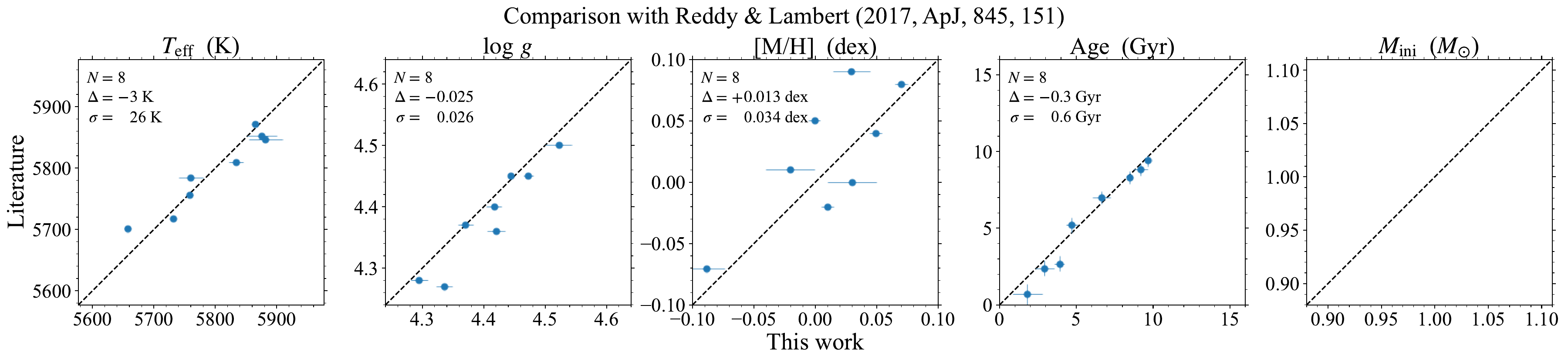}
\includegraphics[width=18cm, page=23]{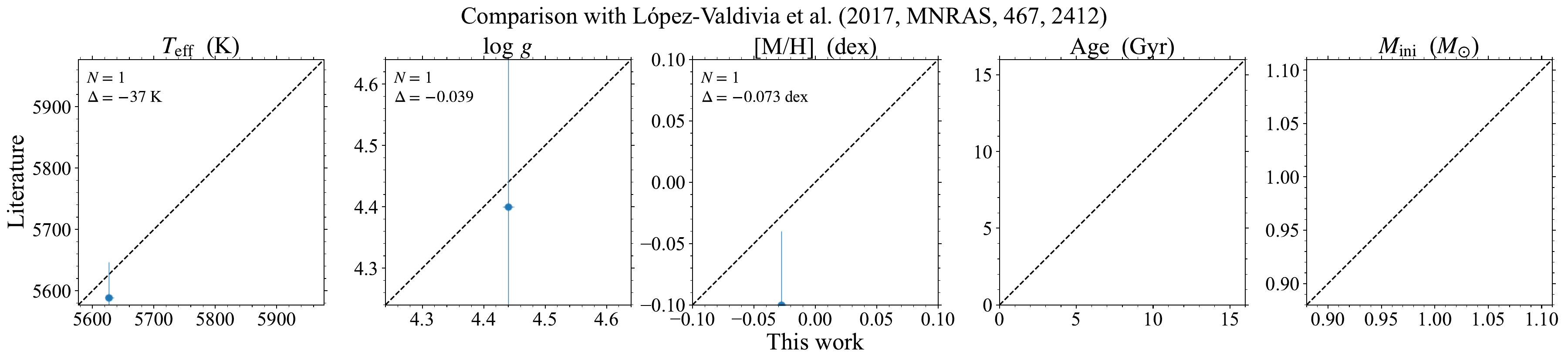}
\includegraphics[width=18cm, page=24]{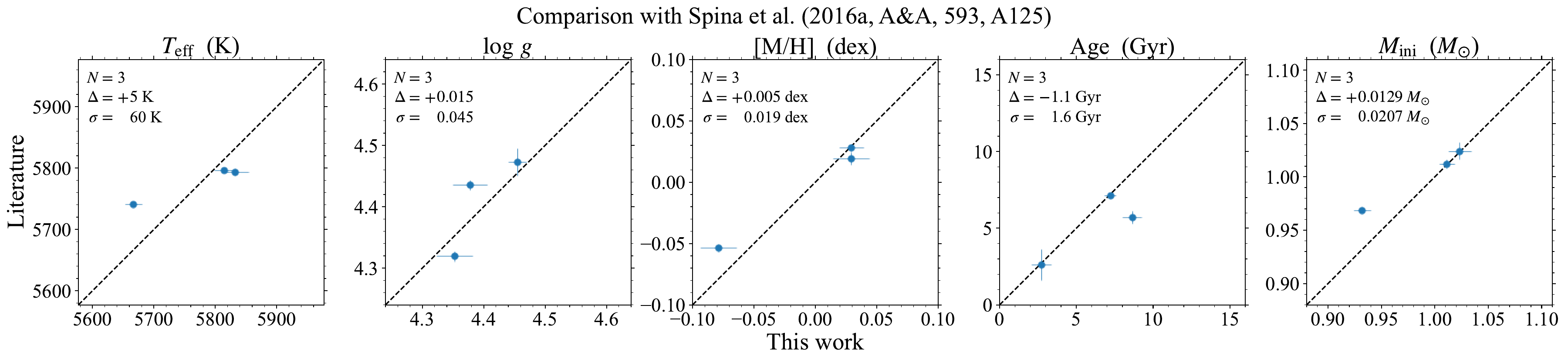}
\includegraphics[width=18cm, page=25]{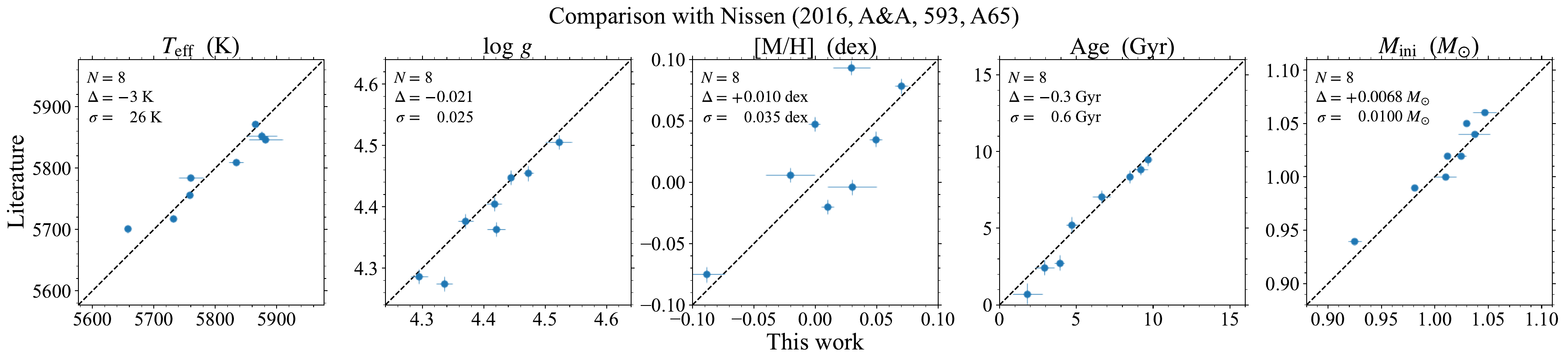}
\caption{Continued. }
\end{figure*}

\begin{figure*}
\ContinuedFloat 
\centering 
\includegraphics[width=18cm, page=26]{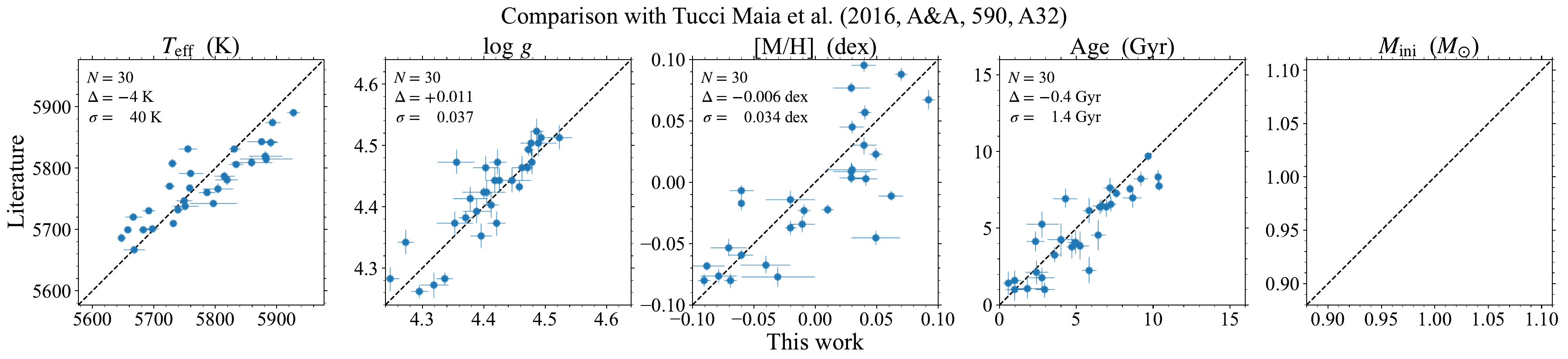}
\includegraphics[width=18cm, page=27]{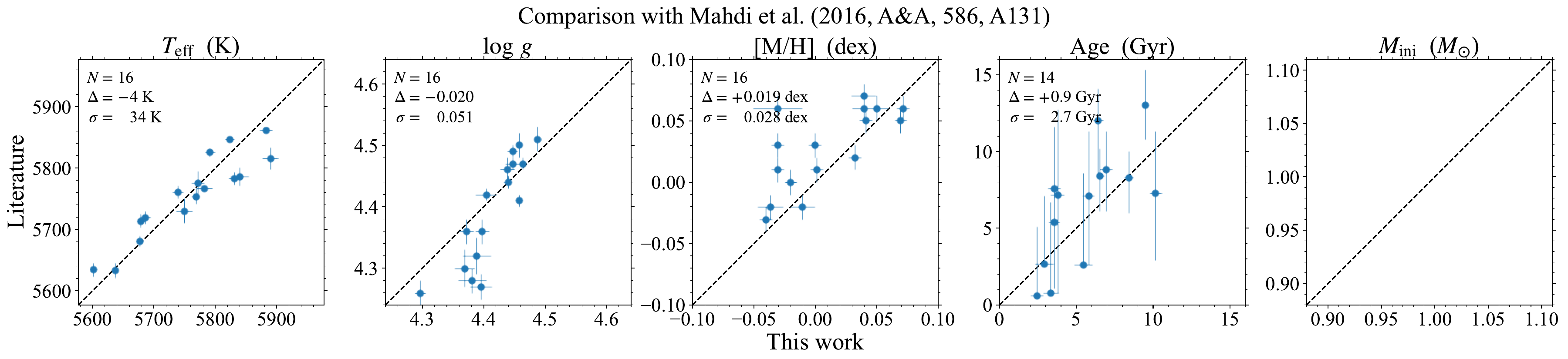}
\includegraphics[width=18cm, page=28]{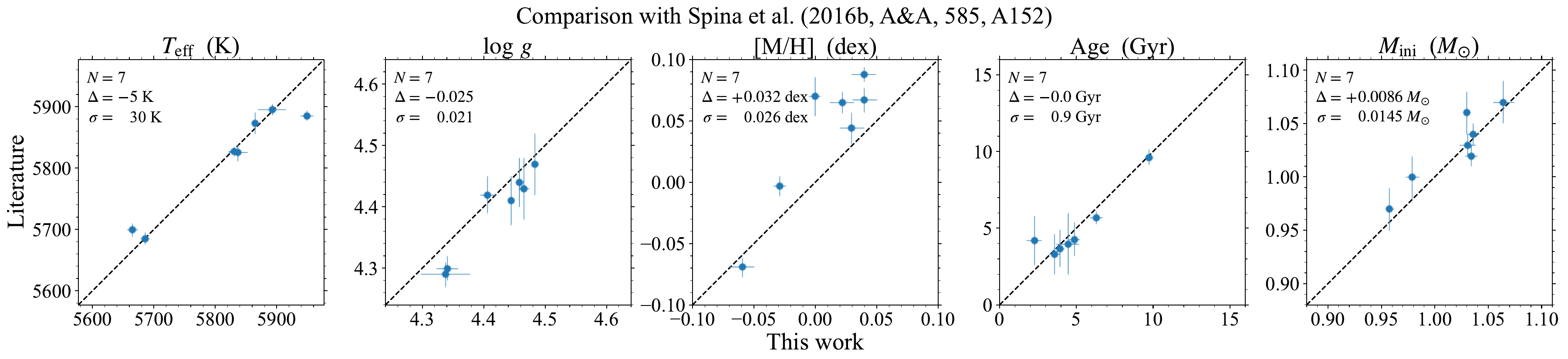}
\includegraphics[width=18cm, page=29]{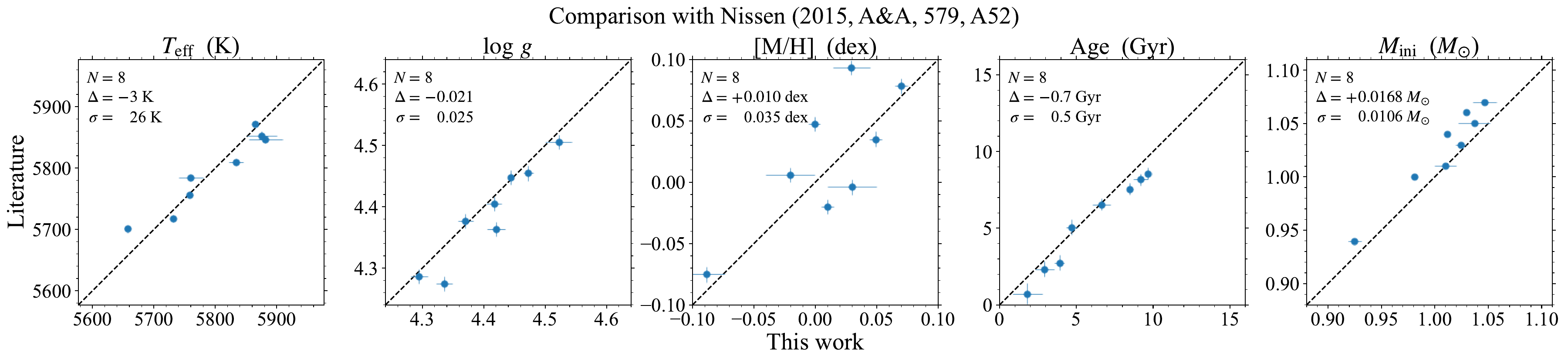}
\includegraphics[width=18cm, page=30]{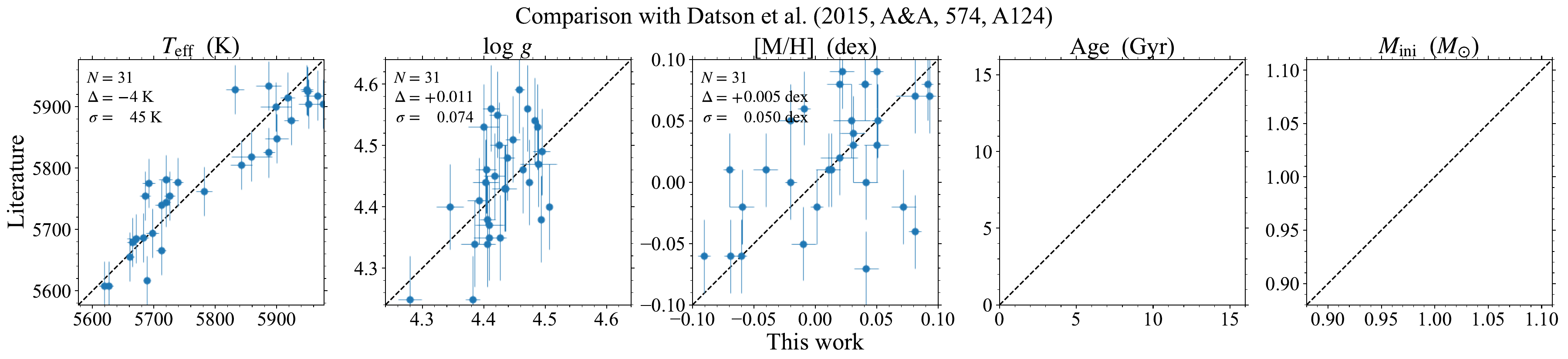}
\caption{Continued. }
\end{figure*}

\begin{figure*}
\ContinuedFloat 
\centering 
\includegraphics[width=18cm, page=31]{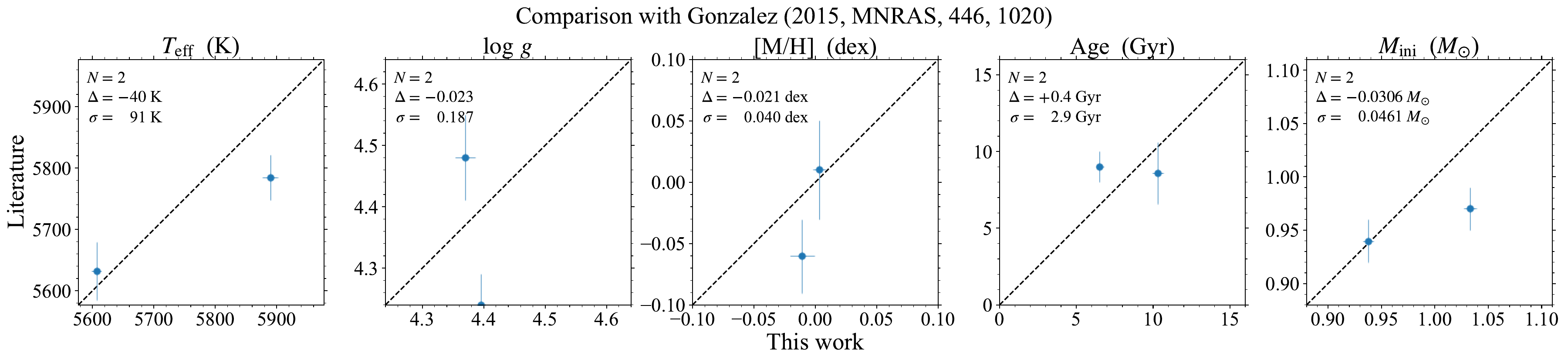}
\includegraphics[width=18cm, page=32]{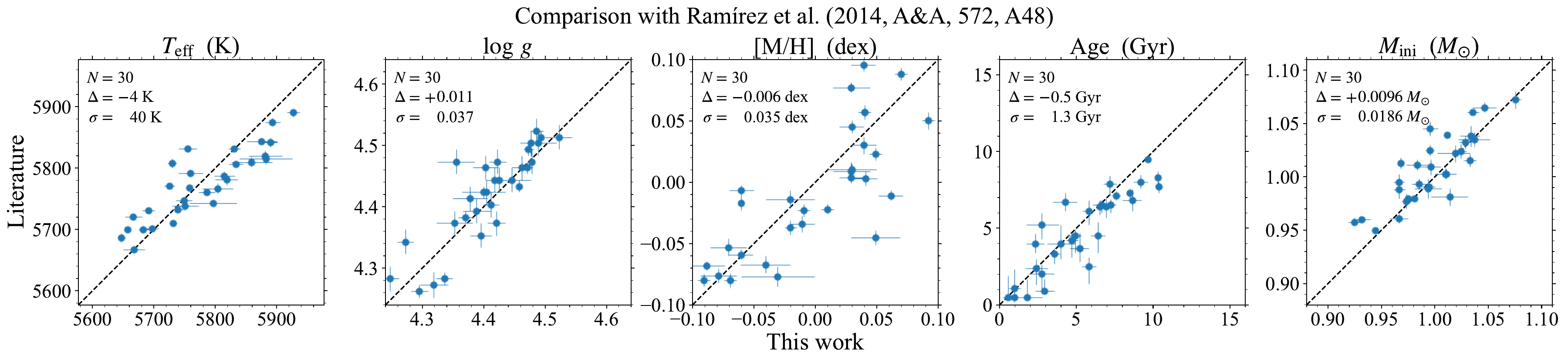}
\includegraphics[width=18cm, page=33]{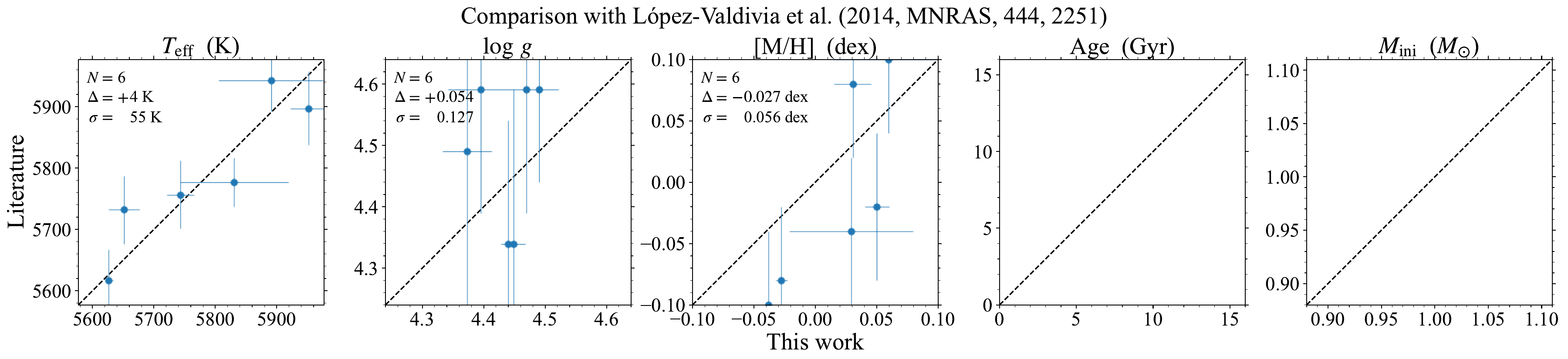}
\includegraphics[width=18cm, page=34]{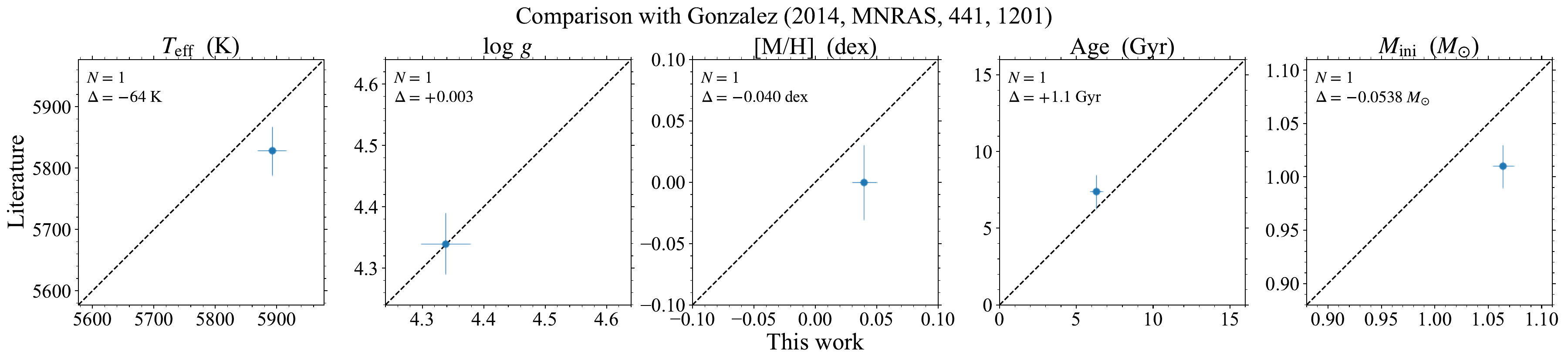}
\includegraphics[width=18cm, page=35]{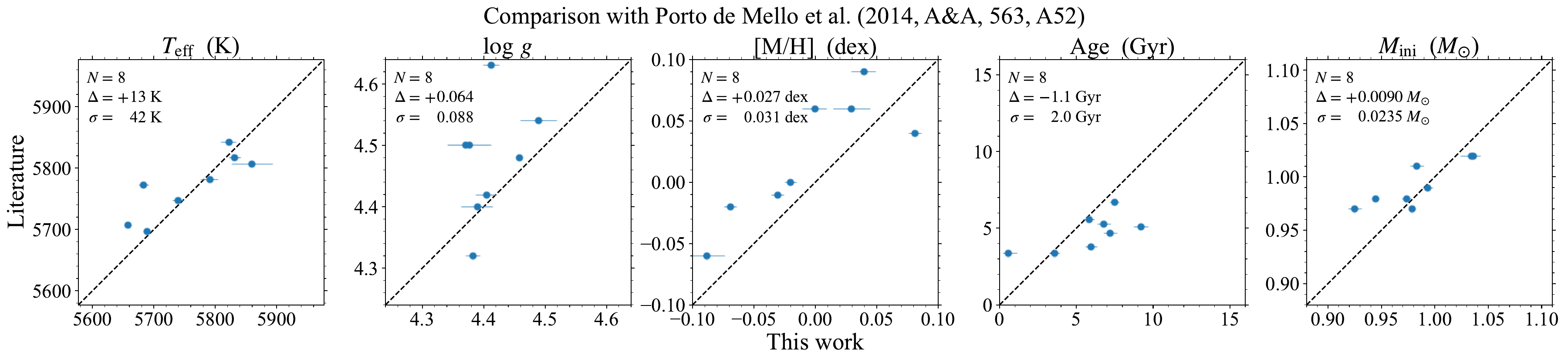}
\caption{Continued. }
\end{figure*}

\begin{figure*}
\ContinuedFloat 
\centering 
\includegraphics[width=18cm, page=36]{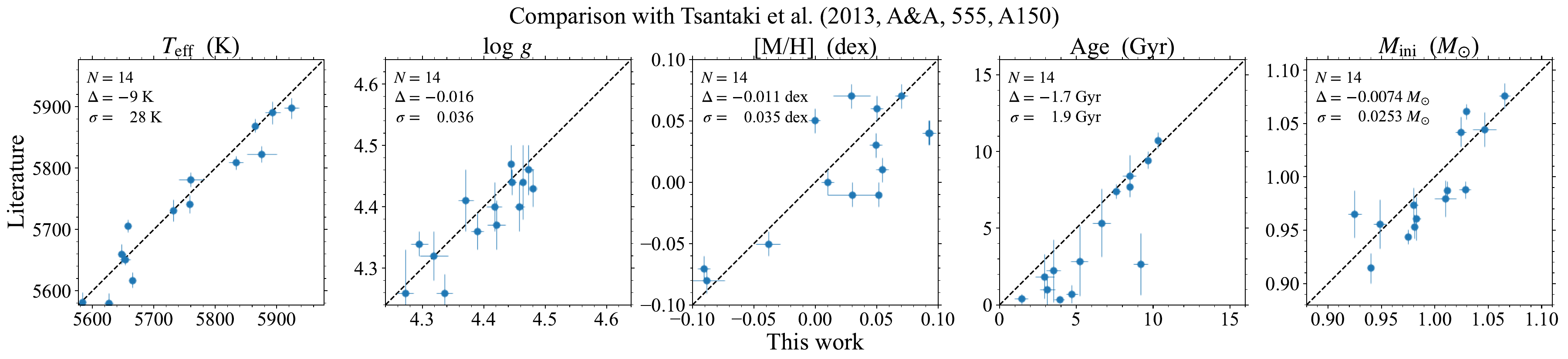}
\includegraphics[width=18cm, page=37]{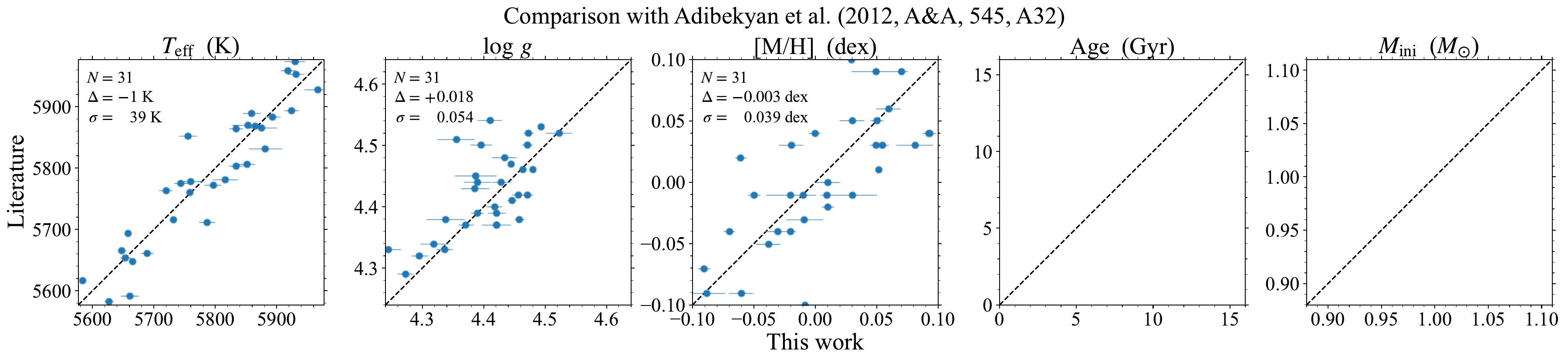}
\includegraphics[width=18cm, page=38]{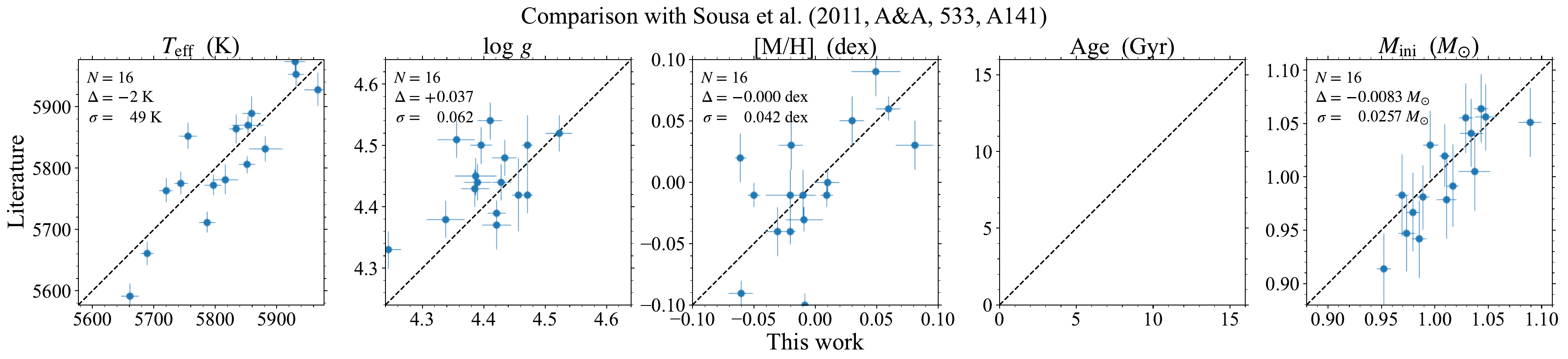}
\includegraphics[width=18cm, page=39]{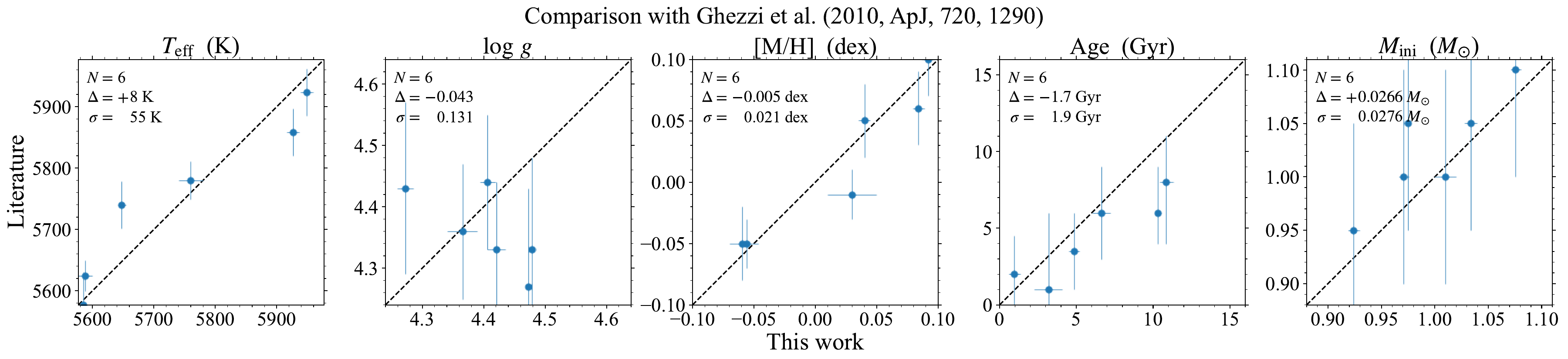}
\includegraphics[width=18cm, page=40]{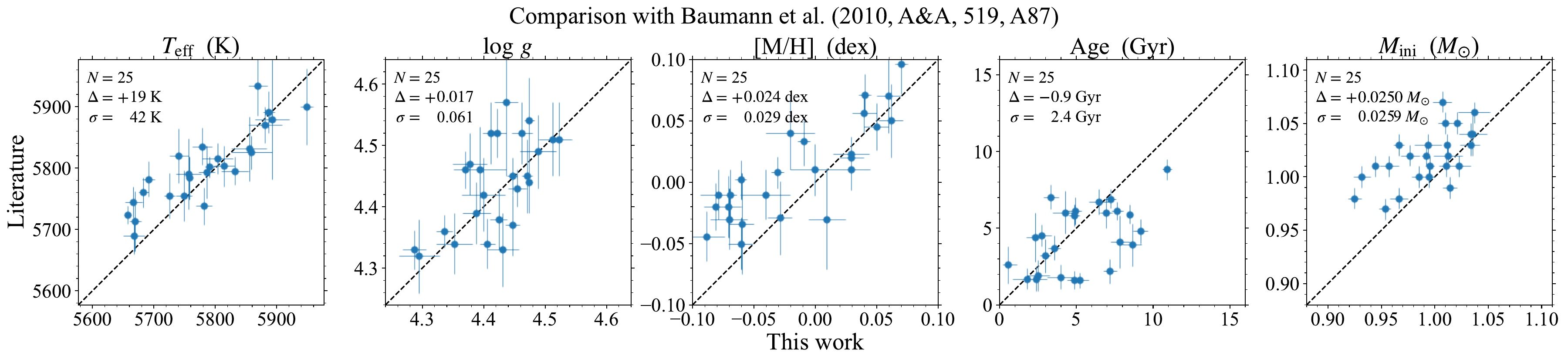}
\caption{Continued. }
\end{figure*}

\begin{figure*}
\ContinuedFloat 
\centering 
\includegraphics[width=18cm, page=41]{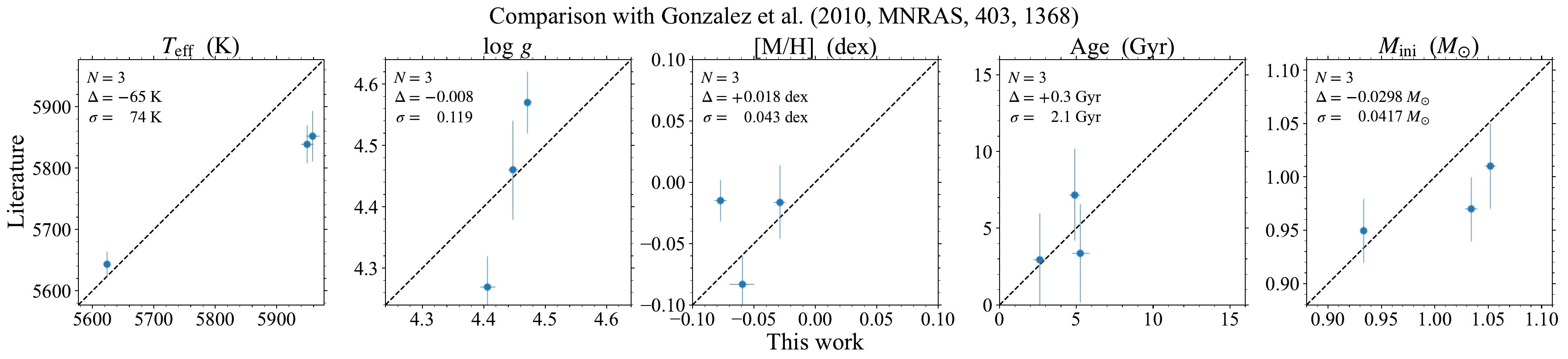}
\includegraphics[width=18cm, page=42]{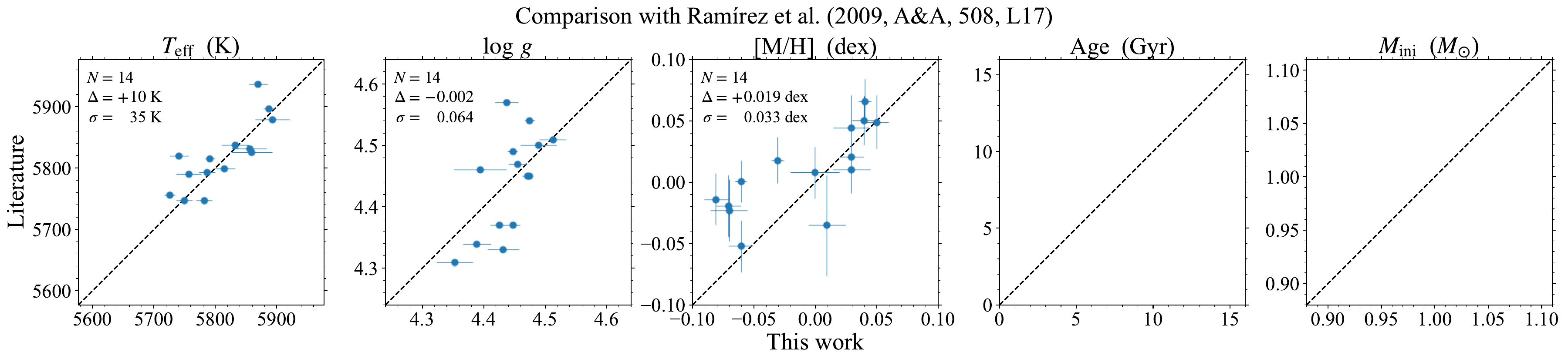}
\includegraphics[width=18cm, page=43]{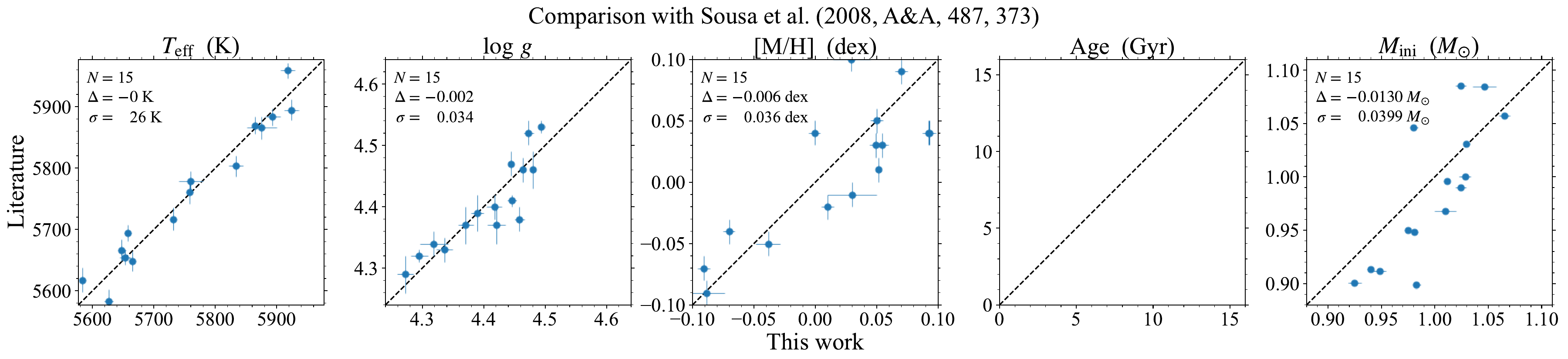}
\includegraphics[width=18cm, page=44]{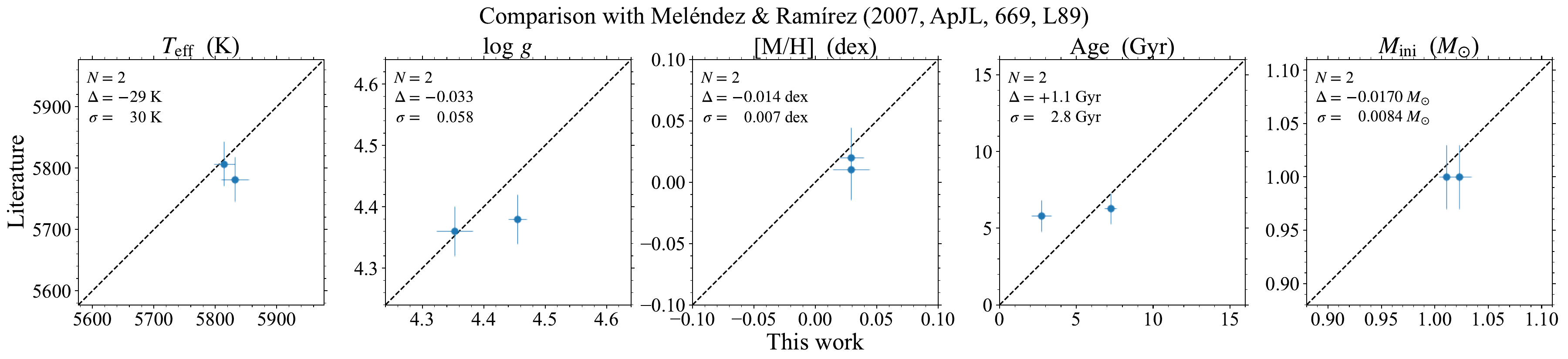}
\includegraphics[width=18cm, page=45]{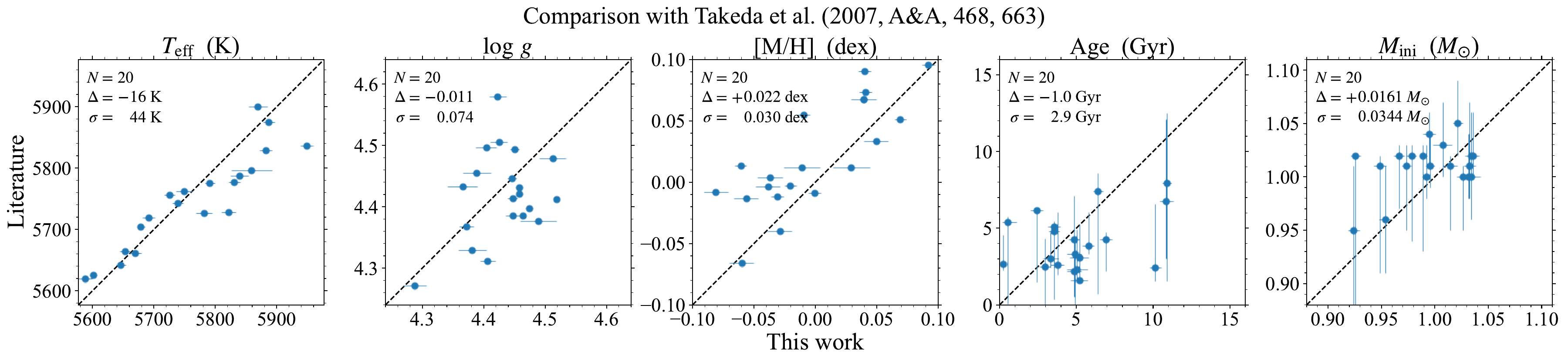}
\caption{Continued. }
\end{figure*}

\begin{figure*}
\ContinuedFloat 
\centering 
\includegraphics[width=18cm, page=46]{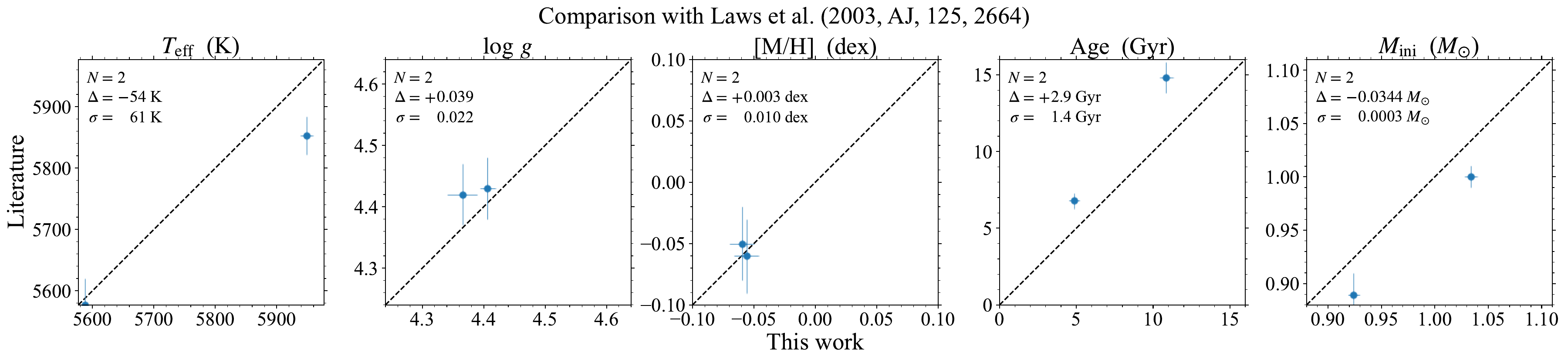}
\includegraphics[width=18cm, page=47]{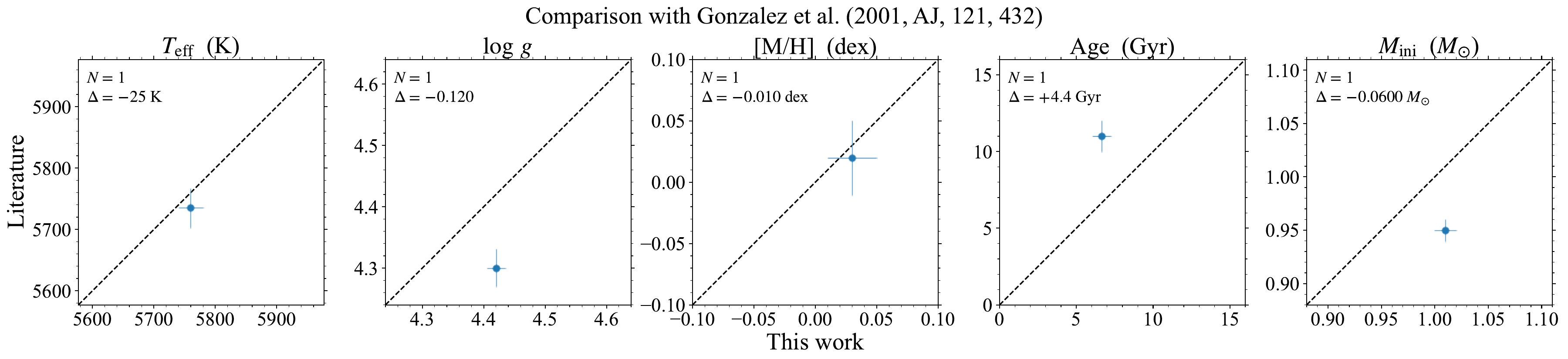}
\caption{Continued. }
\end{figure*}

\end{appendix}

\end{document}